%% file: arxiv_main.tex
\documentclass[11pt, a4paper]{article}

\newcommand{\sol}{FTCircuitBench}
\usepackage[margin=1in]{geometry} 
\usepackage[utf8]{inputenc}
\usepackage[T1]{fontenc}
\usepackage{lmodern} 
\usepackage[parfill]{parskip} 

\usepackage[affil-it]{authblk} 

\usepackage{amsmath} 
\usepackage{amssymb}    
\usepackage{bm}      
\usepackage{braket}  
\usepackage{graphicx}
\usepackage{caption}
\usepackage{subcaption}
\usepackage{booktabs} 
\usepackage{makecell, longtable}
\usepackage{tabularx}
\usepackage{tcolorbox}
\usepackage{xcolor}
\usepackage{booktabs} 

\usepackage{tikz}
\usetikzlibrary{shapes.geometric, arrows, positioning, calc, fit, decorations.pathreplacing, calligraphy, arrows.meta, shapes, patterns, decorations.pathmorphing}
\usepackage{quantikz} 

\usepackage{algorithm}
\usepackage{algorithmic}
\usepackage{listings}

\usepackage{longtable}
\usepackage{makecell}

\usepackage[colorlinks=true,
            linkcolor=blue,
            citecolor=blue,
            urlcolor=blue]{hyperref}
\usepackage{orcidlink}

\definecolor{codegreen}{rgb}{0,0.6,0}
\definecolor{codegray}{rgb}{0.5,0.5,0.5}
\definecolor{codepurple}{rgb}{0.58,0,0.82}
\definecolor{backcolour}{rgb}{0.95,0.95,0.92}

\lstdefinestyle{mystyle}{
    backgroundcolor=\color{backcolour},   
    commentstyle=\color{codegreen},
    keywordstyle=\color{magenta},
    numberstyle=\tiny\color{codegray},
    stringstyle=\color{codepurple},
    basicstyle=\ttfamily\footnotesize,
    breakatwhitespace=false,         
    breaklines=true,                 
    captionpos=b,                    
    keepspaces=true,                 
    numbers=left,                    
    numbersep=5pt,                  
    showspaces=false,                
    showstringspaces=false,
    showtabs=false,                  
    tabsize=2
}
\colorlet{colorX}{red}
\colorlet{colorZ}{blue}
\colorlet{fillX}{red!25}
\colorlet{fillZ}{blue!20}
\colorlet{capFill}{gray!30}
\colorlet{capDraw}{black}
\colorlet{contextGridColor}{gray!40}
\colorlet{activeGridColor}{black}

\title{\textbf{\sol: A Benchmark Suite for Fault-Tolerant Quantum Compilation and Architecture}}

\author[1,2]{Adrian Harkness\,\orcidlink{0009-0001-5518-6442}}
\author[1,3]{Shuwen Kan\,\orcidlink{0009-0004-0726-5260}}
\author[1]{Chenxu Liu\,\orcidlink{0000-0003-2616-3126}}
\author[1]{Meng Wang\,\orcidlink{0009-0008-1749-7929}}
\author[1,4]{John M. Martyn\,\orcidlink{0000-0002-4065-6974}}
\author[5]{Shifan Xu\,\orcidlink{0009-0005-9103-228X}}
\author[6]{Diana Chamaki\,\orcidlink{0000-0002-6374-7609}}
\author[1,7]{Ethan Decker\,\orcidlink{0009-0008-0567-9143}}
\author[3]{Ying Mao\,\orcidlink{0000-0002-4484-4892}}
\author[2]{Luis F. Zuluaga\,\orcidlink{0000-0002-3294-2401}}
\author[2]{Tamás Terlaky\,\orcidlink{0000-0003-1953-1971}}
\author[1,8]{Ang Li\,\orcidlink{0000-0003-3734-9137}}
\author[1]{Samuel Stein\,\orcidlink{0000-0002-2655-8251}}

\affil[1]{Physical and Computational Sciences, Pacific Northwest National Laboratory, Richland, WA, USA}
\affil[2]{Department of Industrial and Systems Engineering, Lehigh University, Bethlehem, PA, USA}
\affil[3]{Department of Computer and Information Sciences, Fordham University, New York, NY, USA}
\affil[4]{Harvard Quantum Initiative, Harvard University, Cambridge, MA, USA}
\affil[5]{Yale Quantum Institute, Yale University, New Haven, CT, USA}
\affil[6]{Department of Chemistry, Columbia University, New York, NY, USA}
\affil[7]{Department of Computer and Information Science, University of Pennsylvania, Philadelphia, PA, USA}
\affil[8]{Department of Electrical and Computer Engineering, University of Washington, Seattle, WA, USA}

\date{\today}

\begin{document}

\maketitle

\begin{abstract}
\noindent Realizing large-scale quantum advantage is expected to require quantum error correction (QEC), making the compilation and optimization of logical operations a critical area of research. Logical computation imposes distinct constraints and operational paradigms that differ from those of the Noisy Intermediate-Scale Quantum (NISQ) regime, motivating the continued evolution of compilation tools. Given the complexity of this emerging stack, where factors such as gate decomposition precision and computational models must be co-designed, standardized benchmarks and toolkits are valuable for evaluating progress. To support this need, we introduce \sol, which serves as: \emph{(1)} a benchmark suite of impactful quantum algorithms, featuring pre-compiled instances in both Clifford+T and Pauli Based Computation models; \emph{(2)} a modular end-to-end pipeline allowing users to compile and decompose algorithms for various fault-tolerant architectures, supporting both prebuilt and custom optimization passes; and \emph{(3)} a toolkit for evaluating the impact of algorithms and optimization across the full compilation stack, providing detailed numerical analysis at each stage. FTCircuitBench is fully open-sourced and maintained on \href{https://github.com/pnnl/FTCircuitBench}{GitHub} \footnote{https://github.com/pnnl/FTCircuitBench}.
\end{abstract}

\vspace{1em}

\newpage 

\tableofcontents

\newpage

\input{text/introduction}

\input{text/related_work}
\input{text/preliminaries}

\input{text/overview}

\input{text/metrics}

\input{text/benchmark_library}
\input{text/evaluation}

\input{text/conclusion}

\section*{Acknowledgments}
This research was supported by PNNL’s Quantum Algorithms and Architecture for Domain Science (QuAADS) Laboratory Directed Research and Development (LDRD) Initiative. This material is based upon work supported by the U.S. Department of Energy, Office of Science, National Quantum Information Science Research Centers, Quantum Science Center (QSC). The Pacific Northwest National Laboratory is operated by Battelle for the U.S. Department of Energy under Contract DE-AC05-76RL01830. This research used resources of the Oak Ridge Leadership Computing Facility (OLCF), which is a DOE Office of Science User Facility supported under Contract DE-AC05-00OR22725. This research used resources of the National Energy Research Scientific Computing Center (NERSC), a U.S. Department of Energy Office of Science User Facility located at Lawrence Berkeley National Laboratory, operated under Contract No. DE-AC02-05CH11231.

\newpage
\bibliographystyle{unsrt} 
\bibliography{refs}

\newpage
\appendix
\section*{Appendix}

\input{text/background}     
\newpage
\input{text/algorithms}     
\newpage

\section{FTCircuitBench Code Example}
\input{tables/code_example}

\newpage
\section{Clifford+T Statistics}
\label{sec:ct_stats_table}
\input{tables/ct_stats_table_alphabetical}

\newpage
\section{PBC Statistics}
\label{sec:pbc_stats_table}
\input{tables/pbc_stats_table_alphabetical}

\newpage
\section{PBC Statistics on 5-Trotter-Step Hamiltonians}
\label{sec:pbc_5trotter_stats_table}
\input{tables/pbc_5trotter_stats_table_alphabetical}

\end{document}

%% file: text/introduction.tex
\section{Introduction}
\label{sec:intro}

Quantum computing promises to revolutionize fields such as quantum chemistry \cite{mcardle2020quantum,cao2019quantum}, materials science ~\cite{DeLeon_2021, Bauer_2020, Bassman_2021}, and cryptography \cite{pirandola2020advances}. However, realizing this potential requires executing algorithms of a scale and complexity far beyond the capacity of current, noisy quantum hardware. The fragility of quantum computation poses immense challenges in demonstrating quantum advantage on problems of practical significance. Fault-tolerant quantum computation (FTQC), realized through quantum error correction (QEC) \cite{knill1997theory}, presents a promising path to resolving this challenge. By encoding logical qubits into many physical qubits, QEC offers a pathway to suppressing logical errors. Despite this promise, the practical realization of FTQC imposes immense overhead on both quantum hardware and the classical control systems. Challenges including real-time error decoding \cite{battistel2023real}, the costly execution of logical non-Clifford gates \cite{bravyi2005universal}, and increasingly challenging compilation \cite{tan2024sat,kan2025sparo} serve as a small subset of the exemplary challenges faced in scaling fault-tolerant quantum computation.

Computing with FTQC requires compiling operations that the underlying codes can implement fault-tolerantly. Two prominent approaches are Clifford+T \cite{fowler2018low} and Pauli Based Computation (PBC) \cite{yoder2025tour}. Clifford+T utilizes the Clifford gate set as well as non-Clifford T-gates to realize universal quantum computation \cite{gidney2025factor}. PBC offers an alternative paradigm, leveraging adaptive non-Clifford Pauli-product operators. The suitability of a computational model relies on the underlying code's ability to implement the logical operators with as little overhead as possible, as well as the hardware's topology. Both models represent promising directions for realizing fault-tolerant computation;  however, translating algorithms into these computational models and optimizing the resource demands remains a critical research area. Consequently, while Clifford+T and PBC are not the only universal FTQC models, they would benefit from efforts similar to those that drove advancements in domains such as compilation for superconducting hardware \cite{murali2019noise,li2019tackling} and digital quantum simulation \cite{li2022paulihedral,mukhopadhyay2023synthesizing,decker2025kernpiler, decker2025f2offlinereinforcementlearning, decker2025symbolichamiltoniancompilerhybrid}.

One challenge in FTQC algorithm exploration and optimization is the lack of standardized, accessible benchmark circuits compiled into these fault-tolerant models. Without such resources, researchers cannot easily compare different optimization strategies, evaluate compilation techniques, or assess the performance implications of architectural choices for specific algorithms. Because there are many parameters to tune throughout the compilation and execution of an algorithm, such as those pertaining to an algorithm's construction (for example, number of Trotter steps) or the precision of its approximate gate decomposition (for example, in translating between different gate sets), it is difficult to predict the ways in which these parameters affect one another. Furthermore, there is little support for an end-to-end toolkit enabling the analysis of high level algorithms compiled to various fault-tolerant computational models. 

To address this gap, we introduce \sol{}, a benchmark suite and Python toolkit comprised of quantum algorithms and their transpiled representations in both Clifford+T and Pauli Based Computation.  The toolkit also allows users to compile their own algorithms into either computational model and introduce their own circuit optimizations. Continuing upon the prior work of QASMBench \cite{li2023qasmbench} and the broader landscape of quantum benchmarking and resource-estimation frameworks surveyed in Section~\ref{sec:related_work}, our suite provides concrete
algorithm instances of various circuit complexities, chosen for their promised utility in the era of fault-tolerance, as well as a toolkit for processing these high-level abstract algorithms into low-level fault-tolerant instruction sets.  These benchmarks serve as a centralized resource, providing baseline representations suitable for understanding transpilation, optimization, resource estimation, and the evaluation of diverse fault-tolerant architectures. We include circuits representing initial, un-optimized compilations, providing a clear starting point for optimization studies. In the \sol{} Python interface, parameters such as precision and recursion allow one to compile into unoptimized Clifford+T and Pauli-based circuits, and thus investigate trade offs in factors such as gate decomposition precision and logical operator error. \sol{} is designed to easily integrate specific layer optimizations for either the benchmark  circuits, or a user-defined algorithm, thus enabling the rapid evaluation of optimization protocols at various layers of the compilation pipeline. In addition, \sol{} provides high-level metrics of compiled algorithms, such as modularity statistics for understanding clustering in circuit interaction graphs, which provide intuition into their compiled logical structure.

\sol{} is deliberately a synthesis in parts and a contribution in parts. We
build on established components rather than reinventing them---Gridsynth
\cite{10.5555/3179330.3179331} and Solovay--Kitaev \cite{dawson2005solovay}
synthesis, benchmark circuits adapted from QASMBench \cite{li2023qasmbench},
and standard graph-theoretic constructions, with the provenance of every
characterization metric labeled individually in
Section~\ref{sec:metrics}---while the contributions we claim as new are the
pre-compiled benchmark library in both the Clifford+T and PBC computational
models, the tableau-based pipeline that \emph{materializes} PBC circuits
rather than merely counting their resources, and the structural
characterization metrics themselves. In scope, \sol{} is
an engineering and infrastructure contribution rather than a theoretical
one: its analyses concern the static, logical-level outputs of compilation,
and it occupies the layer between algorithm-expression frameworks and
physical resource estimators delineated in Section~\ref{sec:related_work}.

\sol{}, comprising circuit data, sample analysis scripts, and benchmark statistics, is publicly available on \href{https://github.com/pnnl/FTCircuitBench}{GitHub} \footnote{https://github.com/pnnl/FTCircuitBench}. The remainder of this paper is organized as follows. Section \ref{sec:related_work} surveys related benchmarking and resource-estimation frameworks, and Section \ref{sec:preliminaries} summarizes the fault-tolerant computational models we target, with an extended review of quantum error correction, Clifford+T, PBC, and logical operations on surface and BB codes deferred to Appendix~\ref{sec:background}. Section \ref{sec:FTCB Overview} introduces FTCircuitBench, describing its overall design goals, software architecture, and usage, and detailing the compilation pipelines used to generate Clifford+T and PBC circuit representations. In Section \ref{sec:metrics}, we define the characterization metrics used to analyze compiled circuits, including gate-level, structural, and Pauli-weight-based statistics for both computational models. Section \ref{sec:library} summarizes the benchmark library, with detailed algorithm descriptions in Appendix~\ref{sec:algos_of_interest}. Section \ref{sec:evaluation} evaluates FTCircuitBench across these benchmarks, presenting empirical results and distilled observations on compilation behavior and resource trends. Finally, Section \ref{sec:conclusion} concludes with a summary of key findings and an outlook on future extensions of the benchmark suite.

%% file: text/related_work.tex
\section{Related Work}
\label{sec:related_work}

\subsection{Quantum Circuit Benchmark Suites}
A rich ecosystem of benchmarking frameworks has emerged and stabilized over
the NISQ era, and this body of work directly informs the design of \sol{}.
QASMBench~\cite{li2023qasmbench} provides a broad library of low-level OpenQASM
application circuits together with characterization metrics for NISQ
evaluation and simulation, and is the most direct predecessor of this work.
MQT~Bench~\cite{quetschlich2023mqtbench} supplies parameterized benchmark
circuits at multiple abstraction levels of the compilation flow, enabling
systematic evaluation of quantum software and design-automation tools.
SupermarQ~\cite{tomesh2022supermarq} defines a suite of application-centric
benchmarks with feature vectors that profile each workload and connect
benchmark structure to hardware performance. The QED-C consortium's
application-oriented benchmarks~\cite{lubinski2023application} measure
end-to-end performance of quantum computers on algorithmic workloads with
volumetric-style analyses, while Benchpress~\cite{nation2025benchmarking}
benchmarks the quantum software stack itself, timing circuit construction,
manipulation, and compilation across SDKs. Arline
Benchmarks~\cite{kharkov2022arline} similarly automates head-to-head
comparisons of NISQ compilation pipelines. Common to these frameworks is a
focus on the NISQ setting: circuits are targeted at physical-qubit gate sets
and near-term devices, and the metrics quantify near-term compiler or hardware
quality. \sol{} adopts their guiding philosophy---standardized circuit
corpora, reproducible pipelines, and workload characterization---but targets
the fault-tolerant regime, where algorithms must be expressed in
discrete fault-tolerant computational models (Clifford+T and PBC) and where
cost is dominated by non-Clifford resources and code-dependent execution
constraints rather than raw gate fidelities.

\subsection{Fault-Tolerant Resource Estimation Frameworks}
A complementary line of tooling addresses fault-tolerant resource estimation.
Qualtran~\cite{harrigan2024qualtran} expresses algorithms as hierarchical
``bloqs'' and derives resource counts (e.g., T~complexity) symbolically from
call graphs; pyLIQTR~\cite{obenland2023pyliqtr} generates circuits for
quantum-simulation workloads via block encodings and qubitization and produces
Clifford+T resource estimates; and BenchQ~\cite{benchq2024} assembles
resource-estimation pipelines developed under the DARPA Quantum Benchmarking
program. At the physical level, the Azure Quantum Resource
Estimator~\cite{beverland2022assessing} maps logical gate and measurement
\emph{counts}---extracted from a program or supplied directly---to physical
qubit numbers, runtimes, and magic-state factory requirements under
configurable qubit, QEC, and error-budget assumptions. These tools are powerful, but each leaves a gap that \sol{}
addresses. Qualtran, BenchQ, and the Azure Quantum Resource Estimator operate
on aggregate counts or symbolic expressions derived from hierarchical
decompositions; they do not materialize compiled logical circuits. pyLIQTR
goes further and does generate concrete circuits with Clifford+T
decompositions, but it targets a specific family of quantum-simulation
constructions rather than serving as a general compilation pipeline for
arbitrary input circuits, does not compile into measurement-based models such
as PBC, and reports resource counts rather than characterizing the
\emph{structure} of the circuits it produces. Across all of these frameworks,
what is missing is a standardized library of pre-compiled logical benchmark
circuits together with structure-level analysis: qubit interaction topology,
the spatial-temporal distribution of non-Clifford gates, and the Pauli-weight profile of
a PBC realization. Indeed, the estimator of
\cite{beverland2022assessing} assumes a Pauli-based execution scheme (Parallel
Synthesis Sequential Pauli Computation) in its accounting yet does not perform
the corresponding compilation, and a companion study of the tool notes that
``the tool does not (yet) analyze the qubit connectivity used in the
algorithm''~\cite{vandam2023using}.

\subsection{Positioning of \sol{}}
\sol{} occupies a layer between these bodies of work. Relative to
NISQ-era suites such as QASMBench, the new contributions are (i) pre-compiled
benchmark instances in the Clifford+T and PBC fault-tolerant computational
models, (ii) an end-to-end, parameterized compilation pipeline
(Gridsynth/Solovay-Kitaev synthesis and tableau-based PBC conversion) that
materializes those instances from arbitrary QASM inputs, and (iii)
structural characterization metrics---interaction-graph statistics,
T-gate spatial-temporal densities, and Pauli-weight distributions---aimed at
fault-tolerant execution costs. Relative to resource-estimation frameworks,
\sol{} produces the concrete compiled artifacts that count-based estimators
abstract away, and can therefore serve as both a consumer of circuits
generated by algorithm-expression tools (via their OpenQASM export paths) and
a supplier of empirically grounded logical-level inputs and structural
context to physical resource estimators. In this sense, \sol{} is
complementary to, rather than in competition with, the existing tool
landscape: it is the compilation-and-characterization layer that connects
high-level algorithm generators to physical-level resource estimation.

%% file: text/preliminaries.tex
\section{Preliminaries}
\label{sec:preliminaries}
 
This work targets two fault-tolerant computational models. In the
\emph{Clifford+T} model, algorithms are expressed over the Clifford group
supplemented by the non-Clifford $T$ gate; arbitrary single-qubit rotations
$R_z(\theta)$ are approximated to a precision $\varepsilon$ by finite
Clifford+T sequences, and each $T$ gate is executed by consuming a distilled
or cultivated magic state, making non-Clifford resources the dominant cost of
fault-tolerant execution. In \emph{Pauli Based Computation}
(PBC)~\cite{bravyi2016trading, litinski2019game}, Clifford gates are commuted
to the end of the circuit and absorbed into the final measurements, leaving
an adaptive sequence of Pauli product rotations $R_P(\pm\pi/4)$ and
multi-qubit Pauli product measurements, each non-Clifford step again
consuming a magic state. The \emph{weight} of a Pauli operator---its number
of non-identity tensor factors---and its \emph{support}---the qubits on which
it acts---are the central structural quantities of a PBC circuit.
 
How these logical operations are physically realized depends on the QEC
code: surface codes execute entangling operations and Pauli product
measurements via lattice surgery between patches~\cite{horsman2012surface,
litinski2019game}, while high-rate qLDPC codes such as the Bivariate Bicycle
family execute logical operations within densely encoded
blocks~\cite{bravyi2024high}. These implementation costs motivate several of
the characterization metrics of Section~\ref{sec:metrics}. A self-contained
review of quantum error correction, magic states, Clifford+T synthesis,
PBC compilation rules, and logical operations on surface and BB codes is
provided in Appendix~\ref{sec:background}.

%% file: text/overview.tex
\section{FTCircuitBench Overview}
\label{sec:FTCB Overview}

\begin{figure}
    \centering
    \includegraphics[width=1\linewidth]{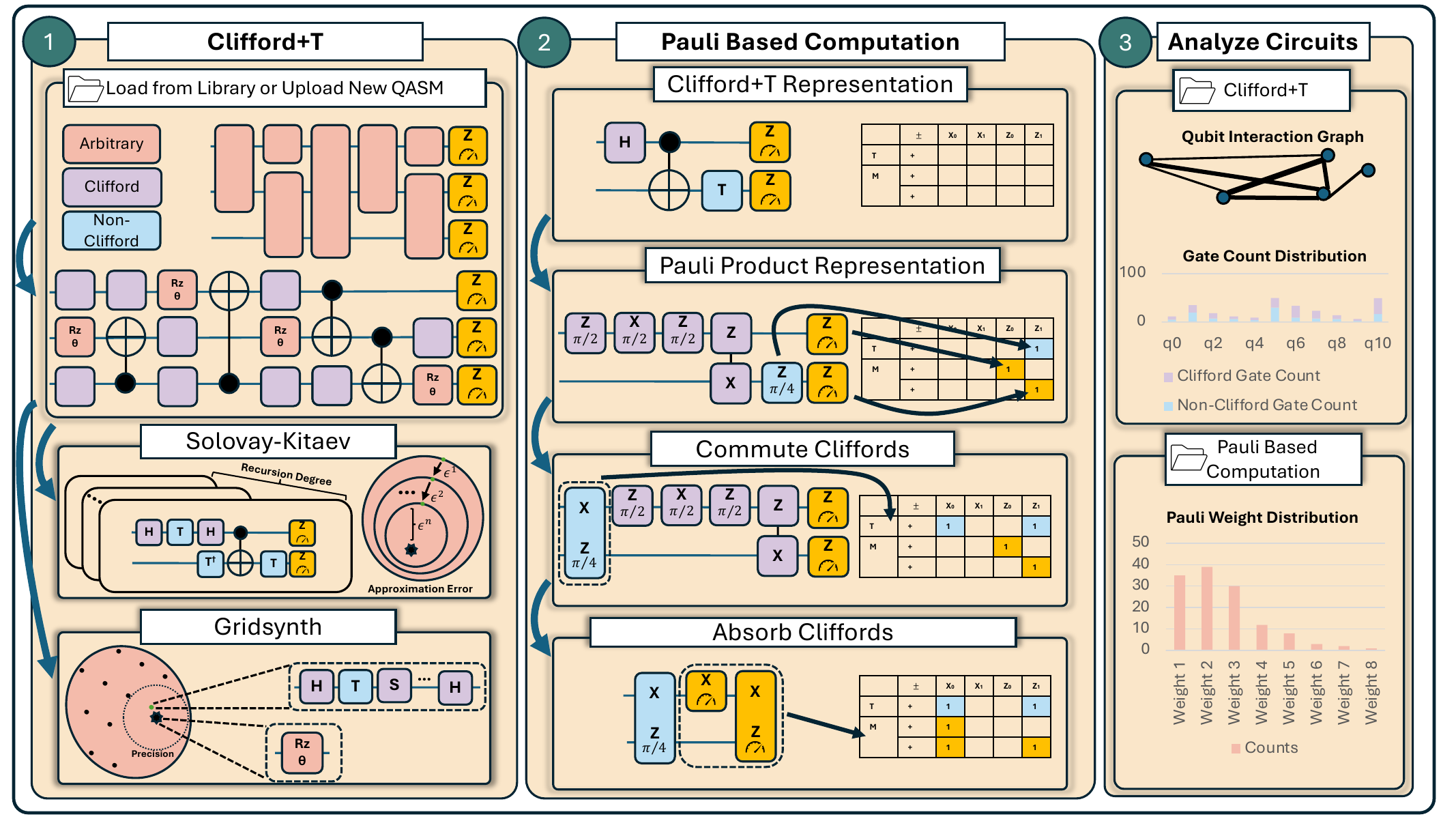}
    \caption{FTCircuitBench overview.  The pipeline begins by loading a quantum algorithm and performing an initial transpilation into a Clifford plus  $R_z(\theta)$ gate decomposition. Two synthesis pathways can be applied: Solovay-Kitaev decomposition with adjustable recursion degree and Gridsynth transpilation with adjustable precision for approximating single-qubit $R_z$ rotations in the Clifford+T basis. The resulting Clifford+T circuit can then be written as a PBC circuit by appropriately commuting all Clifford gates past the T-gates and absorbing them in the measurement basis.  Transpiled circuits and statistics on both circuit representations, as well as any optimization metrics, are then saved.}
    \label{fig:FTCB Overview}
\end{figure}

FTCircuitBench is a comprehensive benchmark suite and compilation framework designed to facilitate the study and evaluation of fault-tolerant quantum computation architectures and compilation strategies.  A main component of FTCircuitBench is a repository of quantum algorithms relevant to fault-tolerant quantum computing, which are processed through the multi-stage pipeline in Figure \ref{fig:FTCB Overview}, and evaluated over multiple characterizing statistics that capture an algorithm's low- and high- level structure. In this section, we discuss the details of the compilation pipelines and analysis tools included in FTCircuitBench.  Section \ref{sec:metrics} covers the metrics we use to analyze algorithms at different stages of the compilation pipeline, and Section~\ref{sec:library} provides an overview of the algorithms of interest included in FTCircuitBench.

Initially, input algorithms are compiled into the Clifford + $R_z(\theta)$ gate set without any additional optimization passes.  Following this, the continuous-angle $R_z$ gates are decomposed into finite sequences of Clifford and $T$ (and $T^\dagger$) gates, either according to the Gridsynth algorithm outlined in \cite{10.5555/3179330.3179331} or with the recursive  Solovay-Kitaev algorithm \cite{kitaev1997quantum, dawson2005solovay}, generating a Clifford+T circuit. This compilation stage accommodates varying precision levels (for Gridsynth) or recursion levels and base approximation depths (for Solovay-Kitaev) to decompose $R_z(\theta)$, allowing for exploration of trade-offs between gate decomposition accuracy and circuit complexity. These Clifford+$T$ circuits can be subsequently translated into PBC representations via the compilation rules outlined in Appendix~\ref{sec:pbc_model}, providing formats suitable for evaluation under different fault-tolerant execution models.

Beyond circuit generation, FTCircuitBench provides metrics and tools for analyzing both global and local features of the compiled circuits. This analysis aims to offer insights into entanglement structure, circuit complexity, and potential bottlenecks relevant to fault-tolerant execution. Metrics such as, but not limited to, T-gate spatial-temporal densities, Pauli operator weight distributions, and circuit depth are provided.

A key feature of FTCircuitBench is ease of integration into research workflows. FTCircuitBench takes algorithm inputs in QASM format, then processes these inputs iteratively, yielding the aforementioned logical circuit representations alongside resource analysis statistics.  It is important to emphasize that the primary objective of FTCircuitBench is not to provide highly optimized circuits. Instead, it aims to generate stable, unoptimized baseline circuits. These baselines serve as a consistent foundation for researchers to: (a) evaluate and compare different fault-tolerant compilation techniques, (b) assess the impact of novel optimization strategies, and (c) conduct architectural co-design studies. FTCircuitBench seeks to provide a stable baseline and seamless tool for performing these complex compilation and analysis pipelines, thereby fostering further research and development in the field.

\subsection{FTCircuitBench Structure}

\subsubsection{Clifford+T Transpilation}
Efficiently decomposing arbitrary quantum circuits into the Clifford+T basis is an active research area, with the twin goals of minimizing resource overhead (especially T-gate count) and achieving high‐fidelity decompositions.  As discussed in Appendix~\ref{sec:Rz to Clifford+T}, arbitrary single-qubit rotations must be decomposed into a finite Clifford+T gate set to enable fault-tolerant execution. FTCircuitBench implements two well-established, ancilla-free decomposition strategies—Gridsynth \cite{10.5555/3179330.3179331} and Solovay–Kitaev \cite{kitaev1997quantum, dawson2005solovay}—to support this stage of the compilation pipeline. 

\paragraph{Solovay-Kitaev}
The Solovay-Kitaev algorithm \cite{kitaev1997quantum, dawson2005solovay} efficiently approximates any single-qubit gate using a fixed finite instruction set (including, but not limited to, Clifford+$T$). FTCircuitBench includes an implementation of the Solovay–Kitaev algorithm primarily as a configurable baseline for approximate single-qubit synthesis. Users specify the recursion depth and, optionally, the base approximation sequence length, allowing exploration of trade-offs between circuit depth and approximation accuracy. While Solovay–Kitaev is neither T-optimal nor as fast as specialized methods, its generality and tunable structure make it useful for benchmarking decomposition behavior across recursion regimes.

\paragraph{Gridsynth}
Gridsynth finds a nearly optimal (i.e., minimal length) Clifford+T sequence approximating any single-qubit $R_Z(\theta)$ rotation. For a user-specified target precision $\epsilon$, it achieves T-counts of $3\log_2(1/\epsilon) +\mathcal{O}(\log\log(1/\epsilon))$ in the typical case with a $\mathcal{O}(polylog(1/\epsilon))$ expected runtime \cite{10.5555/3179330.3179331}. This makes Gridsynth a highly efficient standard for fast, high-precision single-qubit phase rotations without ancillas or measurements. However, Gridsynth is limited to single-qubit $R_z$ rotations, so arbitrary SU(2) decompositions first require a decomposition into $R_z$ rotations using Euler angles.  Furthermore, Gridsynth only decomposes into the Clifford+T gate set, which is sufficient for the FTCircuitBench pipeline. For the majority of users looking for an out-of-the-box Clifford+T transpiler to include in their workflow, the Gridsynth pipeline is the most appropriate because it is both faster in practice and obtains lower T-counts than Solovay-Kitaev for finding approximate decompositions to a given precision.

\subsubsection{Pauli Based Computation Compilation}
\label{sec:pbc_compilation}

In addition to a Clifford+T representation, FTCircuitBench provides a pipeline to convert these circuits into the PBC model \cite{litinski2019game,bravyi2005universal}. The conversion process transforms an $n$-qubit circuit of discrete Clifford and T-gates into a sequence of multi-qubit $R_P(\pi/4)$ Pauli rotations followed by a sequence of varying weight PPMs.

The compilation pipeline, detailed in Algorithm~\ref{alg:pbc-compilation}, leverages the tableau formalism of \cite{PhysRevA.70.052328, wang2025tableau} to track the evolution of Pauli operators in the Heisenberg picture. The transpilation strategy involves processing the input circuit in reverse, absorbing Clifford operations into a measurement tableau, and accumulating non-Clifford T-gates as Pauli Z rotations. These rotations can then be optimized through an iterative layering and merging process to reduce the total T-count and T-depth.

\begin{algorithm}[ht]
\caption{Clifford+T to Pauli-Based Computation (PBC) Compilation Pipeline}
\label{alg:pbc-compilation}
\begin{algorithmic}[1]
\REQUIRE $C$: a \texttt{QuantumCircuit} in $\{\texttt{cx}, \texttt{h}, \texttt{s}, \texttt{t}, \texttt{tdg}\}$
\ENSURE $pbc\_circuit$: compiled PBC circuit, and $stats$: compilation statistics

\STATE \textbf{Initialization:}
\STATE $measurement\_tableau \gets$ Z-basis for all qubits
\STATE $t\_rotation\_tableau \gets \emptyset$

\STATE \textbf{Reverse pass: absorb Cliffords}
\FORALL{$gate \in$ reverse order of $C$}
  \IF{$gate \in \{\texttt{t}, \texttt{tdg}\}$}
    \STATE Create Pauli-Z rotation $R_P(\pm \pi/4)$
    \STATE Append to $t\_rotation\_tableau$
  \ELSIF{$gate \in \{\texttt{cx}, \texttt{h}, \texttt{s}\}$}
    \STATE Apply gate to $measurement\_tableau$
    \STATE Apply gate to $t\_rotation\_tableau$
  \ENDIF
\ENDFOR

\STATE \textbf{Optimize T-rotations}
\STATE $improved \gets \textbf{true}$
\WHILE{$improved$}
  \STATE Partition $t\_rotation\_tableau$ into commuting layers $L$
  \FORALL{layer $l$ in $L$}
    \STATE Merge: $R_P(\pi/4)^2 \rightarrow R_P(\pi/2)$
    \STATE Cancel: $R_P(\pi/4) \cdot R_P(-\pi/4) \rightarrow I$
  \ENDFOR
  \STATE Rebuild $t\_rotation\_tableau$ from $L$
  \IF{no reduction in rotation count}
    \STATE $improved \gets \textbf{false}$
  \ENDIF
\ENDWHILE

\STATE \textbf{PBC circuit assembly}
\STATE $pbc\_circuit \gets \emptyset$
\STATE Append rotations from $t\_rotation\_tableau$ to $pbc\_circuit$
\STATE Append measurements from $measurement\_tableau$ to $pbc\_circuit$
\STATE \textbf{return} $(pbc\_circuit, stats)$
\end{algorithmic}
\end{algorithm}

\paragraph{Tableau Representation and Clifford Absorption} 
The process begins by representing the final computational basis measurements as a tableau of single-qubit Pauli Z operators. The input Clifford+T circuit is then traversed in reverse. As each Clifford gate is encountered, its corresponding symplectic transformation is applied to all Pauli strings in both the measurement tableau and the running T-rotation tableau. This procedure effectively pushes all Clifford operations to the end of the circuit, where they are absorbed into the final measurement basis. When a T or T$^\dagger$ gate is encountered, it is treated as a Pauli Z-rotation of angle $\pm\pi/4$ and added to a separate tableau for T rotations. After this initial pass, the original algorithm is fully described by a list of generalized Pauli rotations followed by a list of generalized Pauli measurements.

\paragraph{T-Rotation Optimization via Layering and Merging}
One optimization layer is included, similar to that in Litinski et al. \cite{litinski2019game}, and serves as a baseline optimization strategy against which users can benchmark their own compilation routines. The optimization occurs in an iterative loop designed to minimize the number of $\pi/4$ rotations. The algorithm consists of two steps:
\begin{itemize}
    \item \textbf{Layering:} First, the compiler partitions the full list of Pauli rotations into a minimal number of layers. Each layer contains a set of rotations whose Pauli operators are mutually commuting. These operations can, in principle, be executed simultaneously.
    \item \textbf{Merging:} Second, within each commuting layer, the compiler identifies and combines identical Pauli rotations. For example, two $R_P(\pi/4)$ rotations on the same Pauli operator $P$ are merged into a single Clifford-level $R_P(\pi/2)$ rotation. A rotation and its inverse, $R_P(\pi/4)$ and $R_P(-\pi/4)$, cancel each other out completely and are removed.
\end{itemize}

This process of layering and merging is repeated until no further reduction in the number of $\pi/4$ rotations is achieved.  The FTCircuitBench implementation leverages the "earliest fit" layering algorithm from \cite{wang2025tableau}.

\paragraph{Final PBC Circuit Generation}
Finally, the non-Clifford tableau is used to construct the output logical circuit. Each remaining Pauli rotation in the tableau is converted into a custom, opaque gate representing a specific multi-qubit Pauli Product rotation (e.g., $R_{XYZ}(\pi/4)$. Each Pauli rotation is written as $P_\phi = e^{-i \phi P}$, where 
$P = \bigotimes_i P_i$ with $P_i \in \{I,X,Y,Z\}$.  The modified measurement tableau is then used to generate a final layer of multi-qubit PPM operators. The resulting circuit is a direct representation of the algorithm in the PBC model, ready for analysis with resource estimators or architecture simulators that support this paradigm. These operators then operate under the conditional operational flow described in Appendix~\ref{sec:pbc_actual_compute}

\subsection{Utilizing FTCircuitBench} 
The FTCircuitBench library provides two primary interfaces for users to analyze quantum circuits through fault-tolerant compilation pipelines. The main entry point for most users is the circuit analyzer, which enables single-circuit analysis with configurable compilation parameters. Users simply provide a QASM file path and optional parameters such as compilation precision (for Gridsynth) or recursion degree (for Solovay-Kitaev), and the tool can automatically transpile the circuit to the Clifford+T basis using either Gridsynth or Solovay-Kitaev algorithms, convert it to Pauli-Based Computation (PBC) format, perform PBC optimization, and generate comprehensive statistics on circuit and optimization metrics. For researchers looking to conduct systematic benchmarking across multiple circuits and parameter settings, the benchmark generator provides automated batch processing that systematically evaluates all circuits in the library's extensive QASM collection across multiple customizable precision levels and recursion degrees, organizing results into a structured directory hierarchy with comparison summaries and metadata. 

Both scripts can be invoked directly from the command line, enabling seamless integration into shell-based workflows and automated pipelines. In addition, all core analysis routines in the circuit analyzer are exposed as regular Python functions, allowing advanced users to import and call them programmatically within their own scripts or notebooks for greater flexibility.  Users can also use the visualization functions provided in FTCircuitBench to plot circuit representations and statistical distributions as described in section \ref{sec:metrics}.  We remark that FTCircuitBench is a standalone Python library, but due to the number of conjugations that need to be performed as Clifford gates are commuted past Pauli operators during PBC transpilation, the native Python implementation can be slow for extremely large circuits. To compile circuits with hundreds of thousands or even millions of gates, FTCircuitBench supports an accelerated C++ implementation of the Clifford+T to PBC transpiler (using the same algorithm) \cite{wang2025tableau} that integrates with FTCircuitBench when it is available, and can be found at \texttt{pnnl/nwqec}\footnote{\url{https://github.com/pnnl/nwqec}}.

%% file: text/metrics.tex
\section{FTCircuitBench Characterization Metrics}
\label{sec:metrics}
Understanding the full scope of a compiled circuit's features is imperative to addressing the challenges and opportunities in algorithm optimization and execution.
Thus, in FTCircuitBench we introduce a set of characterizing metrics and accompanying visualization tools for both high- and low-level circuit analysis, helping users identify the computational model and compilation parameters that best suit their algorithm. This analysis aims to aid in the challenging problem of co-designing algorithms, compilation, error-correction, and computational models.

\input{tables/mini_stats_table}

In this section, we discuss a set of metrics and visualizations that provide quick intuition into a compiled circuit's structure. Tables \ref{tab:mini-ct} and \ref{tab:mini-pbc} are presented as representative statistics for several of the algorithms included in the FTCircuitBench library.  The same set of statistics for all of the FTCircuitBench circuits are included in Appendices \ref{sec:ct_stats_table}, 
\ref{sec:pbc_stats_table}, and \ref{sec:pbc_5trotter_stats_table}, and the full set of precomputed statistics for each algorithm and pipeline can be found in the FTCircuitBench repository.

The metrics in this section deliberately mix standard practice with
adaptations and additions specific to fault-tolerant compilation, and we
distinguish the three explicitly. Gate counts, circuit depth, and synthesis
fidelity are standard measures adopted directly from the compilation
literature~\cite{10.5555/3179330.3179331}. Characterizing benchmark circuits
through their qubit interaction graphs is likewise established practice:
interaction-graph parameters underpin the benchmark profiling of Bandi\'c
et al.~\cite{bandic2023interaction}, and SupermarQ's ``program communication''
feature is a normalized interaction-graph degree~\cite{tomesh2022supermarq}.
We therefore regard interaction-graph construction and its density and degree
statistics as standard, with one adaptation: our edge weights count
lattice-surgery-inducing operations on \emph{logical} circuits, following the
operation costs of surface-code lattice surgery~\cite{horsman2012surface,
litinski2019game}, which also motivates the easy/hard Clifford
classification. Modularity~\cite{10.5555/1809753} and Louvain community
detection~\cite{Blondel_2008} are standard network-science tools, and
partitioning interaction graphs is established within multi-core and
distributed compilation flows; reporting modularity and community counts as
characterization statistics of fault-tolerant benchmark circuits is, however,
an adaptation we have not seen elsewhere. Per-operator Pauli weight is a
standard cost quantity in the PBC and QEC compilation
literature~\cite{kan2025sparo}; our adaptation is to characterize compiled
circuits by their full weight \emph{distributions} and the evolution of those
distributions over the course of circuit optimizations. To our knowledge, three
constructions are original to \sol{} as characterization tools: PBC
interaction graphs built from the co-occurrence of qubits in Pauli-operator
supports, joint spatial-temporal T-gate density profiles (concurrent work
studies temporal-only T-demand traces for delivery-constrained
scheduling~\cite{ye2026tdepth}, underscoring the architectural relevance of
this axis), and the analogous PBC operator-density colormaps. Finally, the
optimization metrics of Section~\ref{sec:opt_metrics} (rotation-count and
average-weight reduction) are the standard objectives of PBC circuit
optimization~\cite{wang2025tableau}. Table~\ref{tab:metric_provenance}
summarizes this classification.
 
\input{tables/metric_provenance}

\subsection{Clifford+T Metrics}

\subsubsection{Fidelity}
Approximating a target unitary \(U\) using only Clifford and \(T\) gates can be carried out with either the Gridsynth or the Solovay-Kitaev algorithm. By specifying an approximation precision \(\varepsilon\), these methods yield a gate sequence \(\widetilde U\) satisfying $
\|U - \widetilde U\| \le \varepsilon,
$
which in turn guarantees a worst-case entanglement fidelity of
\begin{equation}
F = \frac{1}{d^2}\bigl|\mathrm{Tr}\,(U^\dagger \widetilde U)\bigr|
\;\ge\; 1 -\mathcal{O}(\varepsilon).
\end{equation}
However, calculating \(U\) and \(\widetilde U\) is computationally intractable for large circuits. As a scalable alternative for circuits over a user-defined maximum size, we can lower bound the true fidelity by calculating 
$
\tilde{F} = \prod_i F_i
$,
 where $F_i$ is the fidelity of each individual $R_z(\theta)$ decomposition in the circuit, performed using either Gridsynth or Solovay-Kitaev in their respective pipelines.  As each of these are single qubit gates, the cost of calculating $F_i$ no longer scales with the dimension of the Hilbert space.  

\subsubsection{Clifford Gate Counts}
The Clifford gate set \(\{H, S, \mathrm{CNOT}, X, Z\}\) has non‐uniform resource costs for different error correction codes and different physical platform. Specifically for the surface code, \(H\), \(X\), and \(Z\) gates can be applied fault-tolerantly with low overhead in the surface code, whereas \(S\) and \(\mathrm{CNOT}\) gates are performed with lattice surgery. Accordingly, we classify \(H\), \(X\), and \(Z\) as “easy” Cliffords and \(S\) and \(\mathrm{CNOT}\) as “hard” Cliffords. For each quantum algorithm being analyzed, we report the count of each gate type  as well as the overall gate count (Clifford + \(T\)).  We also report the overall circuit depth.  These metrics together provide a comprehensive, global view of the resource requirements for Clifford + \(T\) decomposition to a given precision or recursion degree.

\subsubsection{Interaction Graphs}
We can quantify the entanglement structure of a Clifford+T circuit by using interaction graphs to represent the circuits.  We present an algorithm's structure as a weighted graph $G$, with vertices and edges $\{V,E\}$ where each vertex $V$ represents a logical qubit. Two vertices share an edge if they interact at any point in the circuit.  When two qubits $\{V_i,V_j\}$ interact via a CNOT --- i.e. requiring lattice surgery --- we increase the edge weight $w_{ij}$ between them by $1$.  With interaction graphs, a set of entanglement related metrics are:
\begin{itemize}
    \item  \textbf{Modularity} measures how well a graph can be divided into “communities”, such that there are many edges within each community and few edges between different communities.  The modularity of a graph is defined in \cite{10.5555/1809753} as
    \begin{equation}
    Q = \sum_{c=1}^{n}
        \left[ \frac{L_c}{m} - \left( \frac{k_c}{2m} \right) ^2 \right]
    \end{equation}
    where the sum is over all communities $c$, $m$ is the number of edges,
    $L_c$ is the number of intra-community links for community $c$, and
    $k_c$ is the sum of degrees of the nodes in community $c$. A graph with high‐modularity (close to 1) can be partitioned such that intra‐community connections are significantly denser than what would be expected in a random graph, while a weak modularity (close to 0) reveals mild community structure. On interaction graphs specifically, high modularity indicates "modules" that can be compiled, scheduled, or even mapped to hardware tiles largely independently, with fewer inter-module operations.  Low modularity indicates strongly non-local entangling structure.  In FTCircuitBench we use the Louvain community detection algorithm \cite{Blondel_2008} to find the modularity-maximizing partitioning of a graph.
    \item The \textbf{number of communities} describes how many communities the modularity algorithm detected.  Together with the modularity strength, this can inform the mapping of qubits from algorithm level to hardware level, either on single quantum processors or even on distributed quantum computers.
    \item \textbf{Graph density} is a measure of normalized 2-qubit gate count.  We define it as 
    \begin{equation}
    \text{Density}(G) = \frac{1}{M_{\max}} \displaystyle \sum_{(u,v)\in E} w_{uv}
    \end{equation}
    where \( n = |V| \) is the number of nodes, \( E \) is the set of edges, \( w_{uv} \) is the weight of edge \((u,v)\), and 
    \begin{equation}
    M_{\max} = \frac{n(n-1)}{2}
    \end{equation}
    is the number of edges in a fully-connected unweighted graph with \( n \) nodes. Note that because the edges are weighted to reflect operation counts, the graph density is not upper-bounded by 1, differing from the standard unweighted definition. Since lattice surgery is a relatively expensive operation within surface code fault-tolerant quantum computation, the graph density of the interaction graph can be a good metric to minimize for compilers focusing on CNOT gate reduction and, correspondingly, lattice surgery reduction.
    \item The \textbf{interaction degree} of a node $u$ is the sum of the weights of all edges incident to $u$:
    \begin{equation}
    \mathrm{ID}(u) = \sum_{\substack{v \\ (u,v)\in E}} w_{uv}.
    \end{equation}
    Logical qubits with high interaction degrees function as "hubs" that may benefit from mapping to well-connected or central logical qubits.  The standard deviation of the distribution of interaction degrees over all nodes in the interaction graph can also be used to measure the expected performance gain from smart qubit placement. In particular, circuits with distributions of high standard deviation would generally stand to benefit more from intelligent placement.
\end{itemize}

\subsubsection{T gate Statistics}
T gate statistics quantify the temporal and spatial patterns of T-gate demand.  Since every T-gate consumes a magic-state resource and incurs post-correction latency while awaiting syndrome decoding, we log its precise timestamp and the logical qubit it targets. From this data source, it is possible to derive timing-pattern metrics such as inter-T intervals and peak concurrency, which can reveal tight “T-bursts” that could stall computation due to limited rates of magic state generation. Alternatively, one can calculate per-qubit T-gate statistics (mean, variance, and maximum counts) to map the spatial distribution of magic state demand. These measures expose both the density of T-induced decoding delays and their spread across logical qubits, thereby defining the throughput requirements for magic-state factories in the architecture design.

\subsection{Pauli Based Computation Metrics}

Once a circuit is transpiled from Clifford+T to PBC form, we can further analyze the structure of the algorithm to compare algorithm implementations across the different computational models, analyze Pauli rotation reduction strategies, and gain insight into the Pauli weight statistics of a particular algorithm. Recall that the weight of an n-qubit Pauli operator is defined as the number of non-identity Pauli terms in the operator.  Furthermore, the support of a Pauli operator are the qubit indices for which the operator is non-identity.

\subsubsection{Interaction Graphs}
Mirroring the analysis of Clifford+T circuits, we can form interaction graphs for PBC circuits to understand their entanglement structure.  For a PBC circuit interaction graph $G=\{V,E\}$, each logical qubit is a vertex $V_i$, and an edge $E_{ij}$ exists if qubit $i$ and qubit $j$ are simultaneous supports for any Pauli operator in the circuit.  Each Pauli operator increases edge weight $w_{ij}$ by one for every pair of distinct indices $(i,j)$, such that $i<j$ (to prevent double counting) in the support of the operator.  Because transpiling to PBC commutes out all Clifford gates at the expense of higher-weight operations that simultaneously couple large sets of qubits, the PBC interaction graph is denser than the analogous Clifford+T graph, exhibiting higher weight edges that directly reflect its richer, more global entanglement structure. Statistics such as graph modularity and number of communities can be computed for PBC interaction graphs as they are from Clifford+T graphs.  The distribution of interaction degrees over all logical qubits can also be readily visualized with FTCircuitBench, representing the number of PBC operations each qubit partakes in.  As with Clifford+T circuits, this can inform the mapping of logical qubits to hardware, as well as measure the expected gain from applying a smart mapping strategy rather than a random one.

\subsubsection{Pauli Weight Statistics}
Pauli weight statistics, representing the distribution of Pauli weights
throughout the rotation and measurement operators comprising a PBC circuit,
can be illuminating for many reasons.  Recall that operator weights generally
increase as CNOT gates are commuted past them and absorbed into the
measurement basis during PBC compilation. The resulting Pauli operators with
high weights entangle qubits, so the distribution of Pauli weights, and in
particular the rate of increase of Pauli weights as the circuit progresses,
shows the growth of entanglement and the delocalization of quantum
information.  Pauli weight distributions also show the operational complexity
of running a circuit on hardware, as high-weight Pauli operators can be
prohibitively difficult to execute.
 
Precisely how difficult depends on the underlying code and architecture: the
fault-tolerant implementation of a high-weight Pauli measurement is not an
elementary operation for many QEC schemes, and the cost of executing a Pauli
operator is a function of its weight, its support geometry, and the Pauli
word itself, with a \emph{shape} that differs qualitatively across schemes.
On surface-code lattice surgery, a weight-$w$ measurement merges an ancilla
region spanning all $w$ participating patches: the logical error incurred
grows with the size of the merged region, parallelism is limited by
routing-region disjointness, and $Y$-type support incurs additional cost via
twist defects~\cite{horsman2012surface, litinski2019game, litinski2018lattice}.
On BB codes, by contrast, logical error rates are largely weight-independent
for operators confined to a single code block, with costs instead determined
by whether the required string is reachable through native automorphisms and
whether it spans multiple blocks~\cite{bravyi2024high, yoder2025tour}.
Extractor-based qLDPC architectures flatten the weight-cost curve
entirely---any logical Pauli is measurable in one logical cycle---at the
price of $\widetilde{O}(n)$ additional hardware, leaving support disjointness
as the limit on parallel measurement throughput~\cite{he2025extractors}.
Platforms with long-range connectivity, such as reconfigurable atom arrays,
further relax the dependence on support geometry~\cite{xu2024constant,
zhou2025low}. Count-based resource estimators, which charge one time step
per sequential Pauli measurement regardless of
weight~\cite{beverland2022assessing}, cannot distinguish these regimes. It is for these reasons that the reduction of Pauli operator weights during PBC compilation is an active area of research~\cite{kan2025sparo}. The weight distribution of a compiled circuit thus becomes an architectural diagnostic: the same circuit can be cheap on one code and prohibitive on another, and the distribution reveals which.
 
FTCircuitBench makes these distributions directly accessible, providing
histograms of operator weights for PBC circuits together with summary
statistics.  These can provide intuition about the difficulty of executing an
algorithm in the PBC model.  These can also be used to understand the
performance of a PBC optimization algorithm, as they can visualize the change
in the Pauli weight distribution before and after PBC circuit optimization,
whether through layering-and-merging or through a custom algorithm.
FTCircuitBench can also produce PBC operator density colormaps to visualize
the spacial-temporal distribution of PBC operator supports.  Like the
T-density colormaps, sequential layers are grouped into bins to better
display the temporal grouping of PBC supports on the same qubit.

\subsubsection{Optimization Metrics}
\label{sec:opt_metrics}
A primary goal of FTCircuitBench is to provide a library of unoptimized circuits and corresponding statistics to use for baseline comparisons in fault-tolerant compiler research.  However, we include a basic layering-and-merging algorithm as outlined in Section \ref{sec:pbc_compilation} and described in \cite{wang2025tableau} to compare against other compilers, and to use for illustrating the tradeoffs in objectives that need to be balanced during PBC compilation.  

To measure the impact of our algorithm, as well as any other PBC optimization algorithm, we focus on two metrics: \textbf{Pauli rotation count reduction} and \textbf{average Pauli weight reduction}. Because it is resource-intensive to execute deep circuits with many rotation operations and implement high-weight Pauli operations, minimizing these metrics is desirable, and different algorithms will exhibit tradeoffs between the two.

%% file: tables/mini_stats_table.tex
\begingroup
\scriptsize
\setlength{\tabcolsep}{3pt}

\begin{table*}[t]
\centering
\resizebox{\textwidth}{!}{%
\begin{tabular}{|l|r|r|r|r|r|r|r|r|}
\hline
\makecell{Algorithm \& Pipeline} &
\makecell{Total\\Gates} &
\makecell{Depth} &
\makecell{Clifford\\Gates} &
\makecell{$T$ \& $T^{\dagger}$\\Gates} &
\makecell{Graph\\Density} &
\makecell{Avg $\pm$ Std\\Degree} &
\makecell{Graph\\Modularity} &
\makecell{Number of\\Communities} \\
\hline
        adder-64q-gs-8 & 988 & 369 & 596 & 392 & 0.23 & 2.84 ± 0.91 & 0.84 & 7 \\ \hline        
        qft-29q-gs-8 & 159561 & 22664 & 97377 & 62184 & 2 & 28.00 ± 0.00 & 0 & 1 \\ \hline
        hhl-21q-gs-8 & 4657927 & 3243302 & 2851947 & 1805980 & 157.54 & 19.62 ± 0.90 & 0 & 1 \\ \hline
        fermi-hubbard-1d-72q-gs-8 & 2346060 & 1452260 & 1541180 & 804880 & 14.68 & 4.06 ± 1.32 & 0.69 & 9 \\ \hline
        heisenberg-1d-100q-gs-8 & 2847660 & 2703660 & 1881300 & 966360 & 4.84 & 2.00 ± 0.00 & 0.8 & 10 \\ \hline
\end{tabular}%
}
\normalsize
\caption{Clifford+T circuit statistics for several algorithms compiled using Gridsynth to precision $\epsilon=10^{-8}$.}
\label{tab:mini-ct}
\end{table*}
\endgroup

\begingroup
\scriptsize
\setlength{\tabcolsep}{3pt}

\begin{table*}[t]
\centering
\resizebox{\textwidth}{!}{%
\begin{tabular}{|l|r|r|r|r|r|r|r|r|r|}
\hline
\makecell{Algorithm \& Pipeline} &
\makecell{Raw\\Rotations} &
\makecell{Optimized\\Rotations} &
\makecell{Rotation\\Reduction} &
\makecell{Raw Avg $\pm$ Std\\Pauli\\Weight} &
\makecell{Optimized Avg\\$\pm$ Std Pauli\\Weight} &
\makecell{Avg Pauli\\Weight\\Reduction} &
\makecell{Avg $\pm$ Std\\Degree} &
\makecell{Graph\\Modularity} &
\makecell{Number of\\Communities} \\
\hline
        adder-64q-gs-8 & 392 & 224 & 42.86\% & 4.33 ± 2.75 & 5.60 ± 2.73 & -29.44\% & 10.41 ± 2.60 & 0.78 & 7 \\ \hline        
        qft-29q-gs-8 & 62184 & 61912 & 0.44\% & 12.62 ± 6.29 & 12.82 ± 6.42 & -1.60\% & 28.00 ± 0.00 & 0.03 & 2 \\ \hline
        hhl-21q-gs-8 & 1805980 & 1785138 & 1.15\% & 12.42 ± 4.59 & 12.40 ± 4.48 & 0.19\% & 20.00 ± 0.00 & 0 & 2 \\ \hline
        fermi-hubbard-1d-72q-gs-8 & 804880 & 803360 & 0.19\% & 52.72 ± 7.55 & 52.75 ± 7.51 & -0.06\% & 71.00 ± 0.00 & 0 & 1 \\ \hline
        heisenberg-1d-100q-gs-8 & 966360 & 966360 & 0.00\% & 63.55 ± 33.30 & 63.55 ± 33.30 & 0.00\% & 99.00 ± 0.00 & 0.12 & 2 \\ \hline
\end{tabular}%
}
\normalsize
\caption{PBC circuit statistics for several algorithms compiled using Gridsynth to precision $\epsilon=10^{-8}$, followed by PBC transpilation.  Interaction graph statistics reflect the interaction graph of the PBC circuit post light optimization.}
\label{tab:mini-pbc}
\end{table*}
\endgroup

%% file: tables/metric_provenance.tex
\begin{table}[htbp]
\centering
\begin{tabular}{lll}
\toprule
\textbf{Metric} & \textbf{Provenance} & \textbf{Basis} \\
\midrule
Gate counts, circuit depth        & Standard & Compilation literature \\
Synthesis fidelity ($F$, $\tilde F$)  & Standard & \cite{10.5555/3179330.3179331} \\
Interaction graphs; density, degrees & Standard & \cite{bandic2023interaction, tomesh2022supermarq, li2019tackling} \\
Lattice-surgery edge weighting; easy/hard Cliffords & Adapted & \cite{horsman2012surface, litinski2019game} \\
Modularity, number of communities & Adapted  & \cite{10.5555/1809753, Blondel_2008} \\
Pauli operator weight             & Standard & \cite{kan2025sparo} \\
Pauli weight distributions over circuits & Adapted & This work \\
T-gate spatial-temporal density   & Original & Cf.\ temporal-only \cite{ye2026tdepth} \\
PBC interaction graphs            & Original & This work \\
PBC operator density colormaps    & Original & This work \\
Rotation/weight reduction metrics & Standard & \cite{wang2025tableau, kan2025sparo} \\
\bottomrule
\end{tabular}
\caption{Provenance of the characterization metrics in \sol{}. ``Standard''
metrics are adopted from prior literature; ``adapted'' metrics apply standard
constructions to fault-tolerant compilation in a new way; ``original'' metrics
are, to our knowledge, introduced by \sol{} as characterization tools.}
\label{tab:metric_provenance}
\end{table}

%% file: text/benchmark_library.tex
\section{The Benchmark Library}
\label{sec:library}

The \sol{} library spans three classes of workloads chosen for their expected
impact in the fault-tolerant era: Hamiltonian simulation of canonical
condensed-matter and electronic-structure models, the arithmetic and Fourier
subroutines that recur across larger algorithms, and composite algorithms
(phase estimation, linear-systems solvers, and the quantum singular value
transformation) that exercise deep, structured circuits.
Table~\ref{tab:library} summarizes the library; each algorithm is provided at
multiple sizes and compiled through all pipelines of
Section~\ref{sec:FTCB Overview}. Appendix~\ref{sec:algos_of_interest}
describes each algorithm class, its physical motivation, and its
construction in detail, and Section~\ref{sec:evaluation} states the exact
generation parameters used for the released circuit instances.

\begin{table}[htbp]
\centering
\begin{tabular}{lll}
\toprule
\textbf{Algorithm} & \textbf{Qubits} & \textbf{Construction / source} \\
\midrule
Adder            & 4--64    & Ripple-carry design~\cite{cuccaro2004newquantumripplecarryaddition}, from QASMBench~\cite{li2023qasmbench} \\
QFT              & 4--63    & Design of~\cite{coppersmith2002approximatefouriertransformuseful}, from QASMBench~\cite{li2023qasmbench} \\
Ising model      & 9--100   & 1D, 2D, and 2D-triangular lattices; $J=1$, $h=0.5$ \\
Heisenberg model & 9--100   & 1D, 2D, and 2D-triangular lattices; $J_x{=}J_y{=}J_z{=}1$ \\
Fermi--Hubbard   & 18--128  & 1D, 2D, and 2D-triangular lattices; $U=0$, $t=1$ \\
QPE              & 8--12    & H$_2$ (cc-pVDZ, 6 orbitals) and 4-/5-site Hubbard models \\
HHL              & 4--21    & Implementation of~\cite{11250003}; systems of 2--16 variables \\
QSVT             & 6--13    & Matrix inversion of banded circulant block encodings~\cite{camps2024explicit} \\
\bottomrule
\end{tabular}
\caption{The \sol{} benchmark library. Hamiltonian-simulation circuits are
Trotterized time evolutions provided at both 20 and 5 Trotter steps; all
circuits are released in Clifford+$R_z$, Clifford+T (Gridsynth and
Solovay-Kitaev at two settings each), and PBC representations.}
\label{tab:library}
\end{table}

\paragraph{Extending the library.}
The library is not limited to the algorithms above: \sol{} accepts any
OpenQASM~2.0 circuit, which provides a direct ingestion path from
algorithm-generation frameworks. In particular, pyLIQTR generates concrete
circuits for quantum-simulation workloads (block encodings, qubitization,
and QSP-based constructions) as Cirq objects~\cite{obenland2023pyliqtr},
Qualtran bloqs can be lowered to Cirq
circuits~\cite{harrigan2024qualtran}, and Cirq exports OpenQASM~2.0, so
circuits from either framework can be compiled and characterized through
the full \sol{} pipeline once decomposed to standard gates. The repository
documentation includes a worked recipe for this path. Curating a set of
such generated circuits as first-class library entries, providing
pre-compiled instances of qubitization-based simulation at scale, is a
priority for the next release of the benchmark library.

%% file: text/evaluation.tex
\section{Evaluation}
\label{sec:evaluation}
In the following section, we discuss the parameters used to generate the circuits that realize the algorithms of the benchmark library, together with our evaluation of these circuits using the metrics proposed in Section \ref{sec:metrics}.
\subsection{Circuit Parameters}
\label{sec:circuit_params}
Here, we discuss the specific parameters used to generate the circuits we include in the FTCircuitBench QASM library.  Unless otherwise specified, all statistics and plots in the following section are for circuits compiled with the Gridsynth pipeline to precision $\epsilon=10^{-8}$.

Each circuit in the library is released under four synthesis pipelines,
labeled sk-1 and sk-2 (Solovay-Kitaev at recursion levels 1 and 2) and gs-5
and gs-8 (Gridsynth at precisions $\epsilon = 10^{-5}$ and $10^{-8}$), and
every Hamiltonian-simulation circuit is additionally provided at both 20 and
5 Trotter steps. Full statistics for every circuit and pipeline are
tabulated in Appendices \ref{sec:ct_stats_table},
\ref{sec:pbc_stats_table}, and \ref{sec:pbc_5trotter_stats_table}, with
additional data in the code repository.

\subsubsection{Quantum Simulation}
Hamiltonians have different properties depending on the chosen values for their underlying parameters. However, the complexity of quantum algorithms is generally insensitive to these parameters.

Nonetheless, here we state the exact parameters chosen for our algorithms. 
Our Heisenberg models are anti-ferromagnetic with $J_x=J_y=J_z=1$. 
Our Ising models were compiled with $J=1$ and $h=0.5$. 
These are canonical parameters for magnetic phases of matter within these models. 

For the Fermi Hubbard model, our onsite interaction term was turned off with $U=0$ and $t=1$. 
Our justification is that compilation benchmarks should include circuits of realistic gate complexity but also be verifiable. 
Under some fermionic mappings, hopping terms are high-weight Pauli strings, making them complex for compilation while onsite interaction terms are usually low-weight Z rotations. 
This allows for complex circuits representing physical phenomena, which can be compiled and verified efficiently. 

The Hamiltonians were compiled with 20 Trotter steps, corresponding to a time step of $\Delta t=0.05$. While the Trotter size is relatively large, the resulting circuits remain valuable for characterizing the structural complexity of early fault-tolerant digital quantum simulation.

\subsubsection{Quantum Arithmetic}
The Adder circuits in FTCircuitBench are selected from the QASMBench library \cite{li2023qasmbench} and use a ripple-carry design \cite{cuccaro2004newquantumripplecarryaddition}.

\subsubsection{Quantum Fourier Transform}
The QFT circuits in FTCircuitBench are selected from the QASMBench library \cite{li2023qasmbench} and use the design from \cite{coppersmith2002approximatefouriertransformuseful}.

\subsubsection{Quantum Phase Estimation}
The QPE circuits were generated in Qiskit, using the exact ground state for the initial state preparation. 
In this work we considered QPE circuits of simple systems, namely the H$_2$ molecule and the Hubbard model.
For H$_2$ we used the cc-pVDZ basis set, truncated to 6 spatial orbitals with bond lengths of $0.6$, $0.74$, $1.0$, and $1.5$ Angstroms.
Additionally, we used the 4- and 5-site Hubbard models with the on-site term set to $U=5$ and the hopping term set to $t=1$.

\subsubsection{Quantum Linear Systems Solvers}
We use the HHL implementation referenced in \cite{11250003} and found at \texttt{QCOL-LU/QLSAs}\footnote{\url{https://github.com/QCOL-LU/QLSAs}}. For problem data, we use randomly-generated (unstructured) coefficient matrices $A \in \mathbb{R}^{N\times N}$ and random vectors $\Vec{b} \in \mathbb{R}^N$. For the quantum phase estimation subroutine, we use the same number of phase estimation qubits as the dimension of $\Vec{b}$, leading to circuit widths of $N+\log_2(N) + 1$.  While using $\log_2(N)$ QPE qubits is more standard for large problem sizes, using additional qubits leads to higher precision estimation of the eigenvalues of $A$, making these circuits appropriate for the problems we include for solving linear systems of 2, 4, 8, and 16 variables.

\subsubsection{Quantum Singular Value Transformation}
In this collection, we focus on linear systems solvers (i.e., matrix inversion) as a canonical example. We take as the coefficient matrix a banded circulant matrix $A \in \mathbb{R}^{N \times N}$ whose first row is 
\[
(\alpha, \gamma, 0, \dots, 0, \beta),
\]
for parameters $\alpha, \beta, \gamma$. Each subsequent row obtained by a right cyclic shift of the previous one. For example, when $N=8$, the matrix $A$ takes the explicit form
\[
A =
\begin{pmatrix}
\alpha & \gamma & 0      & 0      & 0      & 0      & 0      & \beta \\
\beta  & \alpha & \gamma & 0      & 0      & 0      & 0      & 0     \\
0      & \beta  & \alpha & \gamma & 0      & 0      & 0      & 0     \\
0      & 0      & \beta  & \alpha & \gamma & 0      & 0      & 0     \\
0      & 0      & 0      & \beta  & \alpha & \gamma & 0      & 0     \\
0      & 0      & 0      & 0      & \beta  & \alpha & \gamma & 0     \\
0      & 0      & 0      & 0      & 0      & \beta  & \alpha & \gamma \\
\gamma & 0      & 0      & 0      & 0      & 0      & \beta  & \alpha
\end{pmatrix}.
\]

We select $\alpha = 3, \ \beta = -1, \ \gamma = -1$, in which case $A$ has eigenvalues  
\[
\lambda_k = 3 - 2\cos\left(\frac{2\pi k}{N}\right)
\]
for $k\in\{0,1,\dots,N-1\}$ and condition number $\kappa=\lambda_{\text{max}}/\lambda_{\text{min}}=5$. An explicit block encoding for $A$ is given in~\cite{camps2024explicit} to prepare $A/4$. The overall circuit width is $\log_2(N)+4$ qubits, with 3 ancillas for block encoding and 1 ancilla for performing $R_z$ rotations used in QSVT. The circuit parameters that produce the polynomial needed for matrix inversion via QSVT are calculated using the \texttt{pyqsp} package\footnote{\url{https://github.com/ichuang/pyqsp}}. In this instance, QSVT achieves a query complexity that scales with $\kappa$ and $\epsilon$ as $\mathcal{O}(\kappa \log(\kappa/\epsilon))$, in contrast to the HHL algorithm’s scaling $\mathcal{O}(\kappa^2/\epsilon)$.

\subsection{Clifford+T Metrics}

\subsubsection{Gate Counts}
\begin{figure}[ht]
    \centering
    \includegraphics[width=1 \linewidth]{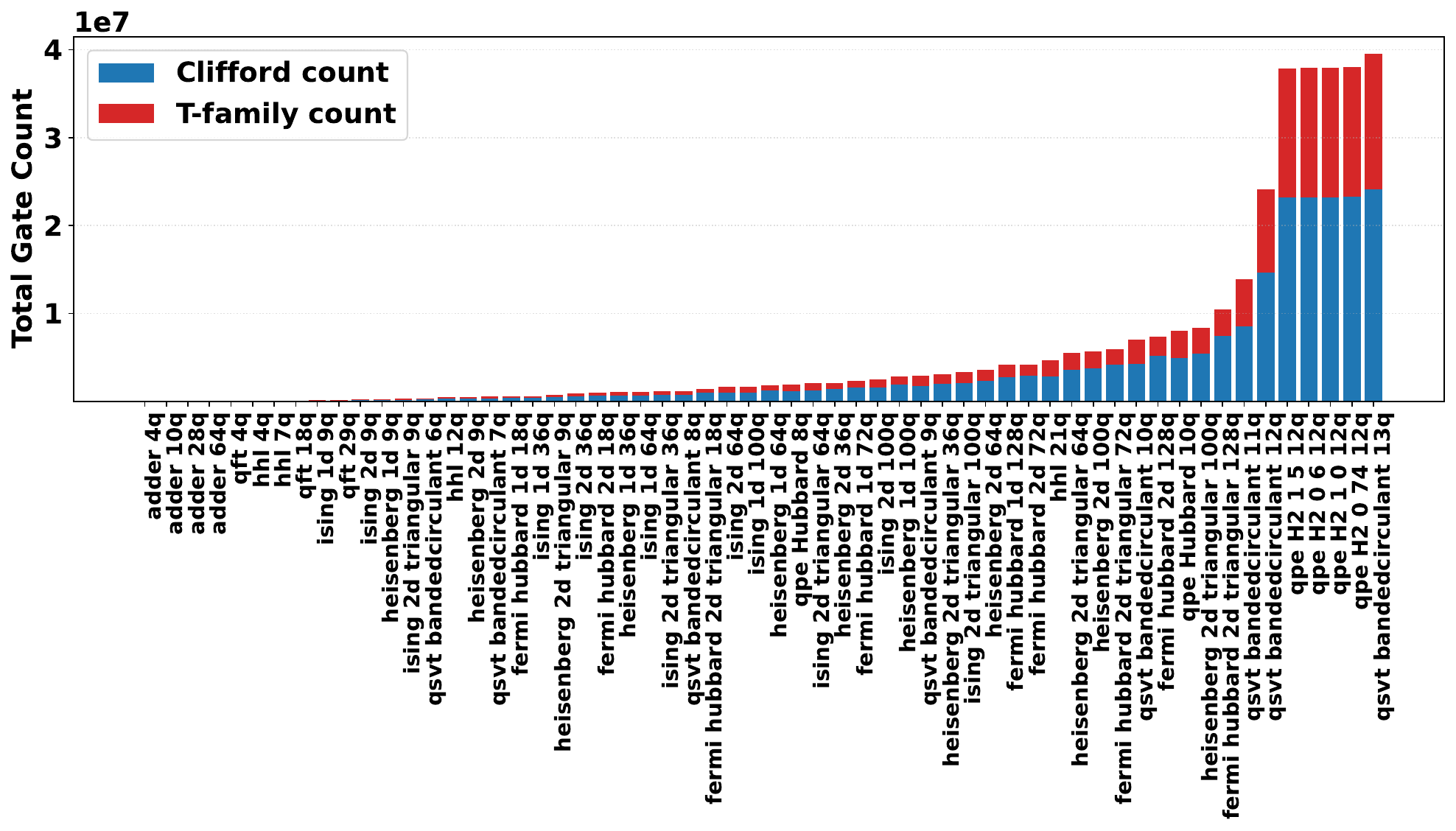}
    \caption{Clifford+T gate counts of FTCircuitBench circuits.}
    \label{fig:gatecounts}
\end{figure}

Gate counts for FTCircuitBench algorithms after compilation to Clifford+T are shown in Figure \ref{fig:gatecounts}. The total gate count for each circuit is partitioned into Clifford gates in blue and T family ($T$ and $T^\dagger$) gates in red. The library contains a large spread of circuit sizes, ranging from as few as 23 gates for the 4 qubit adder to as many as 39,526,510 gates for the 13 qubit QSVT, providing a comprehensive dataset for compiler benchmarking and architecture evaluation.

\subsubsection{Interaction Graphs}
\begin{figure}[ht]
    \centering
    \begin{subfigure}[b]{0.32\linewidth}
        \centering
        \includegraphics[width=\linewidth]{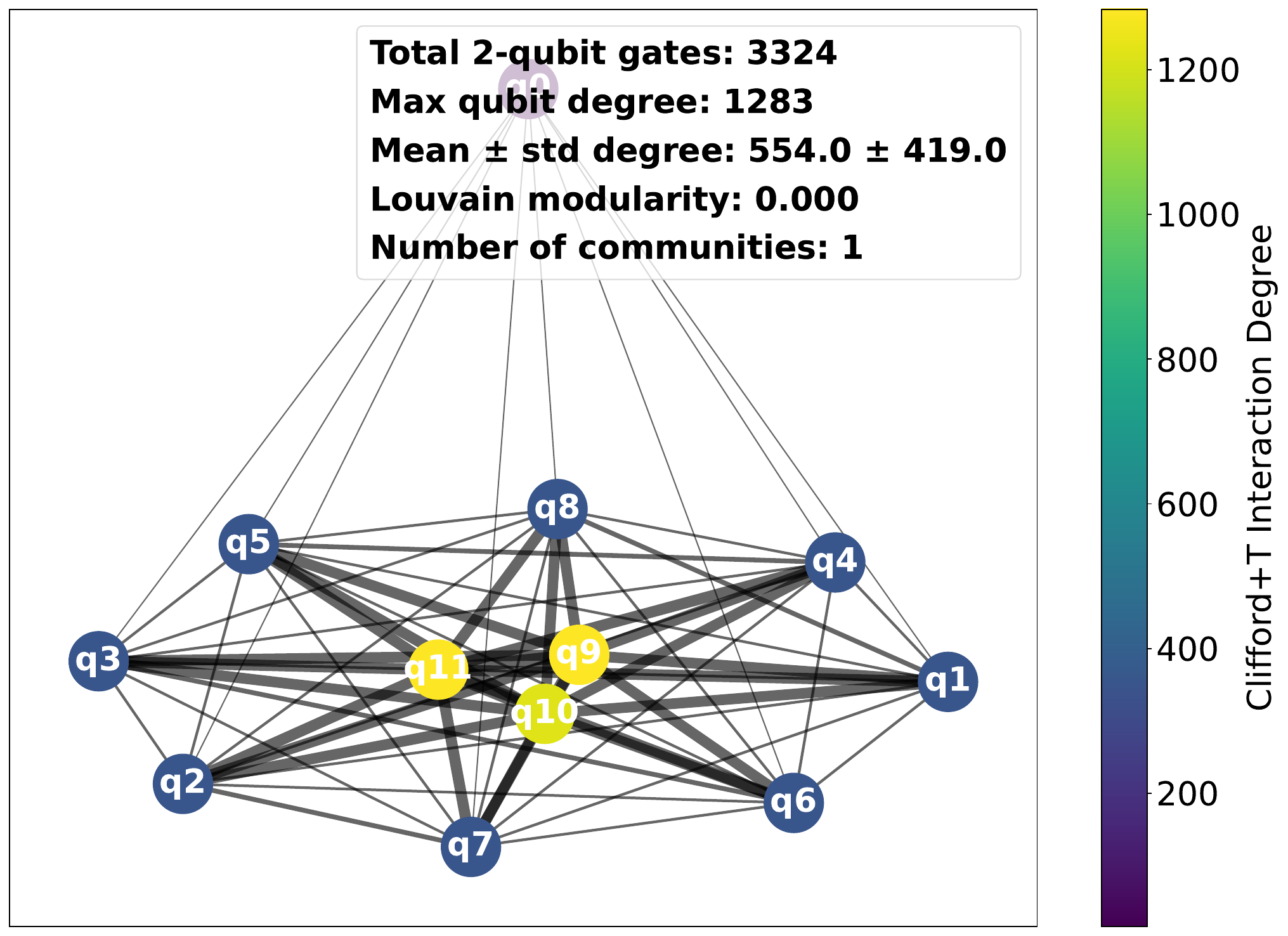}
        \caption{12-qubit HHL}
    \end{subfigure}
    \hfill
    \begin{subfigure}[b]{0.32\linewidth}
        \centering
        \includegraphics[width=\linewidth]{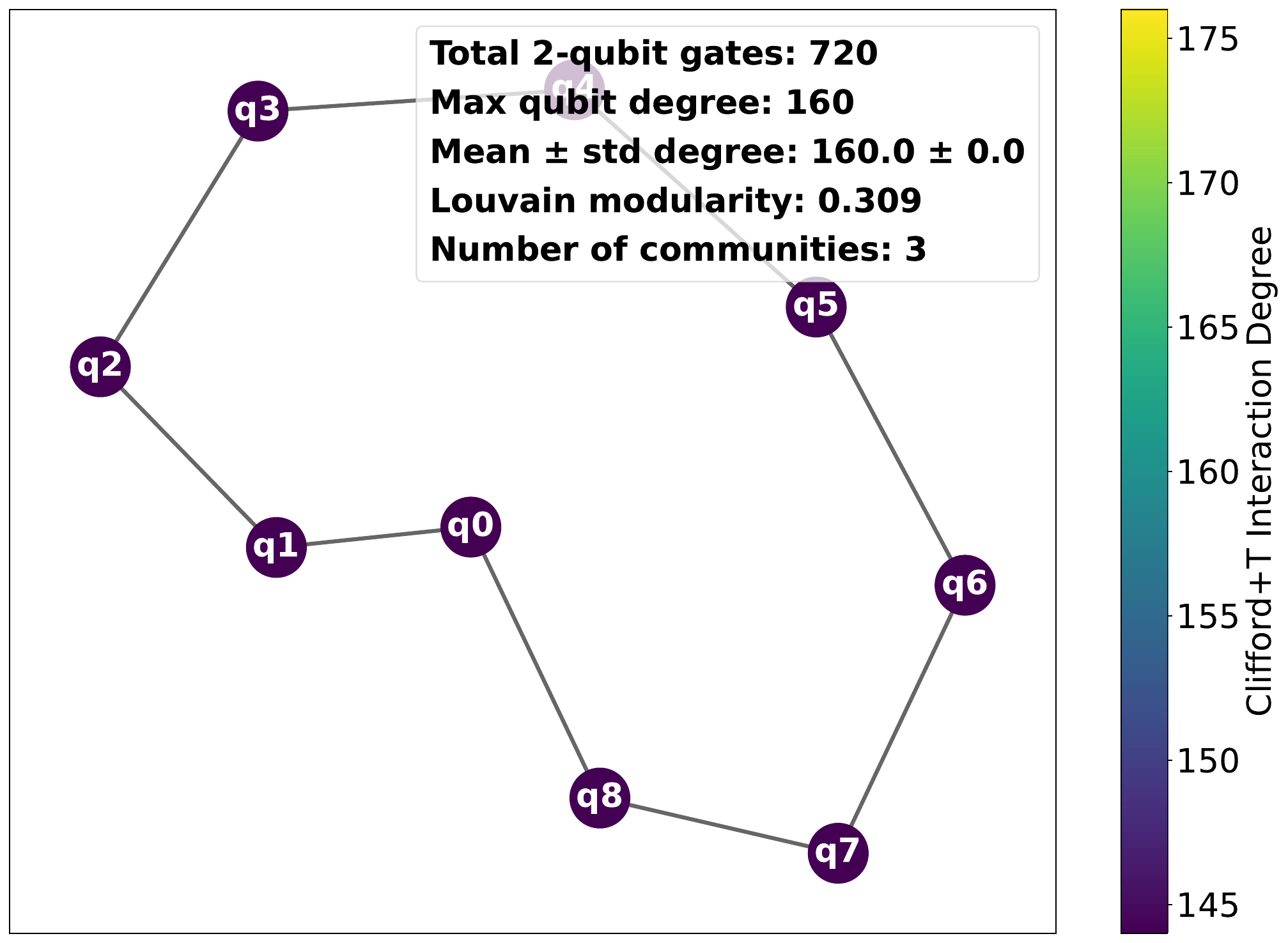}
        \caption{9-qubit 1D Ising model}
    \end{subfigure}
    \hfill
    \begin{subfigure}[b]{0.32\linewidth}
        \centering
        \includegraphics[width=\linewidth]{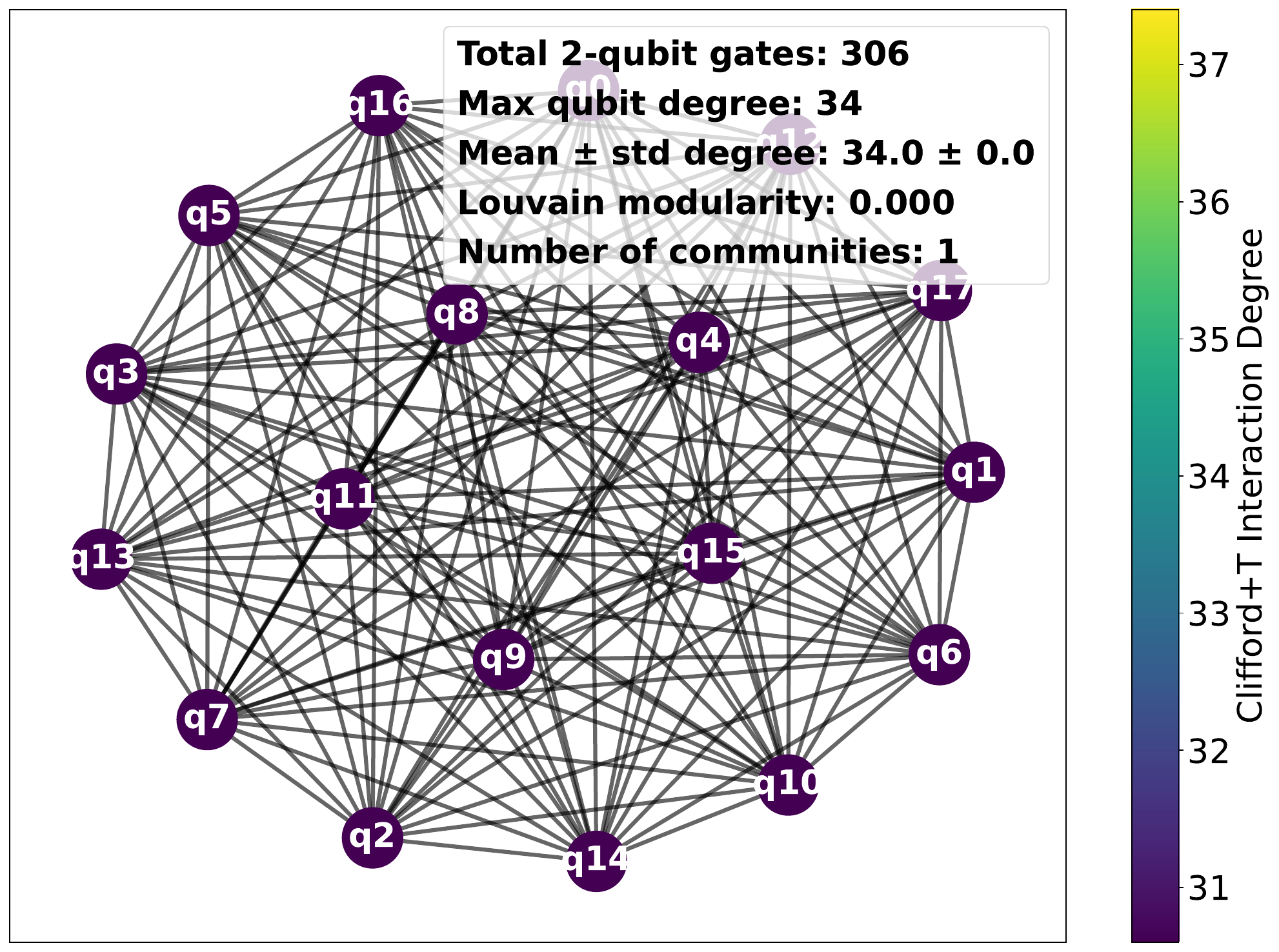}
        \caption{18-qubit QFT}
    \end{subfigure}

    \vspace{0.5em}

    \begin{subfigure}[b]{0.32\linewidth}
        \centering
        \includegraphics[width=\linewidth]{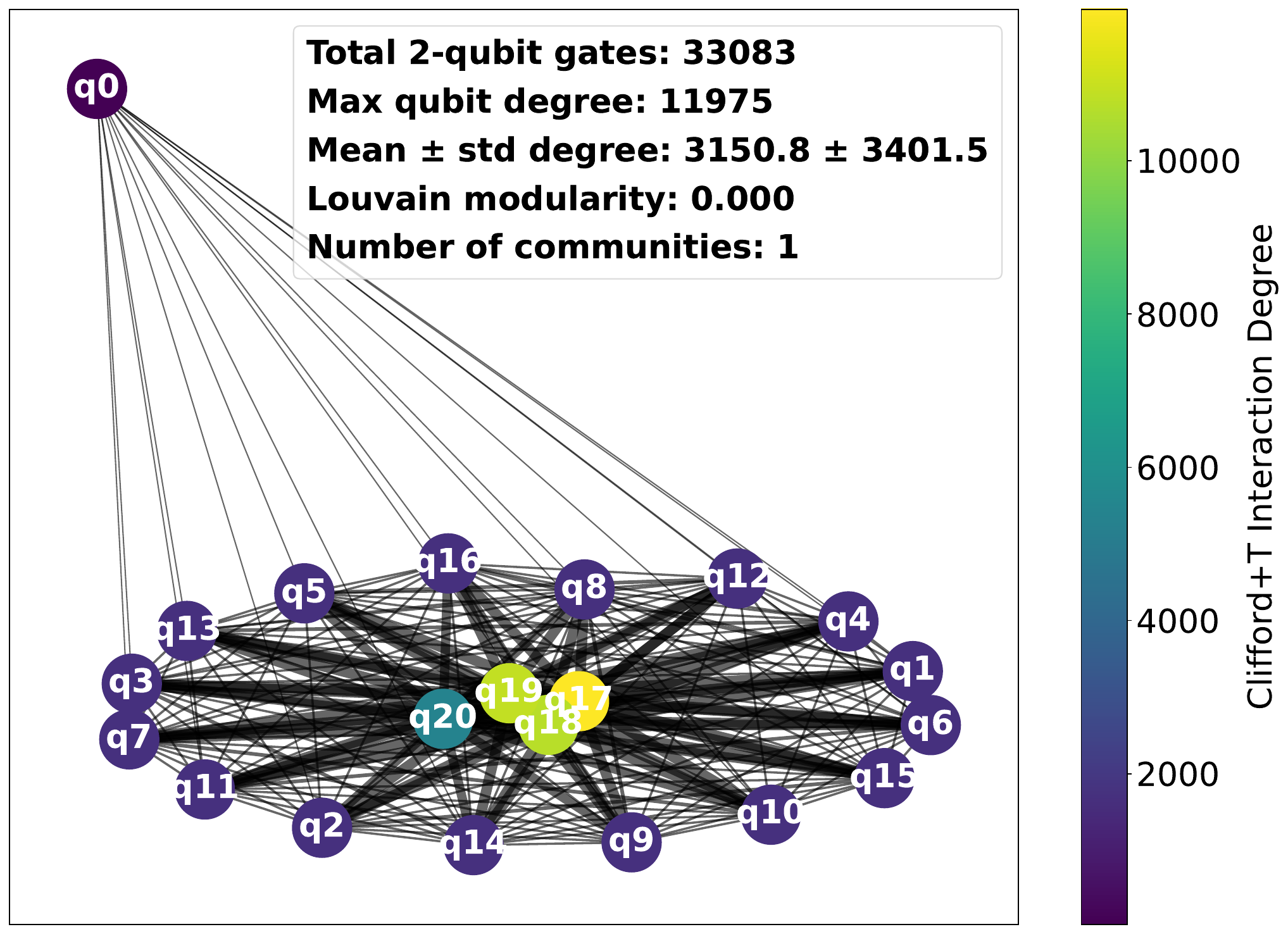}
        \caption{21-qubit HHL}
    \end{subfigure}
    \hfill
    \begin{subfigure}[b]{0.32\linewidth}
        \centering
        \includegraphics[width=\linewidth]{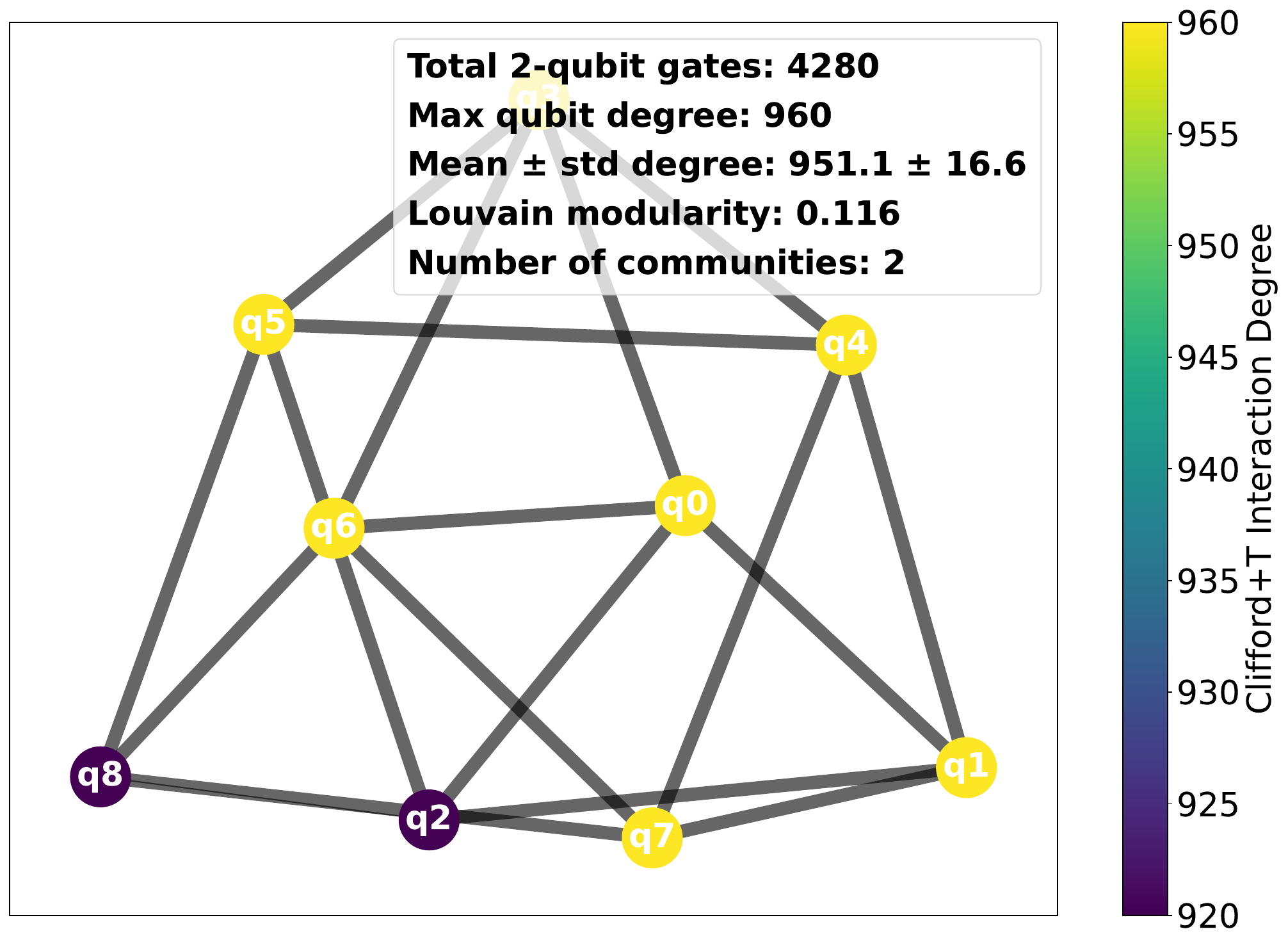}
        \caption{9-qubit 2D Heisenberg model}
    \end{subfigure}
    \hfill
    \begin{subfigure}[b]{0.32\linewidth}
        \centering
        \includegraphics[width=\linewidth]{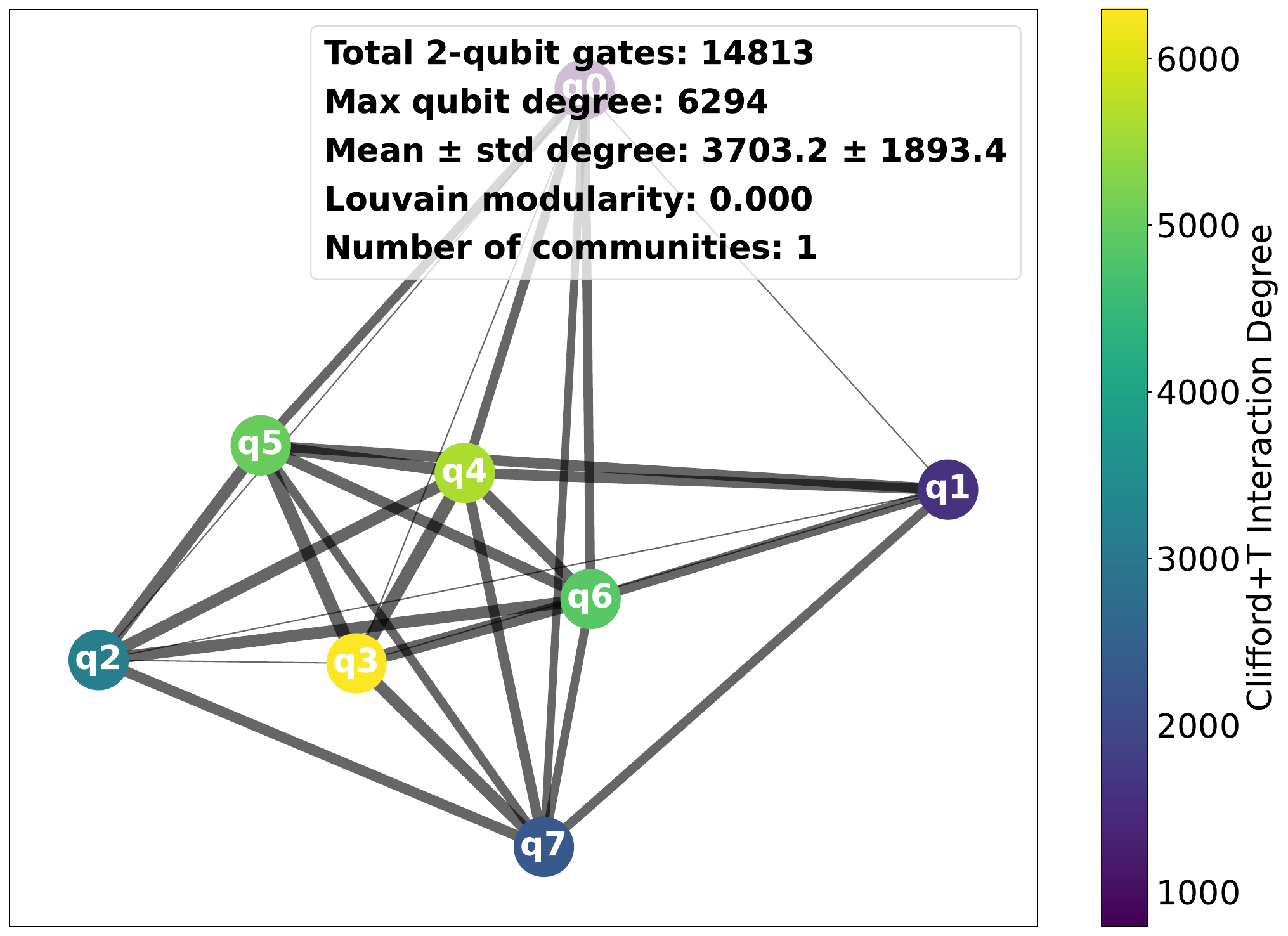}
        \caption{8-qubit Hubbard QPE}
    \end{subfigure}

    \caption{Interaction graphs of several Clifford+T circuits.}
    \label{fig:c+t interaction graphs}
\end{figure}

\begin{figure}[ht]
    \centering
    \includegraphics[width=1 \linewidth]{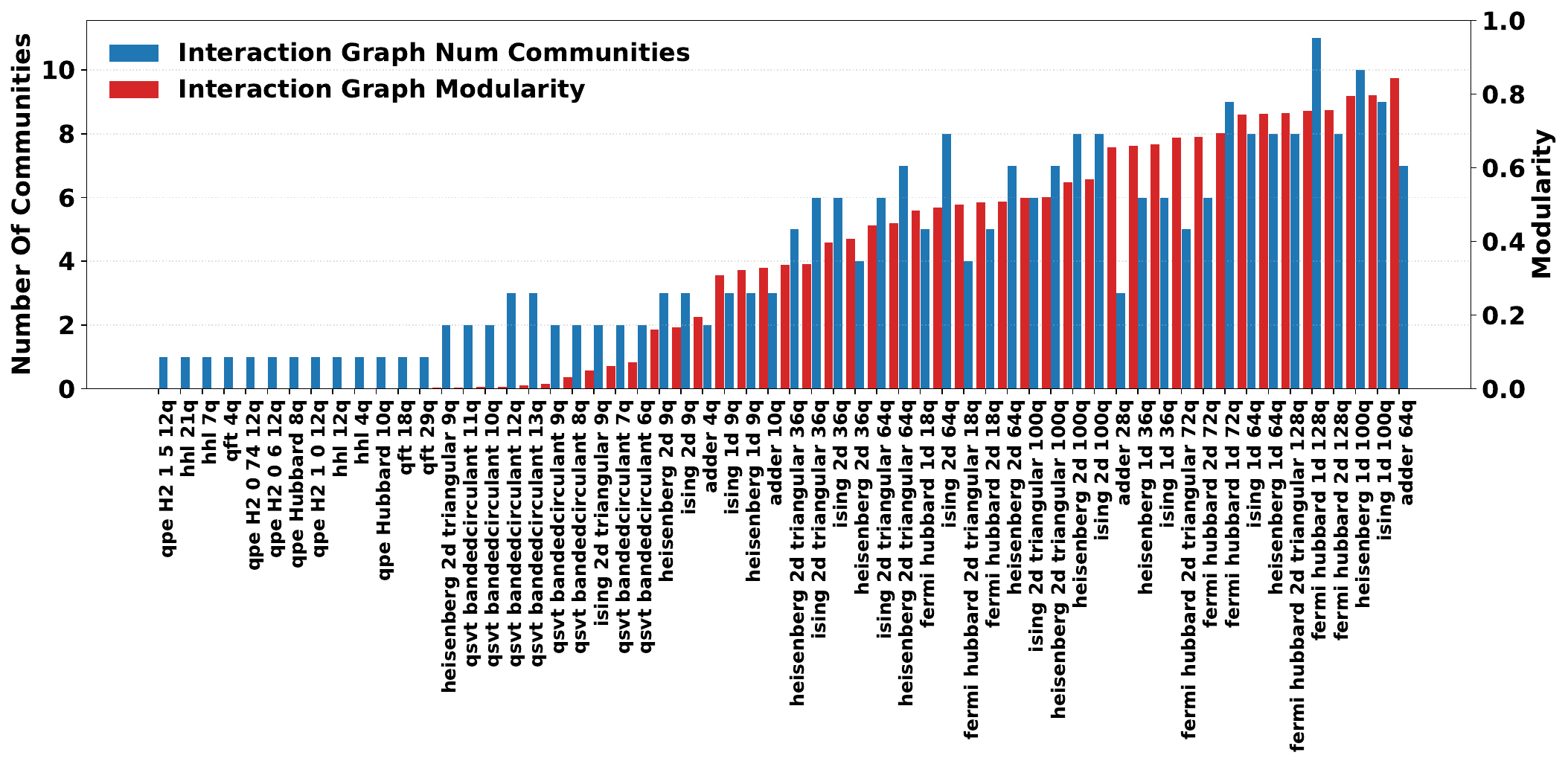}
    \caption{Community structure of Clifford+T interaction graphs.}
    \label{fig:modularity}
\end{figure}

Individual interaction graphs for several Clifford+T circuits are shown in Figure \ref{fig:c+t interaction graphs}, where nodes represent qubits, node colors correspond to their degrees (weighted sum of edges they belong to), and edge thicknesses indicate the number of two-qubit gates between pairs of qubits.  We also plot the Modularity and Number of Communities of each circuit's interaction graph in Figure \ref{fig:modularity}, sorted in ascending order by modularity. While we see that circuits with higher modularity generally also divide into more communities, there are some exceptions to this correlation, such as the 28 qubit adder that has a modularity of .656 while only splitting into 3 communities.

\subsubsection{T Gate Statistics}
\begin{figure}[ht]
    \centering
    \begin{subfigure}[b]{\linewidth}
        \centering
        \includegraphics[width=\linewidth, trim=0 25 0 0, clip]{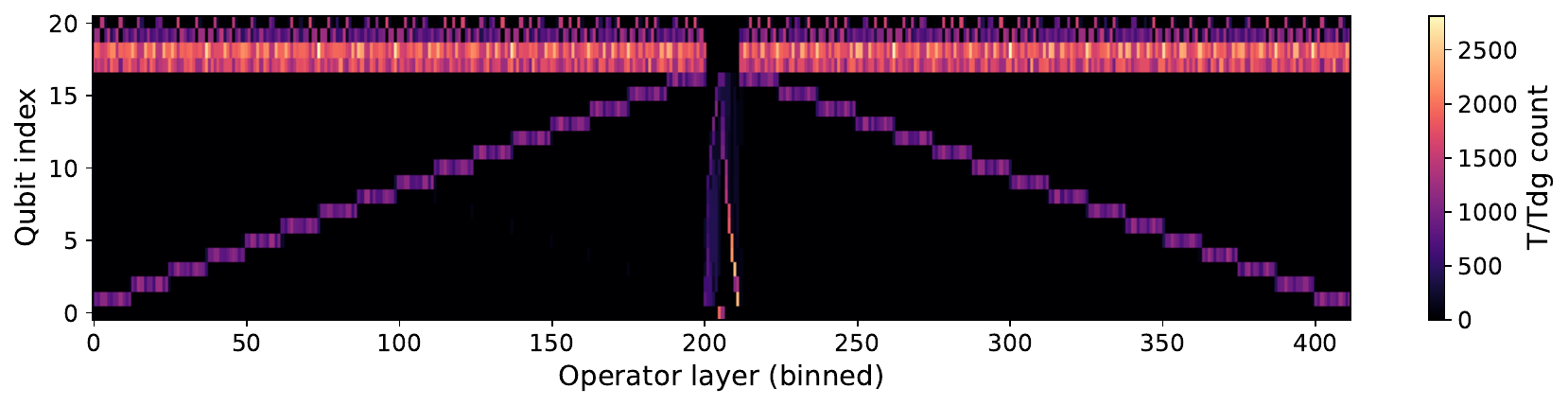}
        \caption{21-qubit HHL}
        \label{fig:hhl_density}
    \end{subfigure}

    \vspace{0.5em}

    \begin{subfigure}[b]{\linewidth}
        \centering
        \includegraphics[width=\linewidth, trim=0 25 0 0, clip]{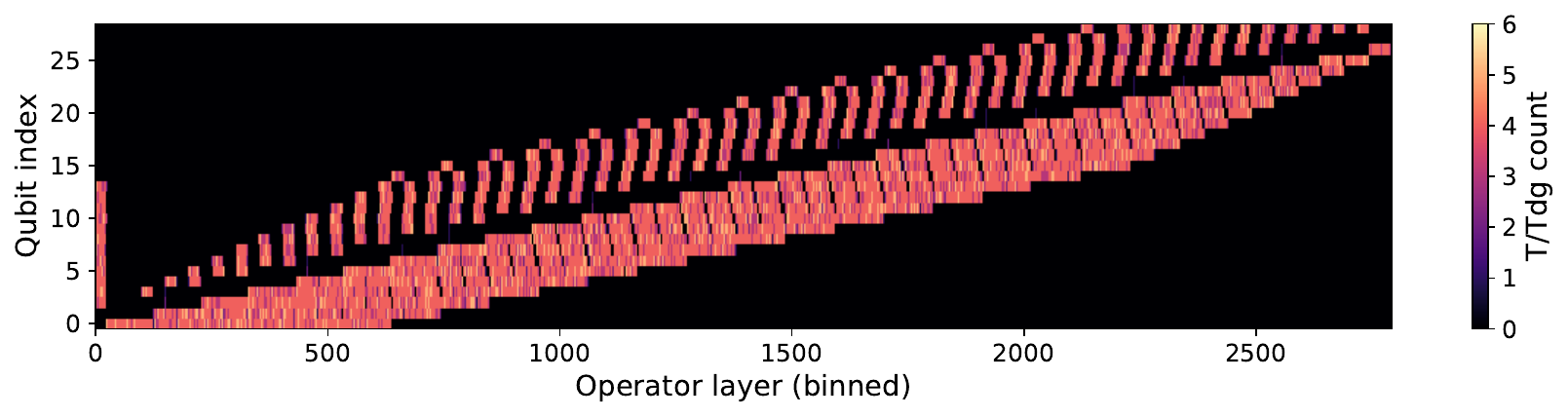}
        \caption{29-qubit QFT}
    \end{subfigure}

    \vspace{0.5em}

    \begin{subfigure}[b]{\linewidth}
        \centering
        \includegraphics[width=\linewidth]{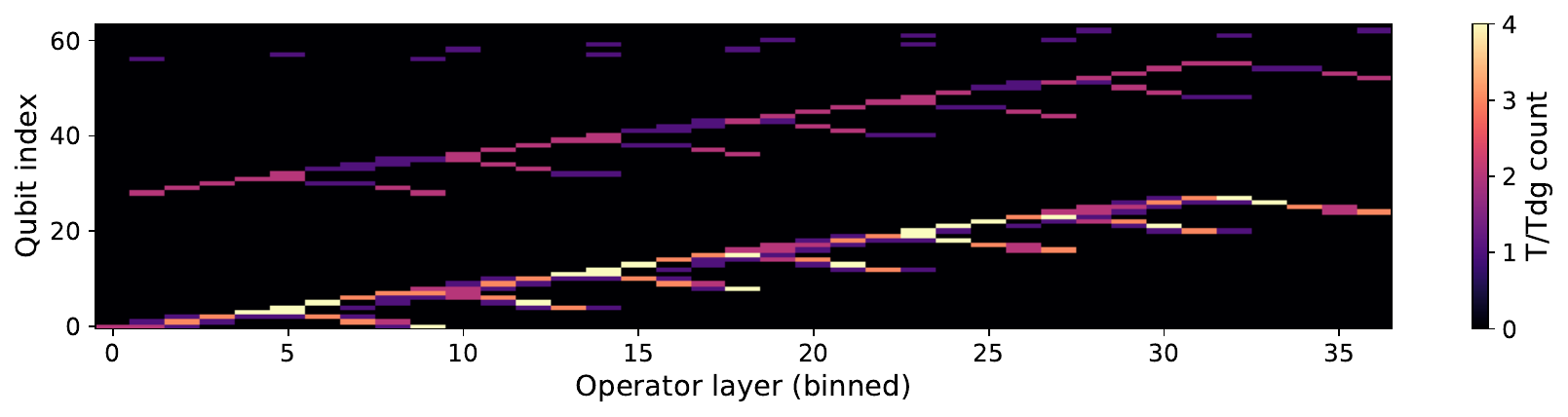}
        \caption{64-qubit Adder}
    \end{subfigure}

    \caption{T-density colormaps for HHL, QFT and ADDER.}
    \label{fig:t_density_colormaps}
\end{figure}

To visualize the distribution of T-gates in a Clifford+T circuit, we plot several T-density colormaps in Figure \ref{fig:t_density_colormaps}. Each plot shows the spatial and temporal distribution of $T$ and $T^\dagger$ gates throughout the algorithm, while providing intuition about the overall structure of the algorithm.  For example, in the 21-qubit HHL colormap, it is clear that the register in which the vector $\Vec{b}$ is initialized and eventually the linear system solution $\Vec{x}$ is encoded (qubits 17-20) has a consistently high demand of T-gates throughout the algorithm relative to the rest of the qubits.

\subsection{Pauli Based Computation Metrics}

\subsubsection{Pauli Weight Statistics}
While the Clifford+T to PBC transpilation algorithm is relatively straightforward to understand, it is unclear how to predict the resulting distribution of Pauli strings representing the PBC circuit.  Thus, we provide several illuminating visualizations of Pauli weight statistics over PBC circuits.  

\begin{figure}[ht]
    \centering
    \begin{subfigure}[b]{0.49\linewidth}
        \centering
        \includegraphics[width=\linewidth]{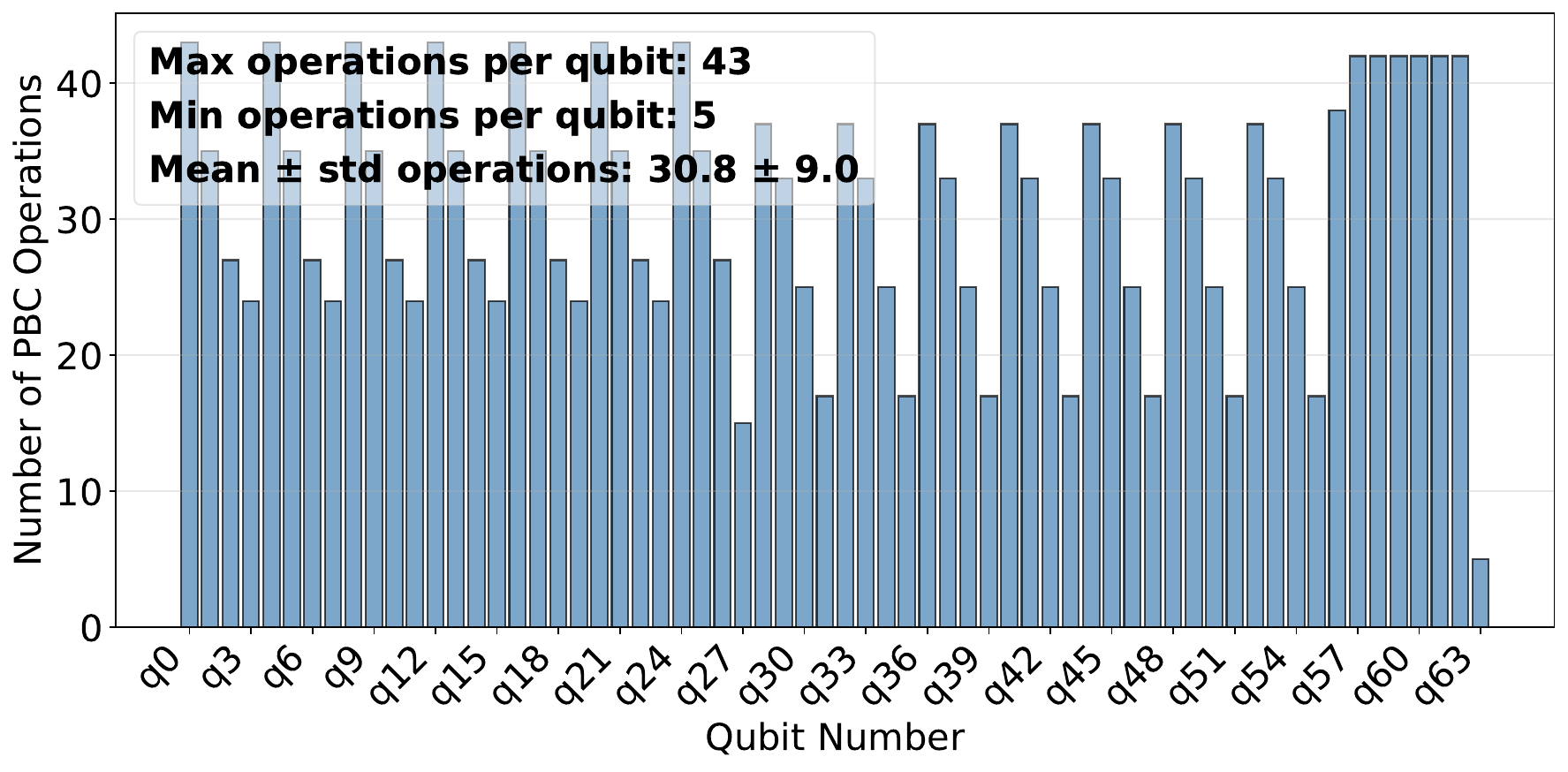}
        \caption{64-qubit Adder}
        \label{fig:adder pbc ops}
    \end{subfigure}
    \hfill
    \begin{subfigure}[b]{0.49\linewidth}
        \centering
        \includegraphics[width=\linewidth]{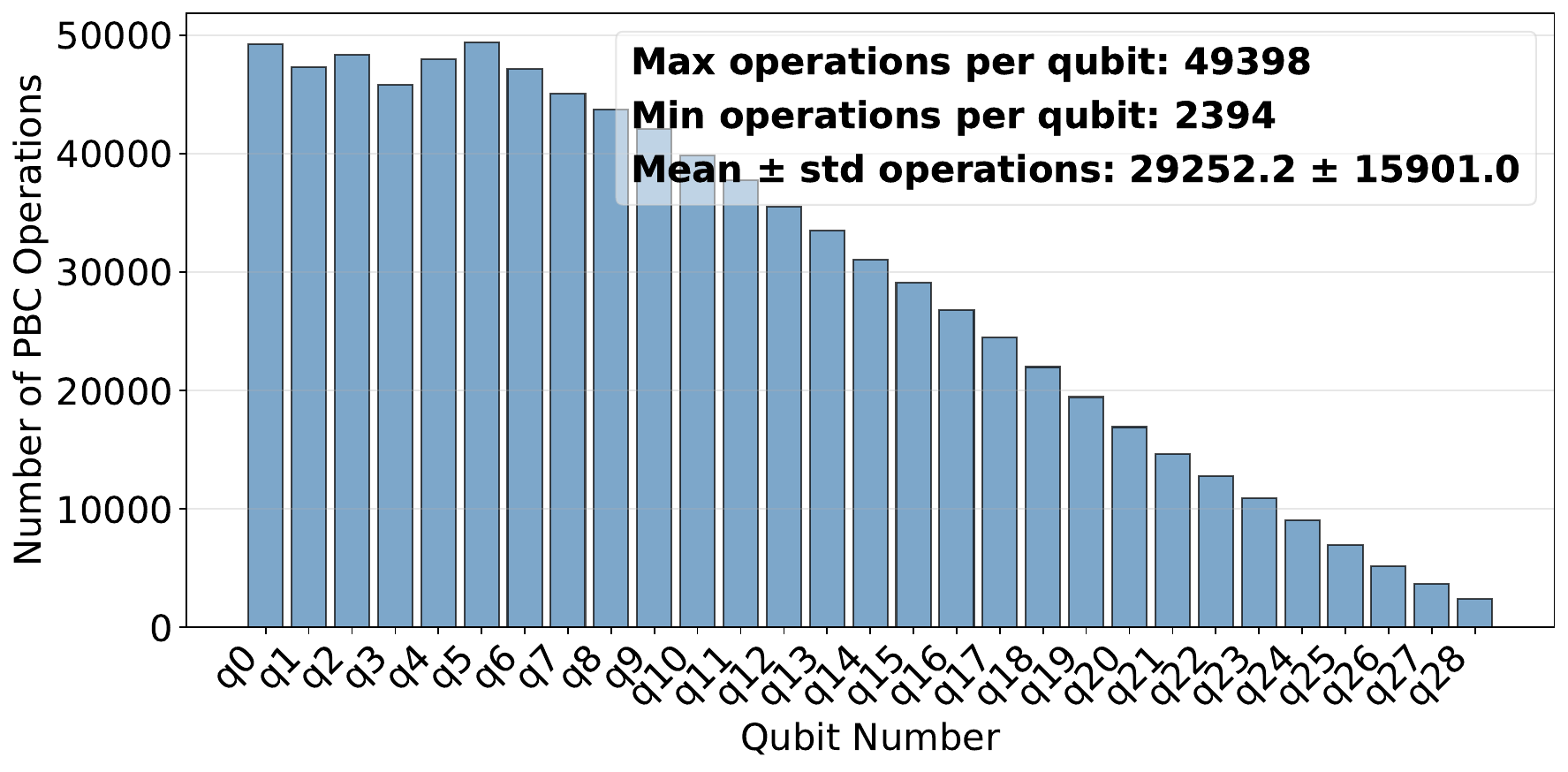}
        \caption{29-qubit QFT}
        \label{fig:ising pbc ops}
    \end{subfigure}
    \hfill
    \begin{subfigure}[b]{0.49\linewidth}
        \centering
        \includegraphics[width=\linewidth]{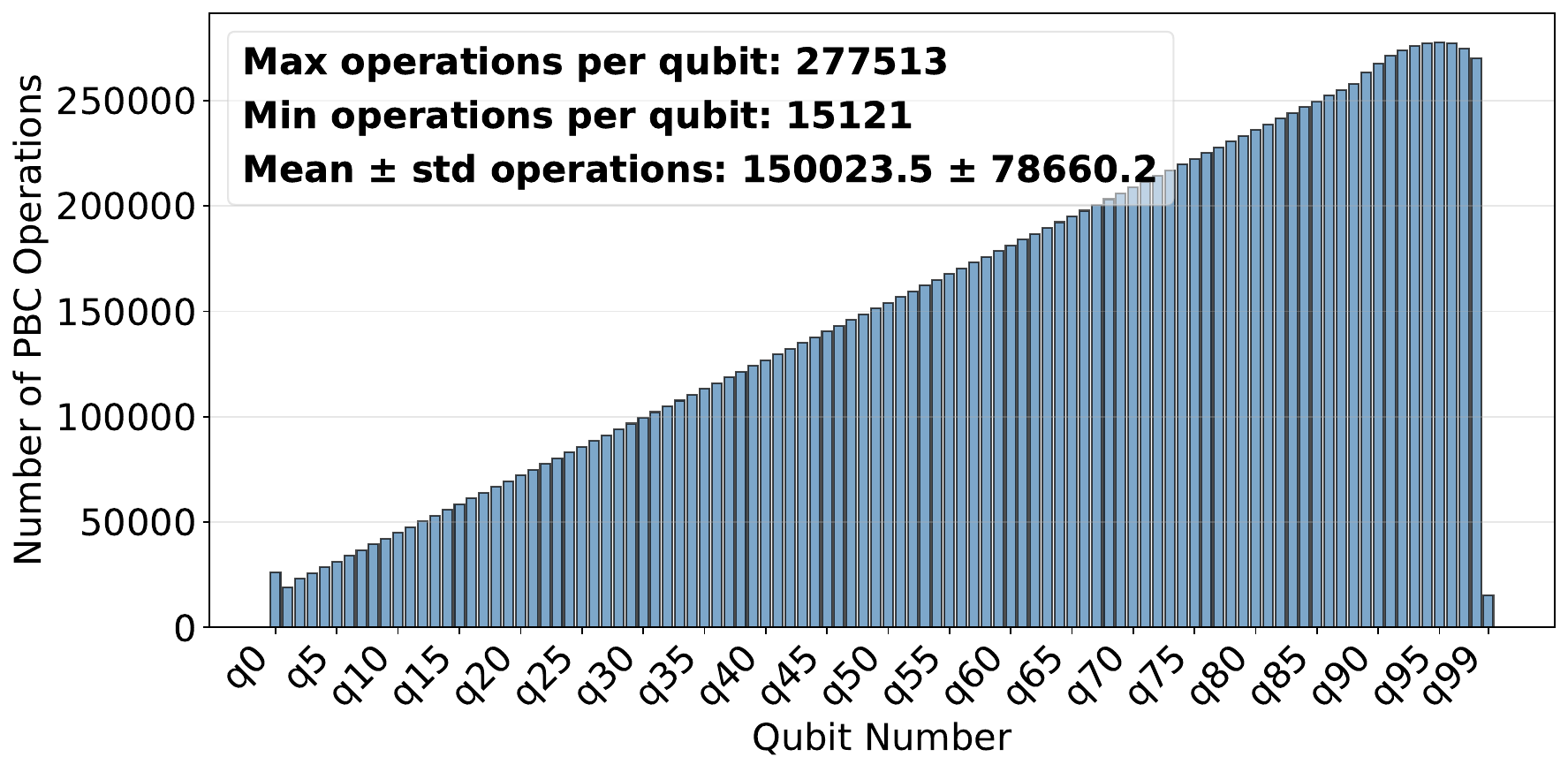}
        \caption{100-qubit 1D Ising model}
        \label{fig:qft pbc ops}
    \end{subfigure}
    \hfill
    \begin{subfigure}[b]{0.49\linewidth}
        \centering
        \includegraphics[width=\linewidth]{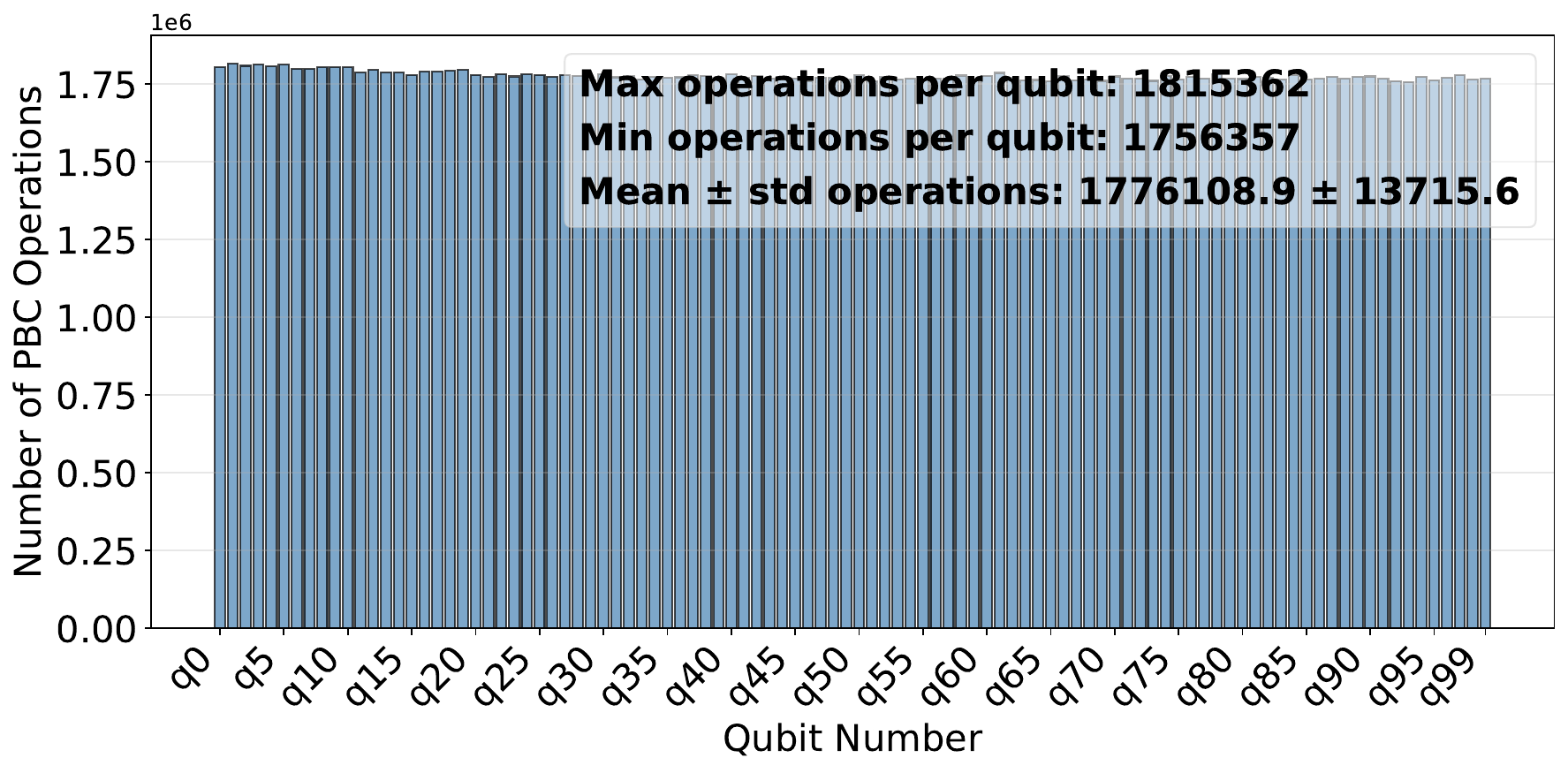}
        \caption{100-qubit 2D Heisenberg model}
        \label{fig:heisenberg pbc ops}
    \end{subfigure}
    \caption{Number of Pauli-Based Computation (PBC) operations per qubit for four quantum circuits. Each bar shows the total number of PBC operations (rotations and measurements) applied on a given qubit.}
    \label{fig:pbc ops per qubit}
\end{figure}

In Figure \ref{fig:pbc ops per qubit}, we show the number of Pauli operators for which each qubit in an algorithm belongs to the support.  With these distributions, we can immediately understand when an algorithm will benefit from intelligent logical qubit placement.  For example, the 64-qubit adder has particularly high PBC operator involvement on every 4th qubit starting at qubit 0 and thus should expect to benefit from mapping those qubits to higher connectivity locations.  On the other hand, the 100-qubit 2D Heisenberg Model has very little relative variance of total PBC operator involvement across all of its 100 qubits, showing that random qubit placement may yield comparable performance to an optimized mapping.

\begin{figure}[ht]
    \centering
    \begin{subfigure}[b]{0.49\linewidth}
        \centering
        \includegraphics[width=\linewidth]{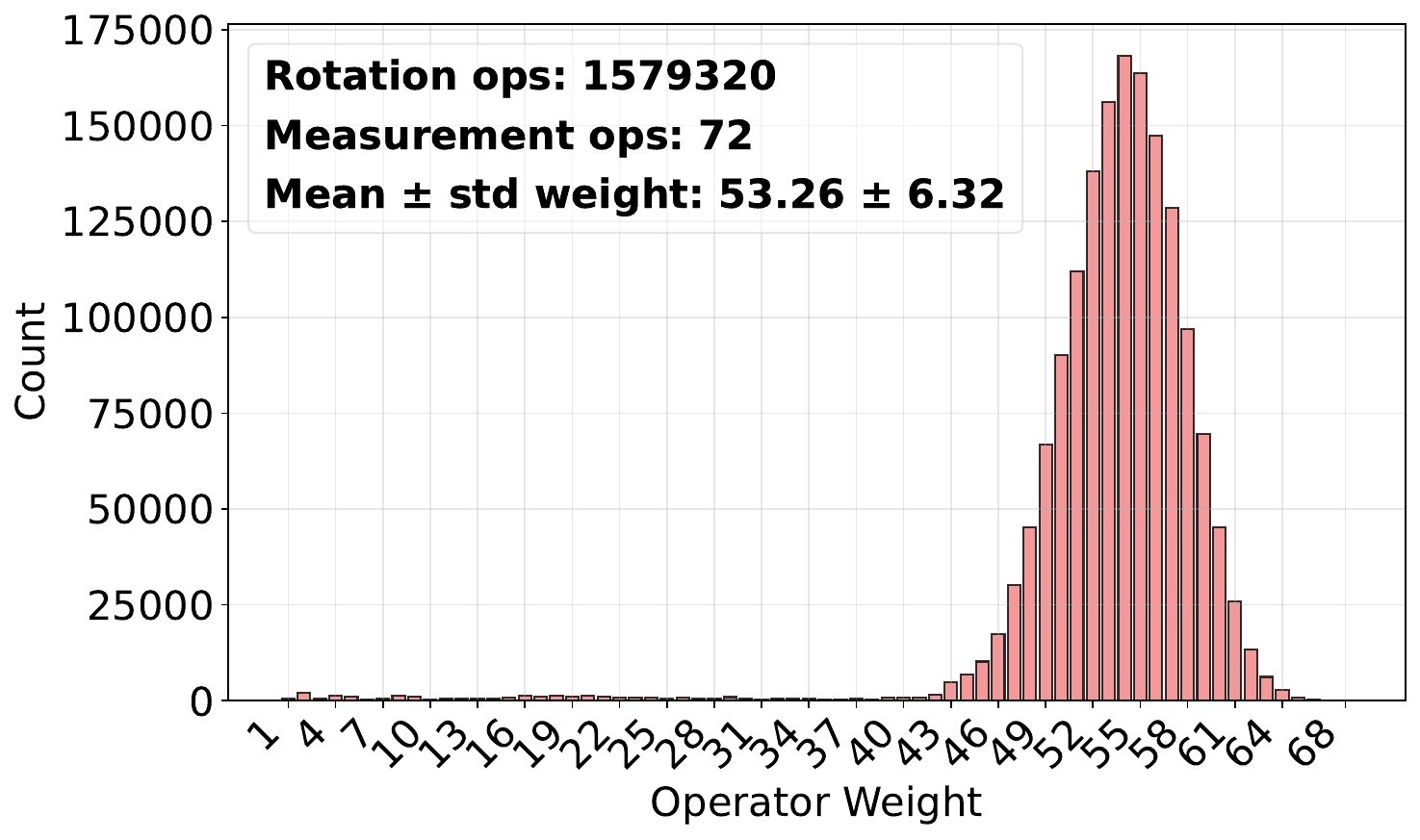}
        \caption{72-qubit 2D Fermi Hubbard model}
        \label{fig:FM 2D weight distribution}
    \end{subfigure}
    \hfill
    \begin{subfigure}[b]{0.49\linewidth}
        \centering
        \includegraphics[width=\linewidth]{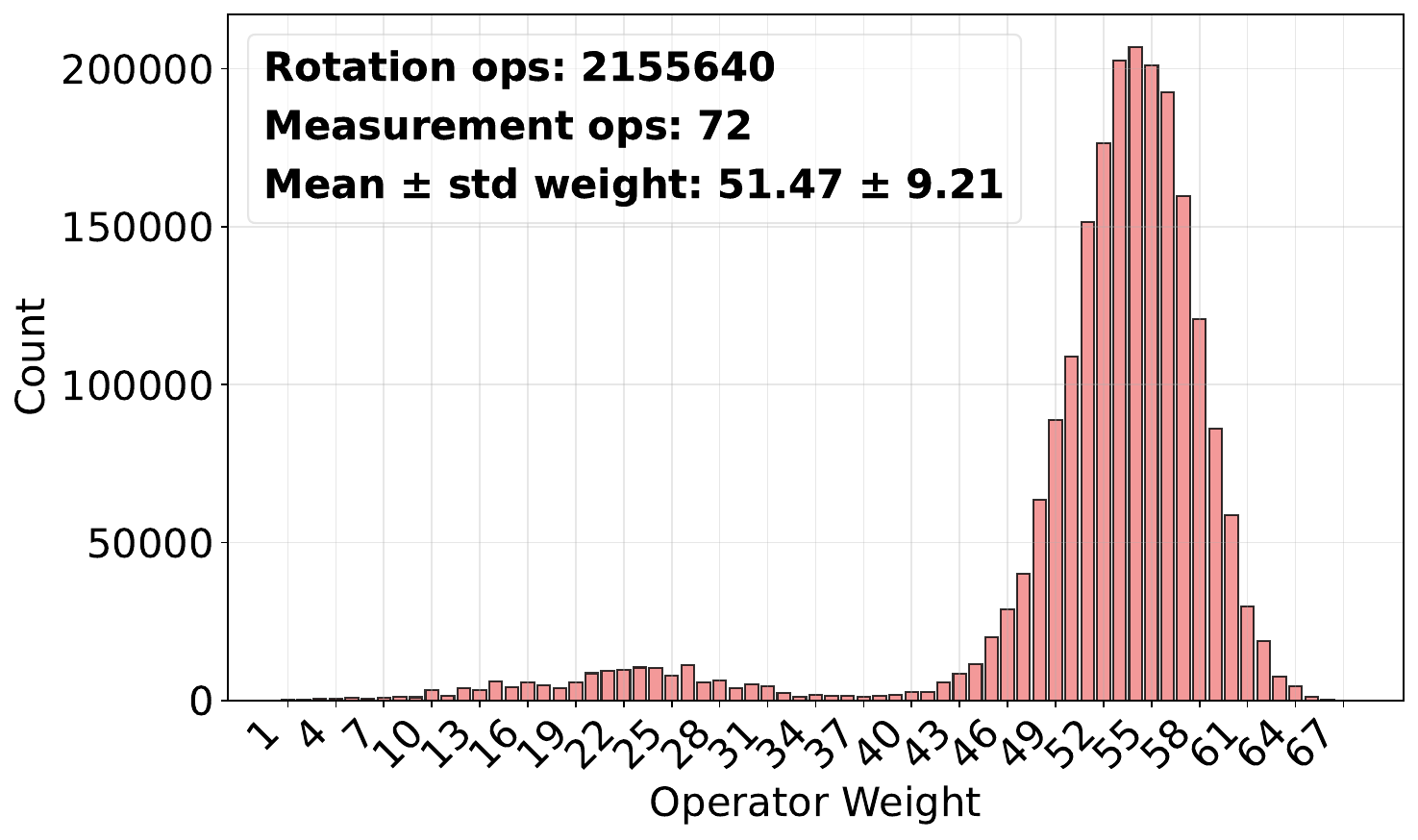}
        \caption{72-qubit 2D Triangular Fermi Hubbard model}
        \label{fig:FM 2D tri weight distribution}
    \end{subfigure}
    \hfill
    \begin{subfigure}[b]{0.49\linewidth}
        \centering
        \includegraphics[width=\linewidth]{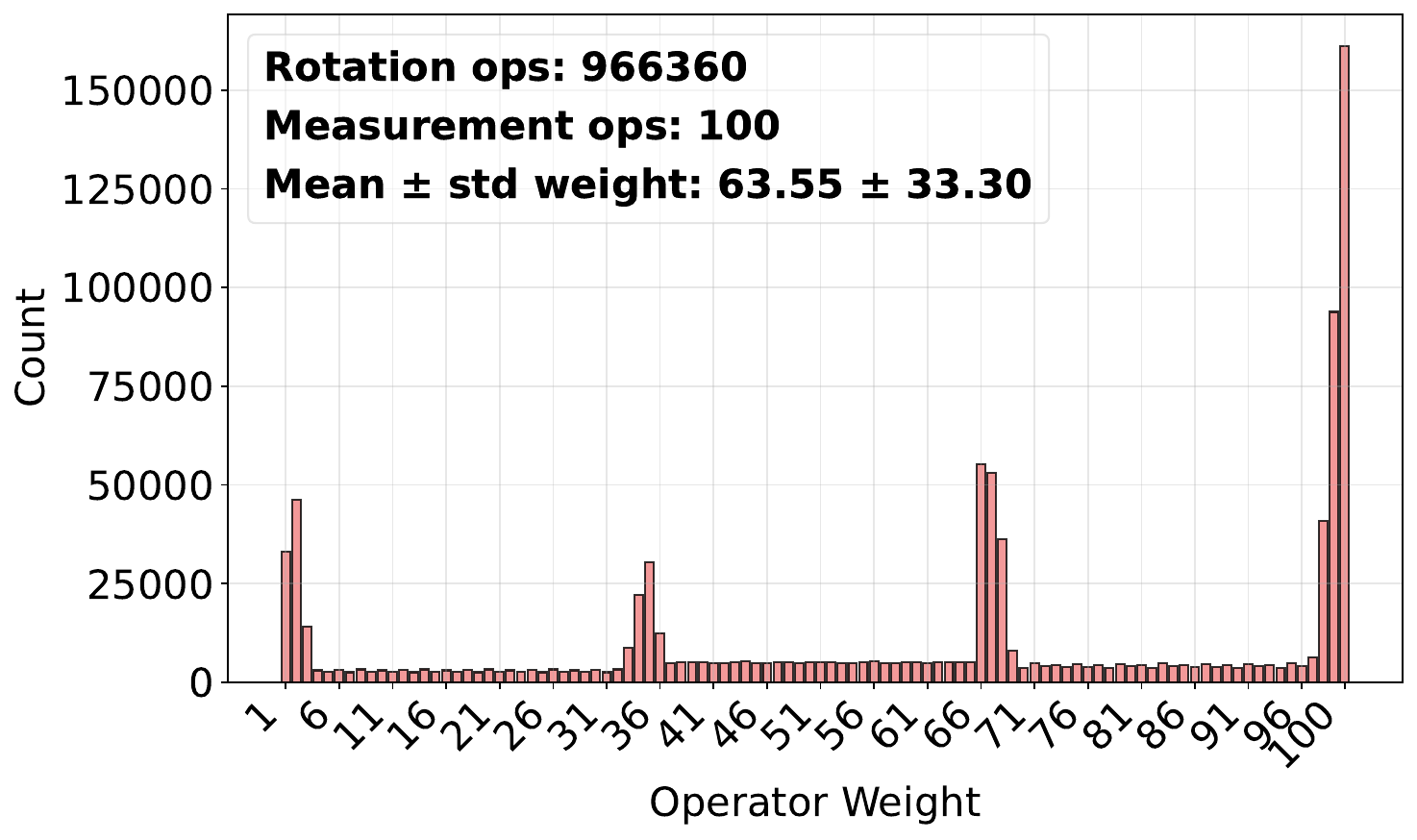}
        \caption{100-qubit 1D Heisenberg model}
        \label{fig:Heisenberg 1D weight distribution}
    \end{subfigure}
    \hfill
    \begin{subfigure}[b]{0.49\linewidth}
        \centering
        \includegraphics[width=\linewidth]{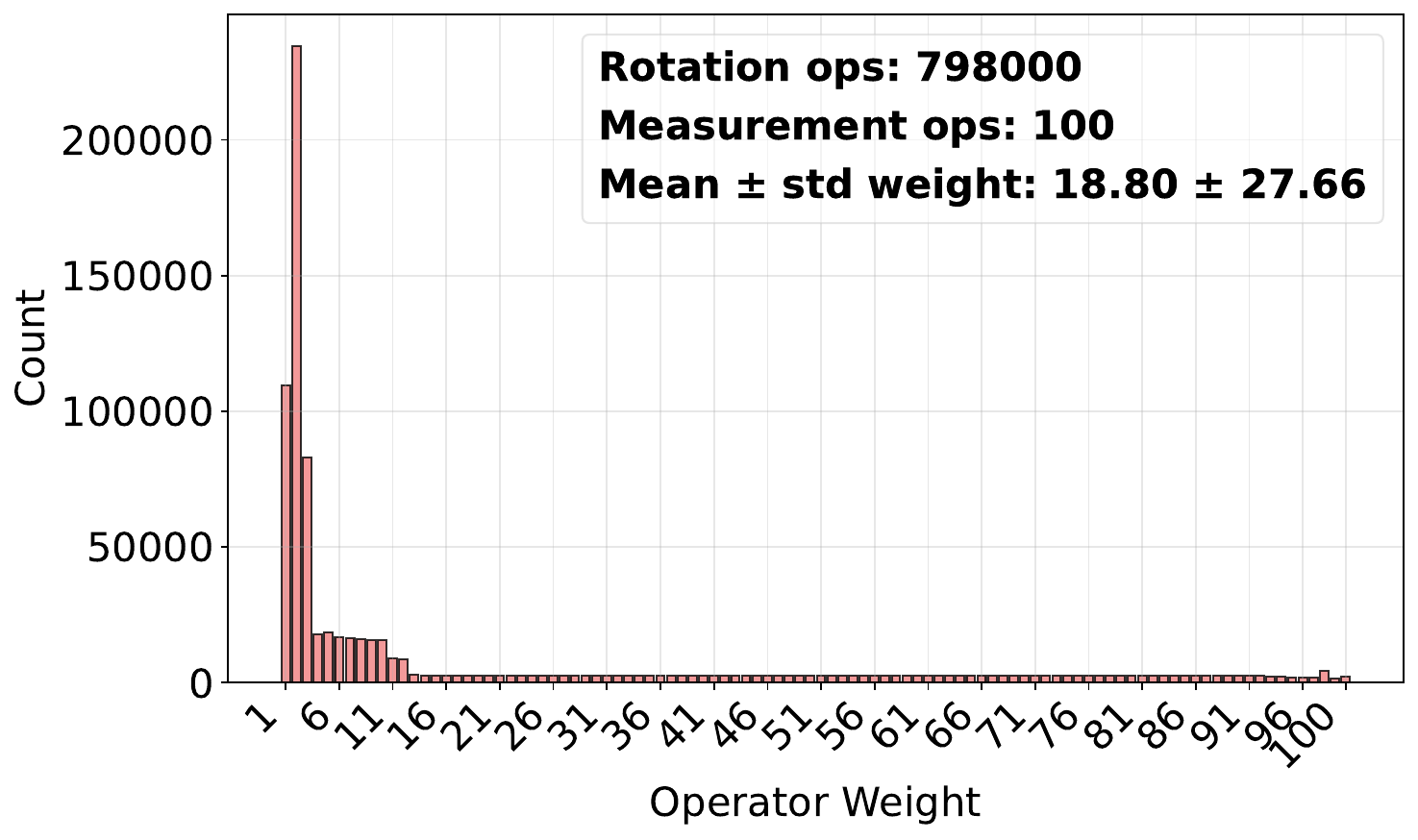}
        \caption{100-qubit 1D Ising model}
        \label{fig:Ising 1D weight distribution}
    \end{subfigure}
    \caption{Pauli operator weight distributions for four Hamiltonian simulation circuits expressed in Pauli-Based Computation form. The weight of each Pauli rotation or measurement operator is defined as the number of non-identity Pauli terms in the operator.}
    \label{fig:weight distributions}
\end{figure}

Figure \ref{fig:weight distributions} shows the counts of Pauli weights over all Pauli operators for several PBC circuits.  Given that high weight Pauli operators can be difficult to implement, these distributions potentially translate to the expected difficulty of executing an algorithm in the PBC model.  In the examples shown, it is clear that a 100 qubit 1D Ising model PBC circuit has primarily low-weight operators with weights concentrated between 1 and 10, and will be relatively easy to implement.  If performing PBC optimization to merge and remove operators, the most meaningful metric to minimize would likely be the total count of operators. Meanwhile, a 100 qubit 1D Heisenberg model PBC circuit has a multimodal weight profile with substantial operator mass at weights near the full
circuit width (mean weight $63.55$), and would pose greater implementation
challenges due to its high weight operators.

\begin{figure}[ht]
    \centering
    \begin{subfigure}[b]{\linewidth}
        \centering
        \includegraphics[width=\linewidth, trim=0 25 0 0, clip]{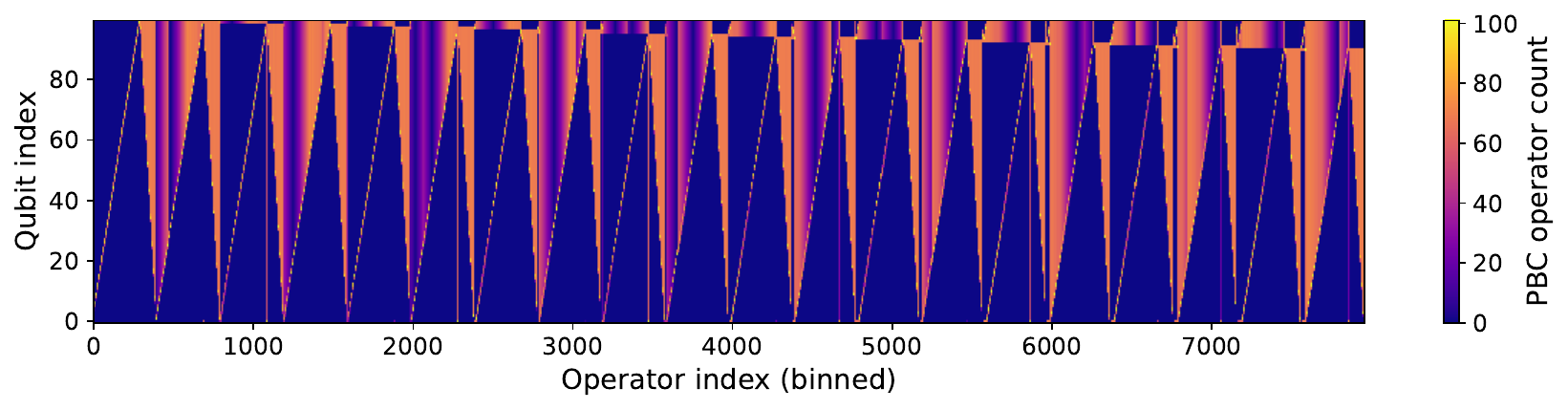}
        \caption{100-qubit 1D Ising model}
    \end{subfigure}

    \vspace{0.5em}

    \begin{subfigure}[b]{\linewidth}
        \centering
        \includegraphics[width=\linewidth, trim=0 25 0 0, clip]{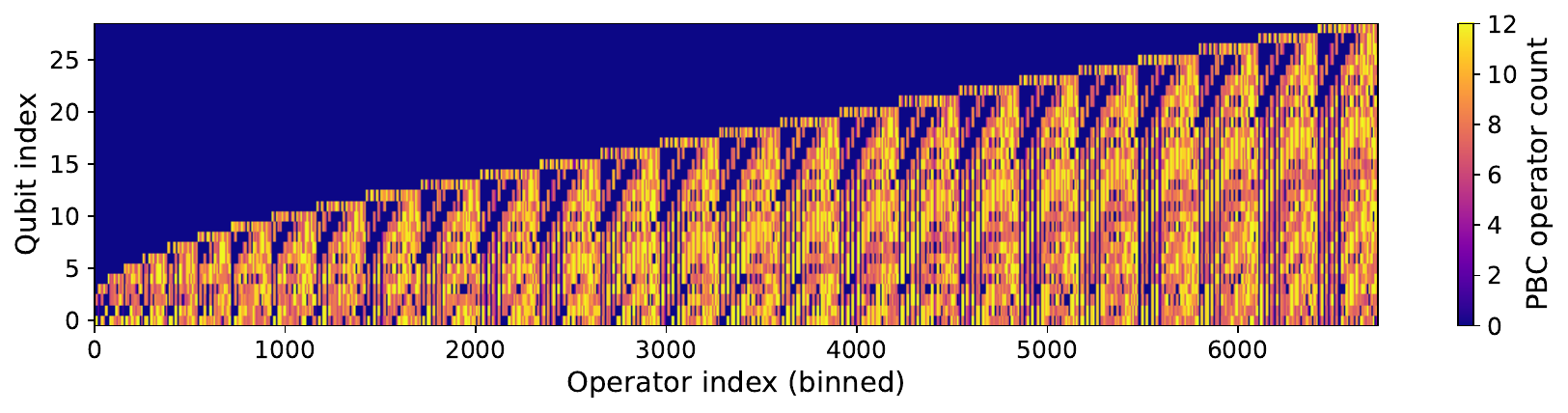}
        \caption{29-qubit QFT}
    \end{subfigure}

    \vspace{0.5em}

    \begin{subfigure}[b]{\linewidth}
        \centering
        \includegraphics[width=\linewidth]{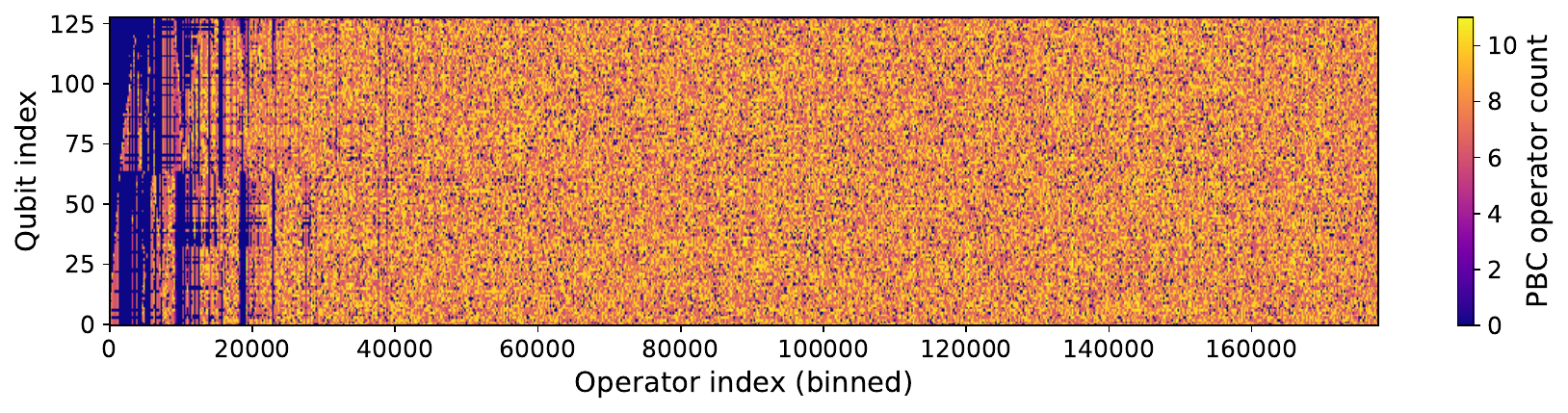}
        \caption{128-qubit 1D Fermi Hubbard model}
    \end{subfigure}

    \caption{PBC operator density colormaps for three circuits. Each plot shows the spatial and temporal distribution of supports for Pauli operations.}
    \label{fig:pbc_colormaps}
\end{figure}

While distributions like those in Figure \ref{fig:pbc ops per qubit} and \ref{fig:weight distributions} can inform qubit placement, execution difficulty, and target PBC optimization metrics, they do not provide a full picture of the PBC circuit structure. To visualize this, we can use colormaps to plot the spatial-temporal distribution of Pauli operator supports (i.e., non-identity Pauli terms).  Much like we plotted T-density colormaps for Clifford+T circuits, Figure \ref{fig:pbc_colormaps} shows these PBC operator density colormaps for several circuits.  

One would expect that colormaps for random Clifford+T circuits, once compiled to PBC form, would exhibit increasingly high-weight Pauli terms towards the back of the circuit, as those Pauli rotations would have many 2-qubit gates commuted past them, while Pauli rotations towards the front of the circuit would have fewer such gates commuted past them during PBC compilation.  While many circuits exhibited this trend, we observe that some circuits such as the 100 qubit 1D Ising model maintained similar Pauli weights throughout the circuit.  While this was not seen in many PBC colormaps for the algorithms in FTCircuitBench, the structured Pauli terms and relative lack of high-weight operators motivates the potential of smart PBC compilation optimization for these classes of circuits.  

Other types of circuits, such as the 29 qubit QFT, did exhibit increasing Pauli weights, but also maintained much structure in the spatial-temporal Pauli term densities, making them potentially of "intermediate difficulty" to further optimize and execute.  

Lastly, we found that many others, particularly for Hamiltonian simulation, quickly developed seemingly random and high-weight Pauli rotation terms after the initial low-weight terms in the beginning of the circuit.  Without any apparent structure to make use of in further compilation passes, these circuits could pose the most difficult to optimize and execute in PBC form; however, they also could be optimized at the Trotterization level.  An example of this is seen in the disordered colormap of the 128 qubit 1D Fermi Hubbard model.  We also see characteristic signatures of these random-looking circuits upon revisiting Figure \ref{fig:FM 2D weight distribution} and Figure \ref{fig:FM 2D tri weight distribution}, where the normal distribution of the Pauli weight counts can be understood to correspond to ensembles of random Pauli rotation operators.

\subsubsection{Optimization Metrics}
\begin{figure}[ht]
    \centering
    \includegraphics[width=1 \linewidth]{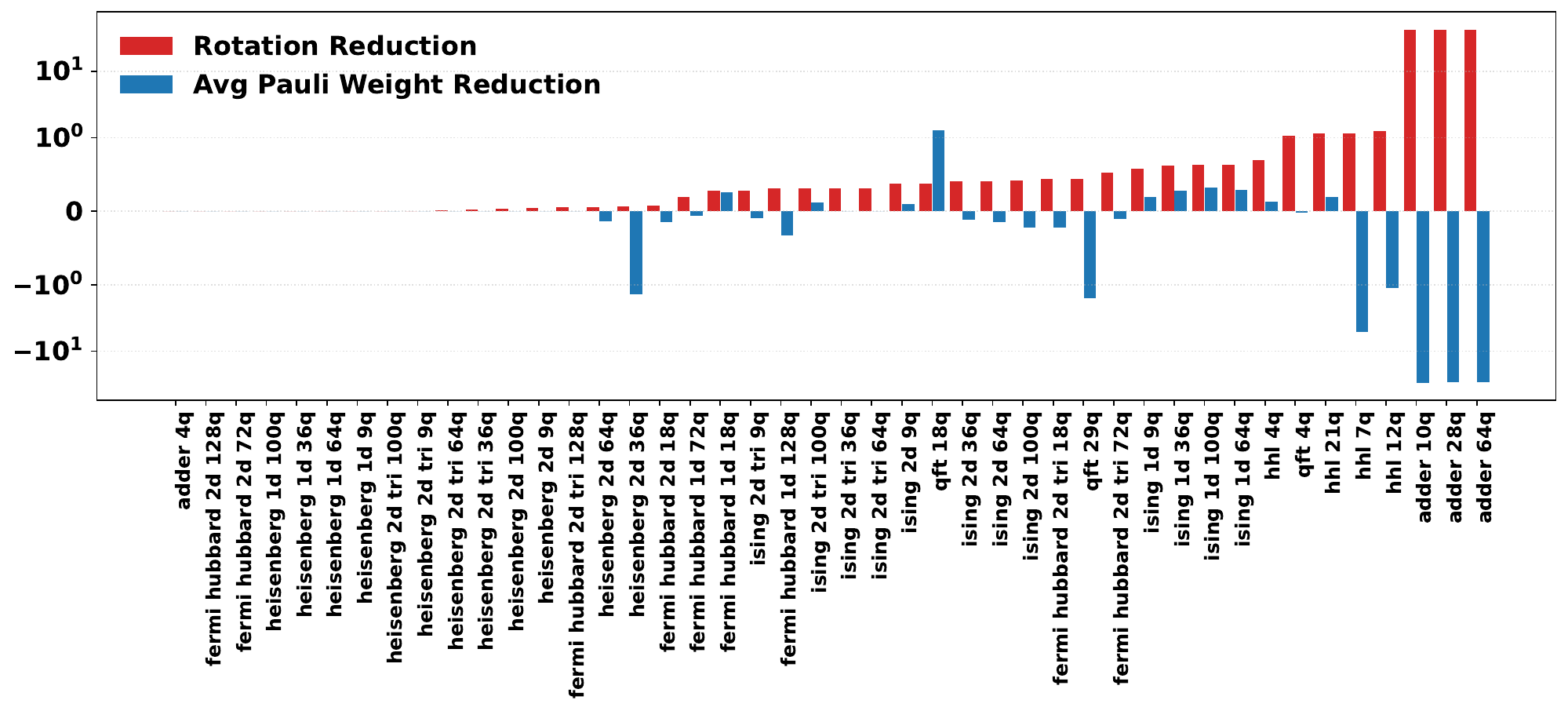}
    \caption{Symmetric log scale plot, scaled logarithmically to both sides of 0, of the change in PBC circuit structure after undergoing circuit optimization via layering and merging of rotation operators. Negative values correspond to increases in average Pauli Weight of the circuit.}
    \label{fig:pbc_opt_histogram}
\end{figure}

While we include a layering-and-merging algorithm for PBC optimization, other algorithms will perform differently depending on the metrics used to grade them.  Thus, in Figure \ref{fig:pbc_opt_histogram} we plot the reduction in Pauli rotation count and the reduction (or increase) in average Pauli weight for circuits in FTCircuitBench.  This can serve as a characterizing fingerprint of the layering-and-merging algorithm, and other PBC optimization algorithms will produce different distributions in these metrics over the FTCircuitBench circuits.  For example, we find that the average Pauli weight of a PBC circuit often increases after applying our layering-and-merging algorithm to reduce Pauli rotation operations, showing that it tends to perform best at reducing low-weight Pauli operator counts.  

\subsection{Parameter Sweeps Across the Compilation Stack}
\label{sec:sweeps}
 
Since the library spans four synthesis pipelines and two Trotter settings
(Section~\ref{sec:circuit_params}), it constitutes a controlled sweep over the two dominant precision parameters: gate-synthesis precision and
algorithm-construction accuracy. Here we trace how these parameter changes
are reflected in the cost metrics of Section~\ref{sec:metrics}.

\begin{figure}[htbp]
    \centering
    \includegraphics[width=\linewidth]{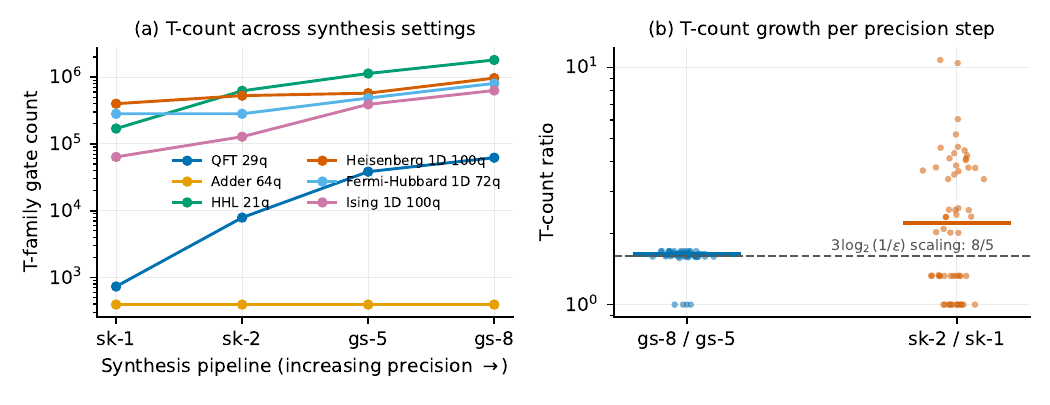}
    \caption{Synthesis-precision sweep over the benchmark library.
    \textbf{(a)} T-family gate counts for representative circuits across the
    four synthesis pipelines (ordered sk-1, sk-2, gs-5, gs-8; note that
    Solovay-Kitaev recursion levels and Gridsynth precisions are distinct
    accuracy scales). Discrete-gate workloads such as the adder are
    invariant, while rotation-dominated workloads grow with precision.
    \textbf{(b)} Distribution of per-circuit T-count growth ratios; horizontal bars mark the median of each distribution. The Gridsynth ratio concentrates tightly at the theoretical $3\log_2(1/\varepsilon)$ value of $8/5$ (dashed), while Solovay-Kitaev growth is angle-dependent and dispersed.}
    \label{fig:precision_sweep}
\end{figure}

\textbf{Synthesis precision and T-cost.}
Figure~\ref{fig:precision_sweep}(a) shows T-family gate counts across the
four pipelines for representative circuits, and
Figure~\ref{fig:precision_sweep}(b) the distribution of T-count growth per
precision step over the full library. For Gridsynth, the ancilla-free
T-count of a single $R_z$ synthesis scales as
$3\log_2(1/\varepsilon)$~\cite{10.5555/3179330.3179331}, predicting a ratio
of $8/5 = 1.6$ between the $\varepsilon = 10^{-8}$ and
$\varepsilon = 10^{-5}$ pipelines for rotation-dominated circuits. The
library reproduces this almost exactly: the median measured ratio is
$1.626$ with interquartile range $[1.60, 1.65]$ across 58 circuits. The outliers at ratio 1.0 are circuits with no rotations to synthesize. The adders, for example, contain T gates only from Toffoli decompositions, so their T-counts do not change with synthesis precision.

These absolute counts are anchored in published arithmetic costs: each
compiled adder's T-count is exactly seven times the Toffoli count of its
ripple-carry construction~\cite{cuccaro2004newquantumripplecarryaddition}
(8, 24, and 56 Toffolis yielding 56, 168, and 392 T~gates for the 10-,
28-, and 64-qubit instances), the cost of the standard ancilla-free
Toffoli decomposition~\cite{nielsen2010quantum}.

Solovay--Kitaev, by contrast, scales far less predictably: the sk-2/sk-1
T-count ratio has median $2.2$ but ranges up to $10.8$, reflecting the
sensitivity of recursive decomposition to the specific angles being
approximated. At the same time, gs-8 attains far tighter approximation error
than sk-2 at a median T-count only $2.4\times$ larger, quantifying, over the
whole library, why Gridsynth is the recommended default pipeline.
 
\textbf{Structural metrics are synthesis-invariant.}
Sweeping synthesis parameters moves gate counts by up to an order of
magnitude, yet the structural characterization of
Section~\ref{sec:metrics} is essentially unaffected: the interaction-graph
density is identical across all four pipelines for 58 of 58 circuits, and
the Louvain modularity is identical for 37 of 58 (with the remainder
differing only through the stochasticity of community detection on equal
graphs). This split cleanly separates the metrics into two classes:
\emph{cost} metrics (gate counts, depth, T-counts) that are properties of a
chosen compilation, and \emph{structural} metrics (interaction graphs,
modularity, density) that are properties of the algorithm itself.
Practically, this means qubit-mapping and partitioning decisions informed by
interaction structure can be made before committing to a synthesis
precision, while resource estimates cannot.
 
\begin{figure}[htbp]
    \centering
    \includegraphics[width=0.7\linewidth]{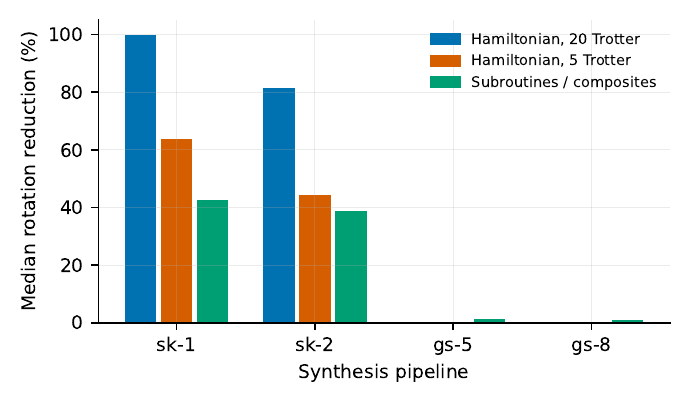}
    \caption{Median Pauli-rotation reduction under the layering-and-merging
    pass, by synthesis pipeline and workload class. Low-precision synthesis
    of high-accuracy (20-Trotter) Hamiltonian circuits trivializes them
    (100\% reduction under sk-1); aligning construction and synthesis error
    budgets (5-Trotter, or Gridsynth) removes the artifact.}
    \label{fig:rotation_reduction}
\end{figure}

\textbf{Interplay between Trotter accuracy and synthesis precision.}
The 20-Trotter-step Hamiltonian circuits have $R_z$ rotation angles of order
$0.1$ radians and smaller, close enough to identity that low-precision
synthesis, whose approximation error is comparable to or larger than the
angles themselves, replaces many of them with identity gates outright. The
surviving T gates arise largely from sequential $R_z$ gates on the same
qubit that sum to Clifford rotations, and are therefore highly cancelable
once translated to Pauli rotations and passed through the
layering-and-merging optimizer.
Figure~\ref{fig:rotation_reduction} quantifies the resulting artifact
across the library: under sk-1 synthesis, all 33 of the 20-Trotter
Hamiltonian circuits undergo a full $100\%$ Pauli-rotation reduction, and
sk-2 still yields a median $82\%$ reduction, while under Gridsynth at
$\epsilon = 10^{-5}$ or $10^{-8}$ the same circuits reduce by less than
$0.3\%$ at the median. Repeating the sweep on the 5-Trotter variants, whose
larger rotation angles survive coarse synthesis, sk-1's median reduction
falls to $64\%$ and no circuit reaches $100\%$. Subroutine and composite
circuits, whose rotation angles are not concentrated near zero, show
intermediate reductions ($\approx 40\%$ under Solovay-Kitaev) with no full
reduction to Clifford gates at any setting. This shows clearly that the error budgets of algorithm construction and gate synthesis must be approximately aligned, since a synthesis error exceeding
the rotation angles renders a benchmark circuit degenerate; and before any
advanced compiler pass, an ``easy'' first pass should merge sequential
$R_z(\theta)$ gates and replace them with their standard representations
(T, S, Z) whenever possible.

\subsection{Case Study: From Weight Distributions to Architectural Verdicts}
\label{sec:case_study}
 
To make concrete how the structural metrics drive architectural decisions,
consider two 100-qubit spin-model circuits from the library under the gs-8
pipeline: the 1D Ising model and the 2D triangular-lattice Heisenberg model.
By conventional aggregate metrics they are siblings: both are 100-qubit Trotterized spin evolutions compiled through the same pipeline. Their PBC weight profiles, however, are opposites. The
Ising circuit's 630{,}520 Pauli rotations have mean weight $19.97 \pm
29.54$, with the bulk of operators at weights between 1 and 10, while the
Heisenberg circuit's 2{,}902{,}360 rotations have mean weight $73.51 \pm
9.22$: the small standard deviation means essentially \emph{every} operator
is high weight (Appendix~\ref{sec:pbc_stats_table}).
 
Reading this contrast through the cost models of Section~\ref{sec:metrics}
yields concrete, and different, verdicts per architecture. On surface-code
lattice surgery, the ancilla region of an average Heisenberg measurement
must span roughly 74 patches, incurring the correspondingly enlarged
logical-error exposure of the merged region~\cite{litinski2019game}, and
because supports of that size on 100 qubits almost surely intersect,
consecutive measurements serialize; the Ising circuit's low-weight,
chain-local operators admit small routing regions and non-intersecting
parallel scheduling. On a BB-code architecture, the verdict can be stated
independently of any qubit-placement choice: an operator of weight $w$ must
touch at least $\lceil w/k \rceil$ code blocks with $k = 12$ logical qubits
per gross-code block~\cite{bravyi2024high}, so the average Heisenberg
operator spans at least 7 blocks and sits deep in the fidelity-degrading
inter-block regime, where joint measurements between modules are simulated at roughly two orders of magnitude higher logical error rates than in-module measurements~\cite{yoder2025tour}, whereas typical Ising operators
fit within one or two blocks, precisely the regime in which BB-code logical
error rates are weight-independent. An extractor-augmented qLDPC
architecture~\cite{he2025extractors} neutralizes the weight difference
entirely, one logical cycle per measurement regardless of weight, but at
that point the binding metric flips from weight to operator \emph{count},
where the Heisenberg circuit is $4.6\times$ more expensive.
 
The architectural decisions that follow are exactly the kind a
count-based estimate cannot make. For the Ising workload, PBC execution is
attractive on any of the three architectures, and the operator count, not
the weight profile, is the metric a compiler should minimize. For the
Heisenberg workload, PBC execution is unattractive on patch-based
architectures at any placement; the options are to execute in the
Clifford+T model instead, where the circuit remains two-qubit-local, to
provision extractor-style hardware that amortizes weight, or to attack the
weights upstream at the Trotterization and synthesis layers. To a count-based estimator, the two circuits differ only by a scalar: one has more operations than the other. Nothing in those counts reveals that one circuit is PBC-friendly on either architecture while the other is not; the weight distributions are what separate them, and they are obtainable only by performing the PBC compilation that \sol{} provides.

\subsection{Bridging to Physical Resource Estimation}
\label{sec:qre_bridge}
 
Because \sol{} materializes logical circuits, its outputs can be handed
directly to physical resource estimators. We demonstrate this with the Azure
Quantum Resource Estimator~\cite{beverland2022assessing}, using its logical
counts interface: for each of the 234 circuit/pipeline combinations in the
library, we supply the compiled qubit and T-family counts and obtain
physical qubit numbers, runtimes, code distances, and magic-state factory
requirements under configurable hardware assumptions. The bridge ships in the repository as a supported subpackage and command-line tool (\texttt{ftcircuitbench.resource\_estimation}), with six
built-in hardware models spanning superconducting, trapped-ion, and
Majorana qubit parameters, and can price either the Clifford+T counts or
the post-optimization PBC operator counts of a circuit; every estimate in
this section is reproducible from the released statistics.
 
\begin{figure}[htbp]
    \centering
    \includegraphics[width=\linewidth]{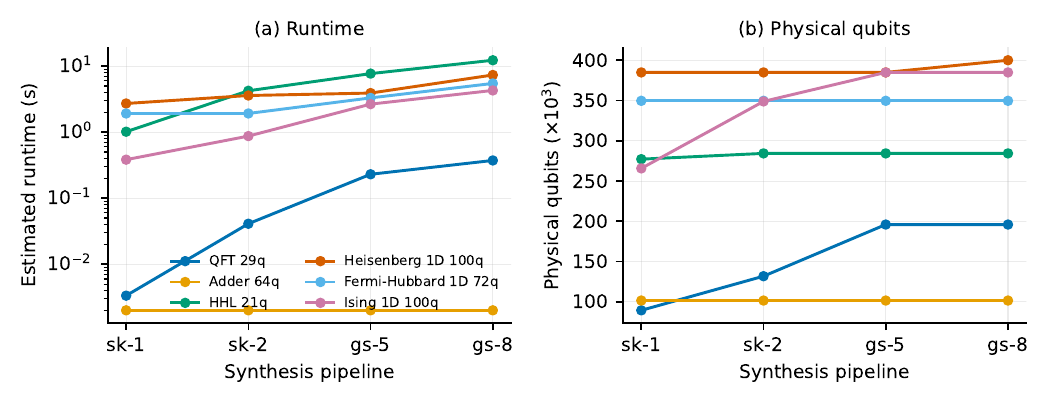}
    \caption{Physical resource estimates for representative circuits across
    the four synthesis pipelines, obtained by supplying \sol{} logical
    counts to the Azure Quantum Resource Estimator under a
    superconducting-style qubit model ($10^{-3}$ physical error rate, 1\%
    error budget). \textbf{(a)} Estimated runtime follows T-count.
    \textbf{(b)} Physical qubit counts are set by width and code distance
    and are nearly insensitive to synthesis precision.}
    \label{fig:qre_sweep}
\end{figure}

Figure~\ref{fig:qre_sweep} extends the parameter sweep of
Section~\ref{sec:sweeps} to the physical level, under a
superconducting-style qubit model ($10^{-3}$ physical error rate, 1\% total
error budget). The sweep exposes a clean division of consequences: synthesis
precision translates almost entirely into \emph{runtime}, which follows the
T-count linearly across the four pipelines, while the \emph{physical qubit}
footprint is nearly flat, set by circuit width and code distance, and moves
only when the tighter error budget implied by a larger T-count forces a
code-distance step (e.g., the 100-qubit Heisenberg model steps from
$d = 17$ to $d = 19$ between gs-5 and gs-8). Discrete-gate workloads such
as the adders are invariant on both axes. In practical terms, tightening Gridsynth from $\varepsilon = 10^{-5}$ to
$10^{-8}$ makes the 100-qubit Heisenberg simulation run $1.9\times$ longer,
while its physical qubit footprint grows only $4\%$, from a single
code-distance step.
 
The comparison also runs in the other direction: the estimator's accounting
assumes the PSSPC execution scheme, one sequential Pauli measurement per
non-Clifford operation, without performing the compilation. Comparing
against \sol{}'s explicitly compiled and merged PBC circuits quantifies
where that assumption is faithful. For Hamiltonian-simulation, QFT, and HHL
circuits under Gridsynth pipelines, count-based accounting is accurate to
within about $2\%$: merging removes almost nothing, and one PPM per T gate
is a good model of the compiled circuit's length. For the arithmetic
circuits, however, layering-and-merging removes $43\%$ of Pauli rotations,
so count-based accounting overestimates the sequential PPM count by
$1.75\times$. And in all cases the count-based model is blind to the weight
structure of the surviving operators, which, as Section~\ref{sec:case_study}
shows, determines whether the one-measurement-per-step assumption is
realistic on a given architecture at all. Explicit compilation and
count-based estimation are thus complementary: the estimator supplies the
physical-level context that \sol{} does not model, and \sol{} supplies the
compiled circuit structure that the estimator abstracts away.
 
\subsection{Distilled Observations of Fault-Tolerant Compilation}
\label{sec:distilled}
The trivialization artifact quantified in Section~\ref{sec:sweeps} carries a
broader lesson for benchmark consumers: entirely unoptimized circuits,
combined with low-precision compilation, can produce metrics that reflect
the interaction of compilation layers rather than the algorithm itself.
While it may seem a reasonable tradeoff to pair low Trotter error with
cost-saving low-precision Clifford+T decomposition, the result is a
trivialized circuit whose non-Clifford content merges and cancels entirely.
Naively mixing precision settings across compilation layers can therefore
yield non-physical or misleading results, and multi-stage pipelines should
validate that per-layer error budgets are mutually consistent.

When analyzing the resource requirements of an algorithm or developing a new one, tools such as those provided in FTCircuitBench are useful for uncovering structure and symmetries.  For example, in Figure \ref{fig:hhl_density}, it is visually clear how the HHL algorithm is constructed via state preparation of the vector $\Vec{b}$ on qubits 17-20, controlled Hamiltonian simulation on the clock register of qubits 1-16 to estimate the eigenvalues of $A$, and then controlled rotations on the ancilla, qubit 0, to effectively invert the eigenvalues.  This is all followed by uncomputation, leading to the symmetric structure of the colormap.  When seen in conjunction with the highly structured interaction graphs of HHL circuits in \ref{fig:c+t interaction graphs}, it is visually intuitive that the $\Vec{b}$ register forms the highly connected "core" of the circuit, and thus compilation researchers can quickly understand the subroutines and components of this algorithm even if they were not previously familiar with it.  In this way, the analysis tools provided in FTCircuitBench can both provide valuable statistics for informing compilation decisions, as well as to serve as quickly informative and mutually complementary visual representations of quantum circuits.

FTCircuitBench also addresses a challenge particular to PBC: while Clifford+T circuits are often straightforward to understand structurally once the algorithmic components are understood, their corresponding PBC circuits are difficult to anticipate. In particular, commuting entangling gates through the T-rotation tableau can lead to unexpected Pauli weight distributions.  This can make it difficult to predict whether a given algorithm is most suited to be run on hardware implementing Clifford+T circuits or the PBC model.  By providing a PBC compiler, summary statistics, and visualization tools, FTCircuitBench serves as a useful tool for analyzing these co-design decisions and characterizing the most predictive metrics for successful execution in either computational model.

\subsection{From Static Circuit Features to System-Level Performance}
\label{sec:system_level}
 
The metrics in this work characterize the \emph{static} outputs of
compilation, whereas delivered performance on a fault-tolerant machine is
ultimately set by dynamic quantities: magic-state delivery, scheduling, and
real-time decoding throughput~\cite{battistel2023real}. The static profiles
are nonetheless direct inputs to those dynamics. Spatial-temporal T-gate
densities bound both magic-state factory demand and its \emph{delivery}
contention: temporally clustered demand determines required factory
throughput~\cite{ye2026tdepth}, while spatially clustered demand concentrates
routing traffic and decode-dependent corrections on specific patches,
localizing decoder reaction-time pressure. Pauli-operator supports likewise
bound achievable parallelism, since simultaneous Pauli product measurements
require non-intersecting routing regions in lattice
surgery~\cite{litinski2019game} and disjoint measurement resources in
extractor-based architectures~\cite{he2025extractors}; and interaction-graph
modularity indicates how cleanly a workload partitions across tiles, chips,
or modules. In this sense, the gap between a count-based estimate and
delivered system throughput is governed largely by exactly the structural
features \sol{} reports.
 
We are explicit about validation status: these metrics are descriptive
statistics of compiled logical circuits. Their cost interpretations are
grounded in published, hardware-derived cost models of the underlying
operations rather than in direct measurements, and we have not yet validated
them against cycle-level architectural simulation or hardware execution.
Doing so---coupling \sol{} circuit artifacts to lattice-surgery schedulers,
decoder-aware simulators, and, as they mature, logical-qubit hardware
demonstrations---is a primary direction for future work, and the benchmark
library is designed to serve as the workload suite for precisely such studies.

%% file: text/conclusion.tex
\section{Conclusion \& Outlook}
\label{sec:conclusion}

Realizing the potential of fault-tolerant quantum computation requires a deep understanding of how algorithmic requirements interact with architectural constraints. The transition from NISQ-era qubit-level computation to fault-tolerant logical execution introduces a complex parameter space where local optimizations can have unforeseen global consequences. We put forward FTCircuitBench to navigate this complexity, offering a standardized environment and modular toolkit to rigorously evaluate the full fault-tolerant compilation stack. By characterizing algorithms through the lens of Clifford+T and Pauli Based Computation, we offer a stable reference point for evaluating the interplay between algorithmic structure and architectural execution.

The necessity of such a global view is exemplified by the artifacts observed when interfacing different compilation layers. For instance, our benchmarks highlight that standard gate decomposition techniques, when set to lower precisions, can interact with PBC compilers to produce trivialized circuits. While this interaction is a natural consequence of the compilation logic, it underscores the critical need to balance local approximation errors against global execution overheads. Without an end-to-end perspective, such sensitivities can lead to distorted resource estimates or unphysical circuit representations, obscuring the true computational costs of an algorithm.

FTCircuitBench supplies the quantitative tooling necessary to navigate these trade-offs. It enables researchers to answer many important questions of the FTQC era: How precise can my gate decompositions be while maintaining algorithmic integrity and minimizing overhead? And where do the actual bottlenecks lie when abstract algorithms are mapped to concrete logical topologies? And how do the structural properties of a compiled circuit reshape the verdicts of count-based physical resource estimation? By making these structural properties and trade-offs visible, from T-gate densities to Pauli weight distributions, we aim to support holistic co-design, where algorithms, error-correcting codes, and computational models are optimized in concert.

%% file: text/background.tex
\section{Background}
\label{sec:background} 

\subsection{Quantum Error Correction}

Achieving reliable large-scale quantum computation requires overcoming the challenge of hardware noise. QEC provides a leading strategy by encoding logical qubits into many physical qubits, and enables an exponential suppression of logical error rates~\cite{nielsen2010quantum,gottesman1997stabilizer}. In the stabilizer formalism, a QEC code defines a logical subspace as the simultaneous \(+1\) eigenspace of a commuting set of stabilizers \(\{S_i\}\)~\cite{gottesman1997stabilizer}. Errors \(E\) are detected through stabilizer measurements: if \(E\) anticommutes with a stabilizer \(S_i\), the outcome of measuring that stabilizer flips to \(-1\), producing a syndrome. This syndrome is then processed by a decoder to infer and correct the error, either by applying a compensating physical operation or by updating a classical record of the Pauli frame.  A broad family of codes arises from the Calderbank–Shor–Steane (CSS) construction \cite{calderbank1996good}, which employs pairs of classical linear codes to define independent \(X\)-type and \(Z\)-type stabilizer checks~\cite{calderbank1996good}.

Quantum error-correcting codes are specified by parameters \([[n,k,d]]\), where \(n\) physical qubits encode \(k\) logical qubits with code distance \(d\). The distance quantifies the minimum weight of an error that can cause a logical error \cite{nielsen2010quantum}. A central code performance metric is the error threshold \(p_{\mathrm{th}}\).  When the physical error rate \(p_{\mathrm{phys}}\) falls below this threshold \((p_{\mathrm{phys}} < p_{\mathrm{th}})\), the logical error rate \(p_{\mathrm{log}}\) decreases exponentially with increasing distance, often approximated as $p_{\mathrm{log}} \propto \left(\frac{p_{\mathrm{phys}}}{p_{\mathrm{th}}}\right)^{\lceil (d+1)/2 \rceil}$\cite{fowler2012surface}. Conversely, for \(p_{\mathrm{phys}} > p_{\mathrm{th}}\), errors proliferate rather than be suppressed, and logical error rates increase. A canonical example is the surface code, which achieves a threshold of \(p_{\mathrm{th}} \approx 0.8\%\) under standard noise models and decoding~\cite{fowler2012surface}, with logical error rates well approximated by $p_{\mathrm{log}} \approx 0.03 \big(100\,p_{\mathrm{phys}}\big)^{(d+1)/2}$~\cite{stein2025hetec}.
For practical QEC, codes with high thresholds, favorable encoding rates (\(k/n\)) and a path toward realizing fault-tolerant logical computation are desired.

Implementing universal quantum computation fault-tolerantly introduces additional challenges, particularly in the realization of non-Clifford gates. Clifford operations such as the Hadamard (\(H\)), phase (\(S\)), and controlled-NOT (CNOT) gates, while fault-tolerantly implementable in many codes, are insufficient for universal computation and can be efficiently simulated on a classical computer~\cite{gottesman1998heisenberg}. To achieve fault-tolerant universal computation, non-Clifford operations must be included, most commonly realized through the injection of specially prepared ancillary states known as magic states. Among these, the T-gate, defined as $T = \mathrm{diag}(1, e^{i\pi/4})$ is the most extensively studied \cite{bravyi2012magic,gidney2025factor}. Together with Clifford operations, the T-gate forms a universal gate set, enabling arbitrary quantum computations \cite{10.5555/3179330.3179331,campbell2017roads}.

Two code families that have attracted significant attention in recent years are the surface code~\cite{fowler2009high,fowler2012surface} and the Bivariate Bicycle (BB) family of high-rate low-density parity-check (qLDPC) codes, proposed in Bravyi et al.~\cite{bravyi2024high}. The surface code enjoys a strictly two-dimensional local connectivity, making it particularly well-suited to platforms such as superconducting qubits~\cite{fowler2012surface}. Universality can be achieved through magic state injection, which can be implemented via lattice surgery, with the requisite \(|T\rangle\) states supplied by distillation or cultivation protocols~\cite{bravyi2005universal,gidney2024magic,fowler2012surface, litinski2019magic}. In contrast, high-rate qLDPC codes promise substantially better logical qubit yield. The encoding rate of the surface code decreases quadratically with distance (\(k/n = \mathcal{O}( 1/d^2)\)), whereas non-local qLDPC codes can potentially attain a constant rate \cite{panteleev2022asymptotically} . One code-specific BB qLDPC code is the gross code, a \([[144,12,12]]\) code achieving \(r = 1/24\), requiring only 288 physical qubits compared to the thousands of qubits needed for surface code patches of comparable distance, while still maintaining a threshold near \(p_{\mathrm{th}} \approx 1\%\)~\cite{bravyi2024high}. The trade-off is that high-rate qLDPC codes necessitate structured non-local connectivity (e.g., thickness-2 graphs), which imposes demands on hardware design and introduces compilation overhead, but in exchange offers compelling resource savings.

Beyond code properties themselves, executing algorithms fault-tolerantly on either code family requires careful consideration of compilation models and resource trade-offs, including mapping and optimization to manage substantial overheads in qubits, runtime, magic states, and measurement rounds~\cite{litinski2019game,beverland2022assessing}. These challenges underscore the importance of co-design between algorithms and QEC architectures, as demonstrated in recent works that explore how algorithmic structure and choice of code jointly impact overall resource costs~\cite{kan2025sparo,stein2025hetec,jin2025iceberg,wang2024optimizing,hua2021autobraid}.


\subsubsection{The Surface Code}

The surface code is a prominent example of a topological quantum error correcting code, valued for its high error threshold, nearest-neighbor connectivity requirement and clear path to fault-tolerant computation ~\cite{fowler2012surface}. A standard rotated distance-\(d\) surface code uses \( d^2 \) physical data qubits to encode \(k = 1\) logical qubit. The surface code is both a type of CSS code and a qLDPC code.

The code is typically defined on a square lattice. In the common rotated variant, data qubits are placed at the vertices of a grid, and stabilizer operators (both X-type and Z-type) correspond to plaquettes (faces) of this grid. These stabilizer generators have weight at most 4 (local) and all mutually commute. Stabilizers at the boundaries of the lattice have weight 2.

\begin{figure}
    \centering
    \includegraphics[width=0.95 \linewidth]{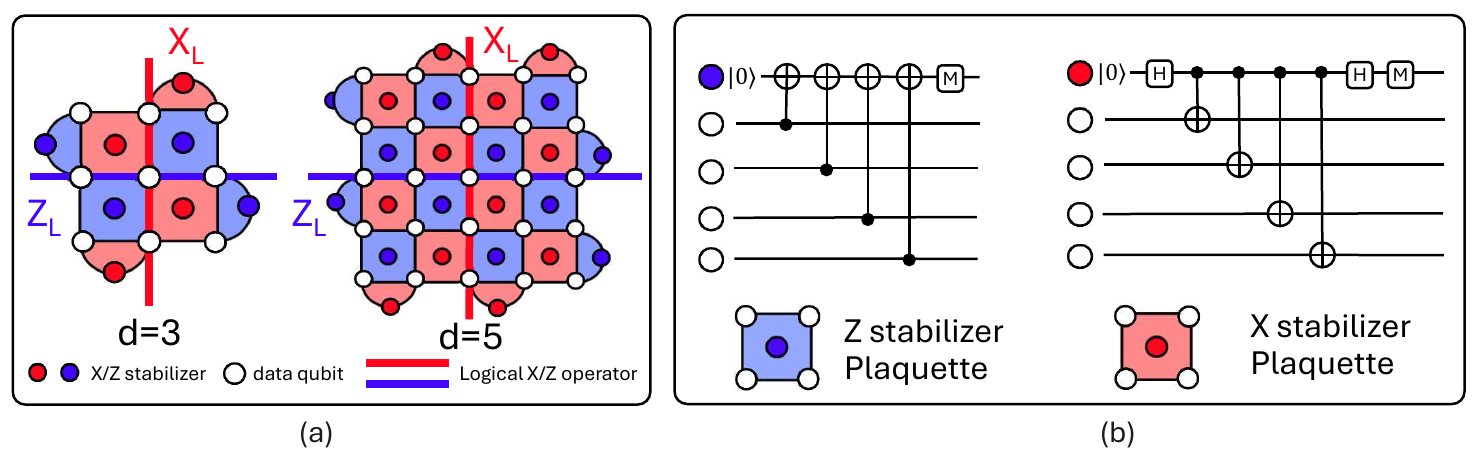}
    \caption{\textbf{Left:} Rotated surface code layout. White nodes denote data qubits; red (blue) nodes denote \(X\)-type (\(Z\)-type) stabilizer measurement qubits. Colored plaquettes indicate stabilizer generators: red corresponds to \(XXXX\) (or \(XX\)) and blue to \(ZZZZ\) (or \(ZZ\)). Logical operators are strings of data qubits running along the highlighted red/blue paths. Qubits are arranged on a 2D grid, requiring only local connectivity between each plaquette’s data-qubit vertices and its central stabilizer qubit. \textbf{Right:} Example syndrome-extraction circuit measuring the indicated \(X\) or \(Z\) stabilizers, assuming all state initialization and measurements are performed in the \(Z\) basis.}

    \label{fig:stabilizer_circuits}
\end{figure}

Error detection proceeds by repeatedly measuring these stabilizer generators. A simultaneous eigenstate measurement of \(+1\) indicates that the underlying state is a valid code space state, while \(-1\) signals an error has occurred. The collection of these outcomes forms the error syndrome. A classical decoding algorithm, such as Minimum Weight Perfect Matching (MWPM) \cite{higgott2022pymatching, Higgott2025sparseblossom}, uses the syndrome to infer the most probable error configuration, and tracks errors in software, removing the need to correct errors in hardware.  An operator representing \(X_L\) or \(Z_L\) (modulo stabilizers) defines the code distance \(d\). Figure~\ref{fig:stabilizer_circuits} shows a circuit for measuring weight-4 stabilizers, as well as the code topology on planar hardware.

Regarding performance, the surface code demonstrates a high error threshold (\(p_{th} \approx 0.8\%\)) under circuit-level uniform depolarizing noise~\cite{fowler2009high}, establishing it as a robust topological quantum error-correcting code. This theoretical robustness has been further supported by recent experimental demonstrations~\cite{google2025quantum}. Universal quantum computation within the surface code framework is typically achieved using the Clifford+T gate set \(\{S, H, T, \mathrm{CNOT}\}\)~\cite{bravyi2005universal, fowler2012surface, gidney2024magic, horsman2012surface}.

\subsubsection{High-Rate qLDPC Codes: The Bivariate Bicycle Family}

Quantum Low-Density Parity-Check (qLDPC) codes are stabilizer codes whose parity checks are sparse: each check involves only a constant number of qubits, and each qubit participates in only a constant number of checks. Unlike the surface code, high-rate qLDPC codes do not have strict two-dimensional geometric locality, but their bounded-degree structure still makes them attractive for implementation on realistic hardware. A variety of qLDPC constructions have been proposed, including BB codes~\cite{bravyi2024high}, hypergraph product codes~\cite{krishna2021fault}, lifted product codes~\cite{guemard2025lifts}, and quantum Tanner codes~\cite{leverrier2022quantum}. Of particular interest are \emph{high-rate qLDPC codes}, which achieve an asymptotic constant encoding rate ($k/n = \Omega(1)$) and, ideally, polynomial or even linear distance ($d = \Omega(n)$). Such codes promise significantly lower qubit overhead compared to topological codes like the surface code, which have vanishing rate \cite{panteleev2022asymptotically}.

For evaluation and benchmarking, it is crucial to consider codes with clear properties and realistic implementation paths. Among high-rate qLDPC codes, BB codes~\cite{bravyi2024high,yoder2025tour} are especially promising, offering a concrete route toward practical QEC. BB codes are CSS codes constructed algebraically using pairs of bivariate polynomials \(A, B \in \mathbb{F}_2[x, y]\) subject to the conditions \(x^l = 1\) and \(y^m = 1\). BB codes are defined using \(2lm\) data qubits, conceptually arranged into two \(l \times m\) blocks, denoted L and R. There are $lm$ X-type stabilizer checks and $lm$ Z-type stabilizer checks, of which $2lm - k$ are linearly independent. The parity check matrices are given by \(H^X = [A|B]\) and \(H^Z = [B^T|A^T]\), where the transposition is defined appropriately for the polynomial representation. The defining polynomials \(A\) and \(B\) are chosen to have low weight (e.g., three terms each), which results in the sparse check matrices characteristic of qLDPC codes. 

One prominent BB code receiving increasing attention, with early hardware demonstrations~\cite{wang2025demonstration}, is the gross code~\cite{bravyi2024high,cross2024improved}. This \([[144,12,12]]\) BB code arises from parameters $l=12, m=6$, using $n=144$ data qubits (or $2n=288$ total physical qubits across the $L$ and $R$ blocks). It encodes $k=12$ logical qubits, yielding a net encoding rate of $1/24$. This represents a significant reduction in qubit count compared to surface code patches achieving similar logical performance~\cite{bravyi2024high}. The code also achieves competitive thresholds ($p_{\text{th}} \approx 1\%$ under circuit-level noise) and supports efficient syndrome extraction, including a depth-8 CNOT measurement circuit~\cite{bravyi2024high}. The gross code’s stabilizers, derived from low-weight bivariate polynomials~\cite{bravyi2024high}, have weight six. Its algebraic structure produces a highly regular Tanner graph: each qubit participates in exactly six checks (three X-type and three Z-type), and each check acts on six qubits. This uniform degree-6 connectivity contrasts with the surface code’s variable, typically lower-weight stabilizers. Although BB codes require non-local connectivity on a simple 2D grid, their Tanner graphs admit a "thickness-2" decomposition into two planar subgraphs~\cite{bravyi2024high}. This suggests possible implementation pathways: in superconducting circuits, the non-local interactions may be realized through multi-layer fabrication and through-substrate vias~\cite{mallek2021fabrication,kosen2022building}, while neutral-atom and trapped-ion platforms, which naturally support long-range entangling operations, can accommodate such connectivity with relatively low overhead.




\subsubsection{The Evolving Landscape of Quantum Error Correction Codes}

While the surface code and Bivariate Bicycle codes (the family that includes the gross code) are prominent candidates for fault-tolerant quantum computation, the search for better error-correcting code continues. A code's quality is not captured by a single metric but is a combination of its abstract properties and physical implementation. While the parameters $[[n,k,d]]$, representing the number of physical qubits, logical qubits, and code distance, respectively, provide an important first-order assessment, a comprehensive evaluation must also consider several practical factors. These include, but are not limited to \textit{(i)} the compatibility of the code's stabilizer graph with the underlying physical qubit architecture \cite{xu2024constant, berthusen2024toward}; \textit{(ii)} the structure and overhead of its fault-tolerant logical gate set \cite{knill1997theory}; and \textit{(iii)} the complexity and latency of its classical decoding requirements, which must operate in real-time \cite{battistel2023real}. The codes emphasized in this work serve as two high-potential codes, but the landscape of QEC is dynamic, with ongoing research into improving and discovering codes. Specifically, significant work has been directed towards non-local qLDPC codes, partly motivated by the Brayvi-Poulin-Terhal (BPT) theorem, which proves that a 2D local code is bound by $kd^2=\mathcal{O}(n)$ \cite{bravyi2010tradeoffs}.

Furthermore, a universal constraint on all such codes is imposed by the Eastin-Knill theorem, which states that no single quantum error-correcting code can provide a universal, transversal fault-tolerant gate set~\cite{eastin2009restrictions}. Realizing universal FTQC requires the support of a universal gate set, which is most commonly implemented via magic state distillation for gates such as the T or CCZ gate \cite{gidney2024magic,gidney2019efficient}. This requirement introduces significant resource overheads and necessitates a sophisticated compilation process to translate high-level quantum algorithms into the specific fault-tolerant instruction set of a given code. The practical viability of any candidate code, therefore, hinges on a quantitative evaluation of these compilation overheads. How well a code can support the required logical computation consequently impacts its potential. 

Prior works have demonstrated the existence of asymptotically "good" qLDPC codes with constant rate \cite{panteleev2022asymptotically,leverrier2022quantum}, while modified surface codes, such as the XZZX surface code \cite{bonilla2021xzzx}, have shown significantly improved logical error suppression under biased noise models. At the same time, the development of Floquet codes has generalized stablizer codes by allowing the stabilizer group, and thus the instantaneous code space, to evolve periodically in time \cite{Hastings_2021}.  This provides benefits such as reduced hardware overhead by only requiring low weight measurements, and techniques for Floquetifying codes \cite{townsend2023floquetifying} can then transmit these underlying benefits of Floquet codes to static stabilizer codes. These developments, among many others, exemplify progress in the evolving landscape of QEC and the pursuit for better fault-tolerant computational models.

\subsection{Fault-Tolerant Computation Models and Non-Clifford Resources}
\label{sec:ft_computation}

\subsubsection{From High-Level Algorithm to Logical Operators}
\label{sec:algo_to_logical}

The process of transforming a high-level quantum algorithm into a sequence of fault-tolerant logical operations involves several stages. Initially, the algorithm, which might be specified as a Hamiltonian evolution problem or a sequence of abstract operations, must be decomposed into a universal set of elementary quantum gates. Compilers such as Kernpiler~\cite{decker2025kernpiler} can map quantum simulation problems directly to circuits consisting of 1-qubit and 2-qubit gates. Alternatively, standard quantum compilation toolkits like Qiskit~\cite{javadi2024quantum} or \(t|ket\rangle\)~\cite{sivarajah2020t} provide routines for decomposing arbitrary unitaries into such elementary gates, typically including single-qubit rotations (e.g., \(R_z(\theta)\)) and CNOTs.

Once the algorithm is expressed in terms of elementary gates (including rotation gates of arbitrary continuous angles), these must be further translated into the fault-tolerant logical gate set supported by the chosen QEC architecture, often the Clifford+T set. Ross and Selinger~\cite{10.5555/3179330.3179331} provide a well-known algorithm that generates near-optimal, ancilla-free Clifford+T sequences for arbitrary \(R_z(\theta)\) rotations. While alternative decomposition methods exist, potentially offering lower T-depth by utilizing ancilla qubits, repeat-until-success circuits, or measurement-based operations~\cite{paetznick2013repeat, bocharov2015efficient}, the Ross-Selinger approach is a common baseline. The Solovay-Kitaev algorithm is an alternative foundational algorithm for performing this decomposition \cite{kitaev1997quantum, dawson2005solovay}, though not being specialized to the Clifford+T gate set, it also does not produce as low T-gate counts as the Ross-Selinger algorithm does for the same $\epsilon$ approximation precision.  After this stage, the entire algorithm is represented as a circuit of Clifford and T-gates.

If the target computational model is PBC, an additional compilation step is required. The Clifford+T circuit is transformed by applying the commutation rules outlined in Appendix~\ref{sec:pbc_model}. This process systematically eliminates all Clifford gates, leaving a sequence of varying weight non-Clifford Pauli product measurements (PPMs). This final circuit is then ready for resource estimation or mapping onto a PBC-compatible architecture. Despite the differences between Clifford+T and PBC, both models require magic states to achieve universal quantum computation.

\subsubsection{Magic States for Universal Computation}
\label{sec:magic_states}

Magic states are specific non-Clifford quantum states that are not efficiently classically simulatable \cite{howard2017application}. When combined with Clifford operations through protocols such as state injection and gate teleportation, they enable the execution of non-Clifford logical gates \cite{bravyi2005universal}. One of the most extensively studied magic states is the T-state, defined as \(|T\rangle = \frac{1}{\sqrt{2}}(|0\rangle + e^{i\pi/4}|1\rangle)\). The ability to implement the logical T-gate, facilitated by the T-state, when combined with a Clifford gate set allows for universal quantum computation. 

A significant challenge is the generation and distillation of these magic states with high fidelity. To achieve the required logical error rates necessitated, protocols known as magic state distillation or cultivation can be employed. We place specific emphasis on the T-state purification protocols.

\begin{itemize}
    \item \textbf{Magic State Distillation} protocols take multiple copies of lower-fidelity (noisy) magic states as input and, through a sequence of Clifford operations and measurements, probabilistically distill fewer magic states at higher fidelity. One commonly referenced protocol is the 15-to-1 T-state distillation protocol \cite{bravyi2012magic}, which consumes 15 noisy T-states with error rate \(p\) to yield one output T-state with an error rate of approximately \(35p^3\),  achieving cubic error suppression~\cite{litinski2019game}. While effective, distillation can incur substantial overhead in terms of both the number of physical qubits and the time required for the multi-round protocol.
    \item \textbf{Magic State Cultivation}, a recently proposed algorithm detailed by Gidney et al.~\cite{gidney2024magic}, offers a cultivation approach. Rather than combining many noisy states, cultivation focuses on iteratively "growing" the reliability and effective code distance of a single encoded magic state. This process involves an initial injection into a small code, followed by rounds of error-detecting cross-checks and potential code size increases (the cultivation stage), and finally an "escape" into a larger, more robust code which contains the high fidelity magic state. Magic state cultivation is a landmark result that drastically reduced the overhead of T-state generation, and prior works continue to highlight the promise of cultivation \cite{claes2025cultivating,sahay2025fold}
\end{itemize}

The synthesis of high-fidelity magic states, whether through distillation or cultivation, represents a major contributor to the overall resource cost of fault-tolerant quantum computation due to the complex, often lengthy, checking and processing procedures involved.

While we focus on the T-state as the magic state of choice due to the alignment with the computational models we describe in this work, other magic states can also enable universal computation. For instance, the CCZ (controlled-controlled-Z) state\cite{gidney2019flexible, gidney2019efficient}, when combined with Clifford operations, also forms a universal gate set. The domain of high fidelity magic state synthesis covers a multitude of magic states and systems, with examples such as qutrit magic state distillation \cite{anwar2012qutrit} or arbitrary-angle distillation \cite{duclos2015reducing}. Regardless of the specific choice, the preparation of high-fidelity magic states is imperative in universal fault-tolerant quantum computing. 

\subsubsection{$R_z$ to Clifford+T Decomposition}
\label{sec:Rz to Clifford+T}
Universal quantum computation requires the ability to perform arbitrary single-qubit unitary operations in $SU(2)$. To ensure fault-tolerance, these operations must be decomposed into a finite universal gate set. Within the Clifford+T framework, this decomposition generally focuses on approximating arbitrary Z-rotations ($R_z(\theta)$) to a given precision $\epsilon$, as other single-qubit rotations can be constructed from Z-rotations and Clifford gates (e.g., $R_x(\theta) = H R_z(\theta) H$). With purely unitary circuit designs, the information-theoretic lower bound on the number of T-gates (T-count) scales as $3 \log_2(1/\epsilon)$ \cite{10.5555/3179330.3179331}. 

A state-of-the-art decomposition method is the Ross-Selinger Gridsynth algorithm~\cite{10.5555/3179330.3179331} - a method based on solving Diophantine equations to decompose arbitrary Z-rotations into Clifford+T sequences. This algorithm is a widely adopted and highly efficient method for generating near-optimal T-depth sequences of single-qubit Clifford+T gates without relying on additional techniques such as ancillas, measurements, or state distillation. While the Gridsynth algorithm can find the T-optimal solution (shortest T-count) with a factoring oracle, it typically yields circuit approximations with a T-count of $3 \log_2(1/\epsilon) +\mathcal{O}(\log(\log(1/\epsilon)))$ for a given precision $\epsilon$, even without such an oracle, under mild number-theoretic hypotheses \cite{10.5555/3179330.3179331}.  Furthermore, the Gridsynth algorithm has an expected runtime of $\mathcal{O}(polylog(1/\epsilon))$ \cite{10.5555/3179330.3179331}.  

A foundational, and more general, alternative is the Solovay-Kitaev algorithm, which can approximate any arbitrary single-qubit gate in any universal gate set by setting base-depth and recursion degree parameters. A sequence of gates generating a unitary $S$ is said to be an $\epsilon$-approximation of a gate $U$ if $||S-U||\le \epsilon$ in operator norm.  The Solovay-Kitaev algorithm works recursively, starting with a base $\epsilon_0$-approximation of the given single-qubit gate $U$ found through exhaustive search of sequences in the desired gate set, up to a bounded depth. At each higher recursion level of the algorithm, an $\epsilon_n$-approximation, $U_n$, is generated from a previous $\epsilon_{n-1}$-approximation, $U_{n-1}$, such that $\epsilon_n \le \epsilon_{n-1}$ and $\epsilon_n \to 0$ as $n \to \infty$.  This is done by finding an $\epsilon_n$-approximation to $\Delta = U U_{n-1}^\dagger$, and then returning $U_n =  \Delta U_{n-1}$.  For a detailed discussion on how $\Delta$ is calculated, we refer to the overview provided by Dawson and Nielson \cite{dawson2005solovay}.  The Solovay-Kitaev recursive method produces gate sequences whose T-counts scale as $\mathcal{O}(\log^c(1/\epsilon))$ where $c$ is approximately 3.97, doing so with a runtime of $O\bigl(\log^{2.71}(1/\epsilon)\bigr)$ \cite{dawson2005solovay}.  While historically significant for proving that efficient approximation is possible, this scaling is less favorable than that of specialized methods for Clifford+T decompositions.

It is important to acknowledge that alternative decomposition strategies exist which can lead to significantly lower T-counts. For example, techniques based on Repeat-Until-Success (RUS) circuits have demonstrated the potential to reduce the expected T-count by 2.5 times compared to the theoretical lower bound for ancilla-free decompositions, at the cost of extra ancilla qubits \cite{bocharov2015efficient}. However, the primary objective of FTCircuitBench is to provide a standardized set of unoptimized, baseline quantum circuits. Consequently, we deliberately avoid more advanced optimization techniques like RUS for these initial benchmarks. This decision is twofold: firstly, to offer a reasonable "raw" starting point for researchers to evaluate their own optimization methods, and secondly, because RUS circuits inherently require ancilla qubits and non-deterministic mid-circuit measurements to achieve their improved T-counts. By focusing on unitary decompositions for this stage, we restrict our attention to baseline circuit structures.

\subsubsection{The Clifford+T Computational Model} %
\label{sec:clifford_t_model}

The Clifford+T model is a leading computational model for universal fault-tolerant quantum computation \cite{selinger2012efficient}. It comprises the use of Clifford logical operators supplemented by the T-gate~\cite{gidney2017slightly, litinski2019game}. The Solovay-Kitaev theorem establishes that any single-qubit unitary operation can be approximated to arbitrary precision using a finite sequence of gates from such a fixed, finite set of gates provided they generate a dense subgroup of SU(2) \cite{kitaev1997quantum, dawson2005solovay}. Consequently, an arbitrary circuit can be compiled into sequences of Clifford and T operations. 

In prior research, the resource cost of algorithms in the Clifford+T model was often characterized predominantly by the T-gate count, as T-gates were substantially more expensive than Clifford gates and required large space-time overheads ~\cite{beverland2022assessing}. However, ongoing research continuously aims to reduce the cost of T-state generation. Significant advances such as magic state cultivation~\cite{gidney2024magic}, improving upon prior works~\cite{lee2024low, litinski2019magic}, are narrowing the relative cost difference between performing certain Clifford operations and implementing T-gates. Despite these improvements, compiling algorithms into the Clifford+T gate set and estimating the associated overhead for a specific QEC code remains a non-trivial task, requiring code-dependent compilation strategies and resource accounting.

\subsubsection{Pauli Based Computation}
\label{sec:pbc_model}

Pauli Based Computation offers a computational model for universal quantum computation, utilizing adaptive Pauli product measurements and magic states, most commonly T-states \cite{bravyi2016trading,litinski2019game}. In PBC, computation proceeds via a sequence of non-destructive, varying weight PPMs. Measurements drive PBC computation, and feed-forward operations conditioned on prior PPM measurements realize the desired computation. A circuit expressed in the Clifford+T gate set can be compiled into the PBC model. Compiling into PBC leverages commutation rules to effectively "push" Clifford gates through the circuit until they are commuted past final measurements. 

Underpinning PBC is the representation of operators as rotations around a Pauli axes, generally of the form \(R_P(\theta) = \exp(-i\frac{\theta}{2}P)\) for a Pauli operator \(P\) and an angle \(\theta\)~\cite{litinski2019game}. For instance, common Clifford and T-gates can be expressed as sequences of such rotations, such as \(H = R_Z(\pi/2) R_X(\pi/2) R_Z(\pi/2)\), \(S = R_Z(\pi/2)\), and \(T = R_Z(\pi/4)\).

\begin{figure}
    \centering
    \includegraphics[width=0.98\linewidth]{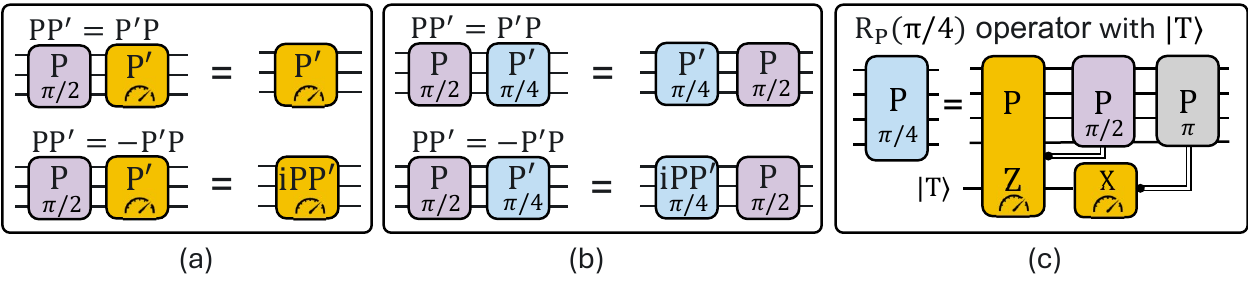}
    \caption{\textbf{(a) Measurement absorption Rule:} A Clifford rotation immediately before readout can be absorbed into the final measurement: if it \emph{commutes} with the chosen measurement basis, it is eliminated; otherwise, the measurement basis is updated accordingly.  \textbf{(b) Commutation Rule:} Moving a Clifford (purple) past a \(T\) rotation (blue) is free when they commute; if not, the \(T\)-frame changes the measured Pauli from \(P\) to \(P' = T P T^\dagger\) (up to a phase, e.g., \(iPP'\)).  \textbf{(c) \(T\) rotation implementation:} Realize \(T\) rotation by consuming a \(|T\rangle\) ancilla state: perform a joint measurement \(P \otimes Z\) between data and ancilla, then apply a conditional Clifford correction (which can be commuted to the end of the circuit). Finally, measure the ancilla and apply the required Pauli correction by recording a Pauli-frame update in software.}
    
    \label{fig:FTCircuitbench_PBC}
\end{figure}

A controlled-\(P_2\) operation on the target qubit, controlled by \(P_1\) on the control qubit (denoted \(C(P_1, P_2)\)), can also be decomposed in this manner. For example, a CNOT gate, \(C(Z,X)\), can be constructed from Pauli rotations. A general \(C(P_1, P_2)\) can be constructed from \(\pm \pi/2\) rotations in the following form \cite{litinski2019game}:
\begin{equation*}
C(P_1, P_2) = R_{P_1}(\pi/2) R_{P_2}(\pi/2) R_{P_1 \otimes P_2}(-\pi/2) R_{P_2}(-\pi/2) R_{P_1 }(-\pi/2).
\end{equation*}
On the other hand, realizing a $R_P(\pi/4)$ operator, i.e. a non-Clifford Pauli product rotation, requires the use of $|T\rangle$ magic states. This is realized by the circuit outlined in Figure~\ref{fig:FTCircuitbench_PBC}~\cite{litinski2019game,litinski2018quantum}

The resultant compiled circuit in PBC allows for the complete removal of Clifford gates from the circuit by repeatedly applying commutation rules. When a Clifford rotation \(R_P(\theta_C)\) encounters a non-Clifford rotation \(R_{P'}(\theta_{NC})\):
\begin{itemize}
    \item If \(P\) and \(P'\) commute (\(PP' = P'P\)), \(R_P(\theta_C)\) can be moved past \(R_{P'}(\theta_{NC})\) without changing \(R_{P'}(\theta_{NC})\).
    \item If \(P\) and \(P'\) anticommute (\(PP' = -P'P\)), commuting \(R_P(\theta_C)\) past \(R_{P'}(\theta_{NC})\) effectively transforms \(R_{P'}(\theta_{NC})\) into \(R_{i PP'}(\theta_{NC})\) The Clifford rotation \(R_P(\theta_C)\) is effectively moved past the non-Clifford
\end{itemize}

Through repeated application of these rules, the circuit is reduced to a sequence of non-Clifford PPMs of varying Pauli weights. PBC thus trades the explicit execution of many Clifford gates for potentially more complex (higher-weight) PPMs and a modified sequence of non-Clifford gates. This approach can offer advantages in resource efficiency for certain QEC codes, algorithms, and hardware architectures, particularly those that natively support or can efficiently implement multi-qubit Pauli measurements~\cite{he2025extractors,kan2025sparo}.

\subsubsection{Optimizing Fault-Tolerant Algorithms}

Minimizing the resource overheads of fault-tolerant computation is a natural goal for reducing overhead. This process typically proceeds through a sequence of distinct stages, from the synthesis of a high-level algorithm into its decomposed gate sequence \cite{li2022paulihedral}, to its compilation onto hardware \cite{wang2024atomique}. Each optimization aims to iteratively reduce the global implementation cost of an algorithm, reducing system overheads.

A fundamental challenge in this multi-stage process is that local optimizations performed at one level of the compilation stack do not guarantee global optimality. Decisions made early in the pipeline can have significant and often unforeseen consequences on the resource requirements of the final implementation. The importance of a more holistic approach is well-documented in the broader field of quantum computing, where co-design and co-optimization strategies that bridge multiple compilation layers have demonstrated substantial performance gains~\cite{decker2025kernpiler,li2022paulihedral,jin2024tetris,murali2020software,murali2019noise,wang2024atomique}. While a single, global optimization across the entire stack is generally intractable, these results highlight the importance of co-designed optimization. 

While the landscape of FTQC compilation is more nascent compared to that of the NISQ era \cite{Preskill2018quantumcomputingin}, it introduces a distinct set of optimization challenges that can supersede those of earlier paradigms. The resource cost is no longer driven primarily by simple gate counts or circuit depth, but by factors unique to fault-tolerant execution. These include the overhead of synthesizing algorithms into a restricted logical gate set and the complex spatial-temporal scheduling of operations required by the underlying error-correcting code. Prior works have begun tackling this optimization and compilation problem, targeting these new bottlenecks. For example, techniques have been developed to optimize the execution of PPMs through commutation-aware scheduling~\cite{litinski2019game}, while others focus on developing schedulers that minimize the spatio-temporal resources for surface code computations~\cite{tan2024sat}. The success of these targeted approaches underscores the impact that compilation strategies have on the ultimate resource cost of a fault-tolerant algorithm, establishing compiler optimization as an emerging and important research area for FTQC.

\subsection{Performing Fault-Tolerant Logical Computation}

 
Once a quantum algorithm is expressed as a sequence of logical operations, it must be executed fault-tolerantly using a quantum error-correcting (QEC) code. Although computational models are code-agnostic, each QEC code admits a specific set of native, fault-tolerant primitives, and implementing logical operations within this framework can incur varying overhead. Such overhead is commonly quantified by the space-time volume, defined as the product of the number of active qubits and the number of code cycles (time steps) during which they are engaged. A practical goal, therefore, is to pair a computational model with a QEC code whose structure and native operations align naturally, thereby minimizing resource costs. As mentioned, we remain in the Clifford+T and PBC computational model \cite{horsman2012surface,fowler2018low,gidney2024inplace,bravyi2024high,cross2024improved,yoder2025tour}.

For the Clifford+T gate set, single-qubit logical operations align well with the surface code structure, as logical qubits can be individually addressed with well defined X, Z, H, and S operators, and ancillary states such as $\ket{T}$ can be prepared using established techniques~\cite{gidney2024inplace,gidney2024magic}. Two-qubit entangling operations are performed via lattice surgery, where the dominant overhead arises from constructing ancilla patches to connect qubits that are physically distant \cite{kan2025sparo}.  Conversely, since the gross code logical qubits are encoded together in the same block, implementing logical Clifford+T operations fault-tolerantly on BB codes presents very different challenges. All logical Clifford gates can often be realized efficiently via code automorphisms, and the gross code supports the full Clifford group~\cite{bravyi2024high, cross2024improved}. Non-Clifford gates are supplied through magic-state techniques. That said, while certain Pauli based computation constructions achieve exactly one Pauli product measurement per T-gate, in practice the PPM cost of both Clifford and T operations depends on the specific Pauli string: automorphisms do not synthesize all required PPMs, and non-native strings must be obtained by conjugating and composing native rotations, which can necessitate multiple PPMs~\cite{litinski2019game,cross2024improved}.

For PBC~\cite{bravyi2016trading,litinski2019game}, the main challenge shifts to the realization of high-weight PPMs. In surface-code architectures~\cite{horsman2012surface}, high-weight PPMs generally require complex lattice-surgery sequences with ancillas that connect all participating data qubits, which can introduce substantial logical-error overhead and even dominate the error budget in practice based on the layout of the logical qubit on hardware~\cite{kan2025sparo}; more broadly, many code families exhibit a trade-off wherein higher-weight logical operators incur higher intrinsic logical error rates.  In contrast, a key advantage of the BB code is that the logical error rate is largely independent of operator weight. Logical operators are accessed not through ancilla path constructions, but rather by performing a sequence of PPMs that yield the desired operator~\cite{cross2024improved}. As long as the logical operator fits within a single code patch, its fidelity remains stable; degradation occurs only when applying inter-block logical operators whose weight exceeds the patch size~\cite{he2025extractors}.  This observation suggests that a primary drawback often attributed to PBC may be less severe for BB codes, positioning them as strong candidates for exploration within the PBC framework.

\subsubsection{Performing Logical Clifford+T Gates on Surface Code}

The state-of-the-art method for implementing logical gates in the surface code is lattice surgery, which enables entanglement between logical qubits by temporarily merging and splitting code patches to realize joint measurements between them \cite{horsman2012surface, fowler2018low, litinski2018lattice}. Although some Clifford gates can be executed transversally with minimal overhead, others must be realized through lattice-surgery–based measurement constructions, leading to non-uniform costs within the Clifford+T gate set.

\textbf{X, Z, and H gates:}
Logical X, Z, and H operations can be applied transversally by performing the corresponding physical gate on every data qubit. From a space–time perspective, these transversal operations can be absorbed into the first or last step of a syndrome extraction round, effectively introducing no additional operational overhead. Moreover, logical X and Z gates can be tracked entirely in software through the Pauli frame, eliminating the need for physical implementation and thereby avoiding additional error sources. In contrast, the transversal Hadamard H gate has the side effect of exchanging the logical X and Z operators, which may require a reorientation of the surface code patch and thus incur extra overhead.

\textbf{S and T-Gates:}
 S and T-gates are realized via teleportation from high-fidelity resource states: a logical $\ket{Y}=\tfrac{\ket{0}+i\ket{1}}{\sqrt{2}}=S\ket{+}$ for the S-gate \cite{gidney2024inplace} and a logical magic state $\ket{T}=\tfrac{\ket{0}+e^{i\pi/4}\ket{1}}{\sqrt{2}}=T\ket{+}$ for the T-gate~\cite{gidney2024magic,litinski2019magic}. 
Concretely, the T-gadget~\cite{zhou2000methodology,bravyi2016improved} applies a $\mathrm{CNOT}$ with the data qubit as control and the $\ket{T}$ ancilla as target, measures the ancilla in the X basis, and then performs the classically controlled correction $S^{m}$ on the data, where $m\in\{0,1\}$ is the measurement outcome (hence a $50\%$ chance to apply S). An analogous gadget teleports the S gate using an ancilla $\ket{Y}$: after the same $\mathrm{CNOT}$ and X-basis measurement, apply a $Z^{m}$ correction on the data, which can be absorbed into the Pauli frame and tracked in software.

\textbf{CNOT gate:} 
To implement a logical two-qubit $\mathrm{CNOT}$ or a single-control, multi-target $\mathrm{CNOT}$ with targets $\{t_1,\ldots,t_n\}$, the protocol typically involves: (1) performing joint $XX$ parity measurements between an ancilla qubit and each target $t_j$ via lattice surgery, followed by a conditional Z gate on the control qubit; (2) performing a joint $ZZ$ parity measurement between the ancilla qubit and the control qubit via lattice surgery, followed by a conditional Z gate on every target $t_j$; and (3) measuring the ancilla in the X basis and applying a conditional Z gate on the control qubit. This process makes the overhead strongly dependent on layout and connectivity.

By contrast, on platforms with all-to-all connectivity (e.g., neutral-atom arrays), a logical $\mathrm{CNOT}$ can be executed transversally directly between the data blocks of two logical qubits, eliminating the need for an ancilla patch and reducing the number of syndrome-extraction rounds, thereby lowering space-time overhead~\cite{PhysRevA.100.012312, zhou2025low}. However, the induced correlated error channels across data qubits necessitate dedicated decoders beyond the standard MWPM approach~\cite{cain2024correlated}.

\begin{figure}
    \centering
    \includegraphics[width= \linewidth]{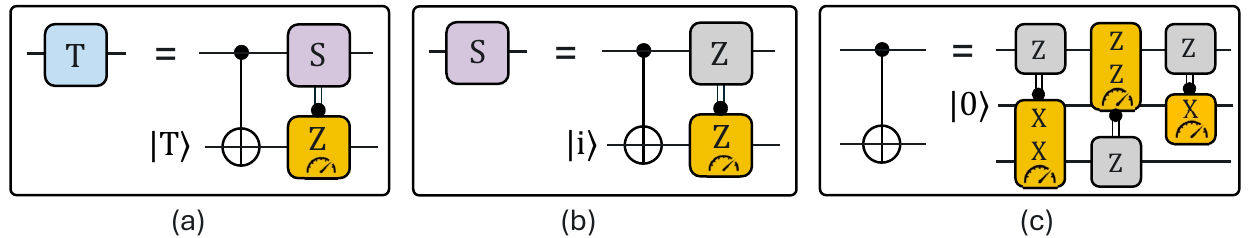}
    \caption{ \textbf{(a) \(T\)-gadget}~\cite{zhou2000methodology,bravyi2016improved}: realizes a logical \(T\) on the surface code by consuming a high-fidelity \(|T\rangle\) state via a CNOT and measurements; with 50\% probability an \(S\) correction is applied to convert \(T^\dagger\) to \(T\).\textbf{ (b) \(S\)-gadget}: analogous construction using a \(|Y\rangle\) resource to implement the logical \(S\) gate. \textbf{(c) Logical CNOT by lattice surgery}: a three-step lattice surgery protocol employing an ancilla patch to set the measurement basis and mediate the entangling operation.}
    
    \label{fig:Logical_T}
\end{figure}

\subsubsection{Performing Pauli Based Computation in the Gross Code}
\label{sec:pbc_actual_compute}

As described previously, Pauli Based Computation involves compiling quantum circuits into sequences of Pauli product rotations, denoted as $R_{\mathcal{P}}(\theta) = \exp(-i\theta \mathcal{P}/2)$, where $\mathcal{P} = P_1 \otimes P_2 \otimes \dots \otimes P_N$ is an $N$-qubit Pauli operator. Each $P_i$ is a single-qubit Pauli operator selected from $\{I, X, Y, Z\}$ acting on the $i$-th qubit. While Clifford rotations (e.g., $\theta = \pi/2$ or $\pi$ for a Pauli operator $\mathcal{P}$) can be implemented without magic states, non-Clifford rotations, such as $R_{\mathcal{P}}(\pi/4)$, are necessary and typically require resource magic states.

One approach to realize these operations is through a measurement-based protocol utilizing an ancilla qubit prepared in a magic state, most commonly the $|T\rangle$ state. The protocol described below aims to implement the $R_{\mathcal{P}}(\pi/4)$ rotation on the data qubits by performing joint Pauli measurements on the data and ancilla system, followed by classically-controlled Clifford corrections. To implement this, let $\mathcal{P} = P_1 \otimes P_2 \otimes \dots \otimes P_N$ be the $N$-qubit Pauli operator acting on the data qubits.
The ancilla qubit is denoted by $\gamma$. The protocol is:

\begin{enumerate}
    \item \textbf{State Preparation:} Prepare the data qubits in an arbitrary state $|\psi\rangle_D$ and the ancilla qubit $\gamma$ in the $|T\rangle$ state:
    $|\Psi_0\rangle = |\psi\rangle_D \otimes |T\rangle_\gamma = |\psi\rangle_D \otimes \frac{1}{\sqrt{2}}(|0\rangle_\gamma + e^{i\pi/4}|1\rangle_\gamma)$.

    \item \textbf{First Joint Measurement:} Measure the operator $M_1 = \mathcal{P}_D \otimes Z_\gamma$. Let the outcome be $m_1 \in \{+1, -1\}$.

    \item \textbf{First Correction:} If the measurement outcome $m_1 = +1$, apply the Clifford rotation $U_{C1} = R_{\mathcal{P}}(\pi/2)$ to the data qubits. If $m_1 = -1$, do nothing.

    \item \textbf{Second Measurement:} Measure the operator $M_2 = I_D \otimes X_\gamma$ (i.e., measure X on the ancilla qubit $\gamma$). Let the outcome be $m_X \in \{+1, -1\}$.

    \item \textbf{Second Correction:} If the measurement outcome $m_X = +1$, apply Clifford operation $U_{C2} = R_{\mathcal{P}}(\pi)$ to the data qubits. If $m_X = -1$, do nothing.
\end{enumerate}

Conditional Clifford corrections, arising from these measurement outcomes, can be commuted through subsequent gates out of the circuit. Rather than needing to apply these Cliffords, we can simply commute them out of the circuit. These logical measurements must themselves be fault-tolerant, as their outcomes potentially modify future operations. This is where the decoder must be in sync with the code. Crucially, realizing non-Clifford rotations via ancilla magic state preparation still requires an auxiliary ancilla system capable of producing them with high fidelity~\cite{stein2025hetec, he2025extractors, swaroop2024universal}. 

%% file: text/algorithms.tex
\section{Benchmark Algorithm Details}
In this section, we describe the algorithms that we include in the FTCircuitBench library. These specific algorithms were chosen in recognition of their expected impact in the era of fault-tolerant quantum computation.

\label{sec:algos_of_interest}

\subsection{Quantum Simulation}

Simulating complex quantum systems stands as one of the core strengths of quantum computing. As quantum computers are naturally well-suited to simulate quantum mechanics, whereas classical approaches often face intractabilities, many current proposals and demonstrations of quantum advantage focus on simulating the time dynamics of interacting quantum systems \cite{Andersen_2025, Kim_2023, King_2025}. In addition, the ability to perform these simulations is of central importance to many fields, notably quantum chemistry~\cite{Cao_2019, mcardle2020quantum, Lee_2021, Santagati_2024} and materials science~\cite{DeLeon_2021, Bauer_2020, Bassman_2021}.

\subsubsection{Quantum Simulation Algorithms}
Quantum simulation is the process of evolving an initial state $|\psi_0\rangle$ under a Hamiltonian $H$ to produce the time-evolved state $e^{-iHt} |\psi_0 \rangle$. This process is used either directly to probe the time dynamics of a system, or as a subroutine in phase estimation to extract the energy spectrum of $H$. A widely adopted approach for time evolution is \emph{Trotterization}, which decomposes $e^{-iHt}$ into a sequence of short-time evolutions generated by the constituent terms of Hamiltonian \cite{trotter1959product, Lloyd_1996_Universal, Childs_2021_Theory}. This affords a conceptually simple and experimentally accessible method for realizing the time evolution.

Although Trotterization is not asymptotically optimal, with more advanced techniques like quantum signal processing~\cite{low2017optimal} and linear-combinations-of-unitaries~\cite{Berry2015Simulating} offering better theoretical scaling, it remains favored in practice due to its simplicity, generality, and significant opportunity for optimization. Accordingly, our benchmark efforts focus on Trotterized simulation as a primary target for optimization and co-design.


\subsubsection{Utility in Quantum Chemistry and Materials Science}

In quantum chemistry, quantum mechanical principles are used to model and predict the structure, energies, and properties of molecules~\cite{mcardle2020quantum}. Many of these tasks ultimately reduce to computing electronic energies with high accuracy, which is particularly challenging in regimes of strong electron correlation, where gold-standard classical methods (e.g., full configuration interaction) scale prohibitively with system size \cite{SzaboOstlund1996, Helgaker2000MolecularElectronicStructure, Scuseria2012StrongCorrelation}. As a result, quantum chemical systems are particularly promising candidates for simulation on quantum computers, with potential utility in battery design and drug discovery.

On the modeling side, electronic structure problems are described by a Hamiltonian of many electrons interacting with nuclei, often treated in a second quantized basis~\cite{Helgaker2000MolecularElectronicStructure}. Strongly correlated electronic behavior can often also be captured by simplified lattice models such as the Fermi-Hubbard and \(t\)–\(J\) models~\cite{Arovas2022HubbardRMP, Zhang1988tJ}. These models capture phenomena beyond mean-field approximations and serve as benchmarks for quantum simulation.

Similarly, condensed matter physics models solid state systems and materials from first principles, often starting from a many-body Hamiltonian that describes the electrons, phonons, and other degrees of freedom~\cite{DeLeon_2021, bauer2020quantum, Bassman_2021}. These models capture important phenomena like magnetism and superconductivity, and remain challenging for classical computational methods at scale. Discretization approaches are used to represent space as a lattice of sites on which the Hamiltonian acts. These models play an important role in demonstrating quantum advantage.

\subsubsection{Canonical Systems}

Given the above applications, we focus on simulating (i) generic second-quantized electronic-structure Hamiltonians and (ii) representative correlated electron models. 
For electronic structure problems describing molecules, we consider the Hamiltonian in second-quantized form:
\begin{equation}
\label{eq:electronic-structure}
H_{\mathrm{el}}
\;=\;
\sum_{p q} t_{p q}\,
c^{\dagger}_{p} c_{q}
\;+\;
\frac{1}{2}
\sum_{p q r s}
V_{p q r s}\,
c^{\dagger}_{p} c^{\dagger}_{q} c_{r} c_{s},
\end{equation}
where $c^{\dagger}_{p}$ ($c_{p}$) creates (annihilates) an electron in
spin-orbital $p$, and $t_{pq}$ and $V_{pqrs}$ encode the single-particle and two-body terms, respectively. After mapping this system to qubits, and using Trotterization to approximate time evolution, we obtain the digital quantum-simulation algorithm of the target Hamiltonian.

Although ~\eqref{eq:electronic-structure} provides an accurate description of an electronic system, it often becomes overly complex, which warrants use of simpler models that capture a subset of the physics at a fraction of the complexity. A first simplification is to imagine the atoms comprising the molecule placed on a lattice and restrict dynamics to two basic processes: an electron may hop to a neighboring atom, and two electrons on the same atom incur an energy penalty \cite{martin2004electronic}. This simplified picture captures the competition between delocalization and onsite repulsion, leading to the \emph{Fermi-Hubbard model}
\begin{equation}
\label{eq:fermi-hubbard}
H_{\text{FH}}
  = -t \sum_{\langle i j\rangle,\sigma}
      \bigl(c^{\dagger}_{i\sigma} c_{j\sigma} + \text{h.c.}\bigr)
    + U \sum_{i} n_{i\uparrow} n_{i\downarrow} ,
\end{equation}
where $t$ describes the hopping strength, and $U$ the onsite energy penalty.  

In the limit of strong interactions $U \gg t$, and an average density of one electron per atom, the electrons effectively stop hopping between sites. The resulting dynamics are described by the spins of the electrons~\cite{auerbach1994interacting}, which is captured by the \textit{Heisenberg model}:  
\begin{equation}
\label{eq:heisenberg}
H_{\text{Heis}}
  = J \sum_{\langle i j\rangle}
      \mathbf S_i \!\cdot\! \mathbf S_j, 
\end{equation}
Here, $\mathbf S_i$ is the spin-$\tfrac12$ operator on site $i$, and $J$ sets the strength of interaction between neighboring spins. Despite its simplicity, the Heisenberg model captures strongly-correlated behavior seen in real materials and spans regimes that range from classically tractable to computationally challenging, making it a key benchmark for quantum simulation.

This model can be further simplified by focusing on only the spin along a single axis, which is traditionally chosen to be the Z-axis. In this limit, and upon introducing an external magnetic field, the system reduces to the \textit{Ising model}:
\begin{equation}
\label{eq:ising}
H_{\text{Ising}}
  = J_{z} \sum_{\langle i j\rangle} S^{z}_{i} S^{z}_{j}
    - h \sum_{i} S^{z}_{i},
\end{equation}
where $J_z$ describes the strength of the nearest-neighbor interaction, and $h$ is the strength of the external magnetic field. Similar to the Heisenberg model, the Ising model and its variants support a rich landscape of phases and dynamics in many-body systems~\cite{Kim_2023}.  



\subsection{Common Quantum Subroutines}

Beyond quantum simulation, quantum algorithms can address a variety of other computational problems. In developing these applications, certain subroutines appear frequently, notably arithmetic operations and the Quantum Fourier transform.

\subsubsection{Quantum Arithmetic}

Quantum adders implement summation of numbers represented by quantum states, typically represented in binary in the computational basis. This capability serves as a subroutine in more advanced computations, such as multiplication, modular exponentiation (as used in Shor's algorithm), and other algebraic manipulations used in quantum simulation and optimization \cite{gidney2025factor, su2021fault}. 

Thus, quantum adders provide important benchmarks for several reasons. For one, their performance reflects the overhead of implementing essential operations such as CNOTs and the non-Clifford resources required for Toffoli gates. In addition, because adder circuits are relatively compact, they allow for straightforward immediate comparisons of resource trade-offs, commonly evaluated in terms of their required logical qubits, circuit depths, and T-gate counts. 
For these reasons, \sol{} include various adder designs to explore these trade-offs in fundamental arithmetic operations.

\subsubsection{Quantum Fourier Transform}

The Quantum Fourier Transform (QFT) is a foundational subroutine that underlies many quantum algorithms. Analogous to the ordinary Fourier transform, the QFT maps states in the computational basis to corresponding states in the frequency basis. On an $n$-qubit state, this can be achieved using only $\mathcal{O}(n^2)$ primitive gates, offering an exponential improvement over the classical Fast Fourier Transform, which incurs a complexity $\mathcal{O}(N\log N)$ when acting on $N = 2^n$ amplitudes~\cite{nielsen2010quantum}. 

The QFT plays a crucial role in several larger quantum algorithms. Most notably, it is used in phase estimation to extract the eigenphases of a unitary operator. Through phase estimation, it also serves as a key component of Shor's factoring algorithm, where it is used to find hidden periodicities in modular exponentiation. In addition, the QFT can be used in Hamiltonian simulation, where working in the momentum basis simplifies evolution under kinetic energy terms. Beyond these instances, the QFT and its variants are used for solving discrete logarithms, hidden subgroup problems, and even quantum arithmetic routines in Fourier space~\cite{nielsen2010quantum, childs2017lecture, draper2000addition, Ruiz_2017_Quantum}. Given the broad utility of the QFT, we include various implementations of this subroutine in FTCircuitBench.

\subsection{Other Quantum Algorithms}

Having introduced commonly used quantum subroutines, we can now discuss three more advanced quantum algorithms: phase estimation, linear system solvers, and the quantum singular value transformation.

\subsubsection{Quantum Phase Estimation}
Quantum phase estimation (QPE) aims to find the eigenphase of a unitary operator $U$, given access to one of its eigenstates $|u\rangle$. That is, given $U|u\rangle = e^{2\pi i \theta } |u\rangle$, the goal is to determine $\theta$~\cite{nielsen2010quantum}. In systems in physics and chemistry, phase estimation is used to estimate the ground state energy of a Hamiltonian, provided access to an approximate ground state~\cite{mcardle2020quantum, Kang_2022_Optimized}. In these settings, the unitary $U$ is often the time evolution operator $U = e^{-iHt}$ or the quantum walk operator $U=e^{i\arccos(H/\lambda)}$ (for a norm $\lambda \geq \| H\|$), both of which encode the spectrum of the Hamiltonian in their eigenvalues. 

The standard form of QPE uses repeated controlled applications of $U$, a quantum Fourier transform, and a set of ancilla qubits to read out the phase $\theta$ in binary. However, several alternative approaches exist, such as Kitaev's iterative phase estimation~\cite{kitaev1995quantum, kitaev2002classical}, statistical approaches to phase estimation~\cite{Lin_2022_Heisenberg, Wiebe_2016_Efficient}, and a variety of randomized techniques~\cite{Wan_2022_Randomized, Campbell_2019}. For simplicity, we include the standard implementation of QPE in FTCircuitBench.



\subsubsection{Quantum Linear Systems Solvers}
Linear systems of equations are ubiquitous across science and engineering, and improvements in solving them can provide far-reaching benefits. Classically, solving a general $N \times N$ linear system $A\Vec{x} = \Vec{b}$ has a worst-case complexity of $\mathcal{O}(N^3)$. In contrast, the quantum algorithm for this problem, initially proposed by Harrow, Hassidim, and Lloyd (HHL) \cite{Harrow_2009}, prepares a quantum state $\ket{x}$ with amplitudes proportional to the solution vector $\Vec{x}$ and attains a complexity $\mathcal{O} \left( \mathrm{polylog}(N)\frac{s^2\kappa^2}{\epsilon} \right)$. In this expression, $s$ is the sparsity of $A$ (maximum number of non-zero entries per row/column), $\kappa$ is the condition number of $A$ (ratio of largest to smallest singular value), and $\epsilon$ is the target precision. Consequently, this polylogarithmic dependence on $N$ offers the potential for an exponential speedup over classical algorithms, particularly for large, sparse, and well-conditioned systems.

However, the HHL algorithm is best used to evaluate an expectation value or statistic on the solution state $\ket{x}$, rather than read out all of its entries, which would require an expensive tomography procedure that could erase its speedup.  Moreover, the HHL algorithm itself is a composition of several fundamental quantum subroutines, namely Hamiltonian simulation, the quantum Fourier Transform, and quantum phase estimation. This composite structure makes it a complex yet informative benchmark for quantum compiler performance, and thus we include an HHL implementation in FTCircuitBench.

\subsubsection{Quantum Singular Value Transformation}

The quantum singular value transformation (QSVT)~\cite{gilyen2019quantum} provides a framework to implement arbitrary polynomial functions of operators (e.g., a Hamiltonian $H$). More specifically, given a unitary $U_A$ that block-encodes a matrix $A$ (i.e., $A$ is encoded in a sub-block of $U$), QSVT transforms the singular values of $A$ by a tunable polynomial $P(x)$. This capability is very general and affords a unified perspective of quantum algorithms~\cite{martyn2021grand}. Remarkably, many quantum algorithms constructed through QSVT achieve near-optimal complexities and broad utility, including algorithms for Hamiltonian simulation~\cite{low2017optimal}, linear systems solvers~\cite{gilyen2019quantum}, Gibbs sampling~\cite{gilyen2019quantum}, state preparation~\cite{obrien2025quantum}, topological data analysis~\cite{berry2024analyzing, Hayakawa2022quantum}, and amplitude amplification~\cite{yoder2014fixed}. We include an implementation of QSVT in FTCircuitBench, with its functionality demonstrated through applying QSVT to a linear system solver.

%% file: tables/code_example.tex
\begin{lstlisting}[language=Python, caption={Example FTCircuitBench Pipeline.}]
from ftcircuitbench import (
    show_clifford_t_interaction_graph,
    show_operator_weight_histogram,
)
from ftcircuitbench.api import PipelineConfig, run_analysis_for_file
from ftcircuitbench.resource_estimation import (
    estimate_circuit,
    logical_counts_from_stats,
    resolve_hardware_models,
)

# Compile a circuit to Clifford+T with Gridsynth (precision 1e-8) and with
# Solovay-Kitaev (recursion degree 2), then to PBC with layering-and-merging
configs = [
    PipelineConfig(pipeline="gs", gridsynth_precision=8, optimize_pbc=True),
    PipelineConfig(pipeline="sk", sk_recursion=2, optimize_pbc=True),
]
result = run_analysis_for_file("qasm/hhl/hhl_12q.qasm", configs)

# Analyze the Gridsynth Clifford+T circuit
gs = result.pipelines["gs"]
print(gs.clifford_stats["total_t_family_count"])
print(gs.clifford_stats["qubit_interaction_degree"])

# Visualize circuit interaction graph
show_clifford_t_interaction_graph(gs.clifford_t_circuit)

# Analyze the PBC circuit
print(gs.pbc_stats["pbc_rotation_operators"])
print(gs.pbc_stats["pbc_avg_pauli_weight"])

# Visualize Pauli weight distributions
show_operator_weight_histogram(gs.pbc_circuit)

# Bridge to the Azure Quantum Resource Estimator (requires the `qre` extra)
counts = logical_counts_from_stats(
    {**gs.clifford_stats, **gs.pbc_stats}, label="hhl-12q-gs-8"
)
models = resolve_hardware_models(["superconducting 1e-3"])
estimate = estimate_circuit(counts, models)
print(estimate.estimates["superconducting 1e-3"].physical_qubits)
\end{lstlisting}

%% file: tables/ct_stats_table_alphabetical.tex
\begingroup
\fontsize{6.5}{7.5}\selectfont 
\setlength{\tabcolsep}{3pt}

\begin{longtable}{|l|r|r|r|r|r|r|r|r|}
\caption{Clifford+T Circuit Statistics}\label{ct_table}\\

\hline
\makecell{Algorithm \& Pipeline} &
\makecell{Total Gates} &
\makecell{Depth} &
\makecell{Clifford Gates} &
\makecell{$T$ \& $T^{\dagger}$ Gates} &
\makecell{Graph\\Density} &
\makecell{Avg $\pm$ Std\\Degree} &
\makecell{Graph\\Modularity} &
\makecell{Number of\\Communities} \\ \hline
\endfirsthead

\multicolumn{9}{c}{{\bfseries \tablename\ \thetable{} — continued from previous page}}\\
\hline
\makecell{Algorithm \& Pipeline} &
\makecell{Total Gates} &
\makecell{Depth} &
\makecell{Clifford Gates} &
\makecell{$T$ \& $T^{\dagger}$ Gates} &
\makecell{Graph\\Density} &
\makecell{Avg $\pm$ Std\\Degree} &
\makecell{Graph\\Modularity} &
\makecell{Number of\\Communities} \\ \hline
\endhead

\hline \multicolumn{9}{r}{{Continued on next page}} \\ \hline
\endfoot

\hline
\endlastfoot

        adder-10q-gs-5 & 142 & 99 & 86 & 56 & 1.44 & 2.60 ± 1.02 & 0.33 & 3 \\ \hline
        adder-10q-gs-8 & 142 & 99 & 86 & 56 & 1.44 & 2.60 ± 1.02 & 0.33 & 3 \\ \hline
        adder-10q-sk-1 & 157 & 102 & 101 & 56 & 1.44 & 2.60 ± 1.02 & 0.33 & 3 \\ \hline
        adder-10q-sk-2 & 157 & 102 & 101 & 56 & 1.44 & 2.60 ± 1.02 & 0.33 & 3 \\ \hline
        adder-28q-gs-5 & 424 & 189 & 256 & 168 & 0.52 & 2.79 ± 0.94 & 0.66 & 3 \\ \hline
        adder-28q-gs-8 & 424 & 189 & 256 & 168 & 0.52 & 2.79 ± 0.94 & 0.66 & 3 \\ \hline
        adder-28q-sk-1 & 463 & 192 & 295 & 168 & 0.52 & 2.79 ± 0.94 & 0.66 & 3 \\ \hline
        adder-28q-sk-2 & 463 & 192 & 295 & 168 & 0.52 & 2.79 ± 0.94 & 0.66 & 3 \\
\hline
\end{longtable}
\normalsize
\endgroup

%% file: tables/pbc_stats_table_alphabetical.tex
\begingroup
\fontsize{6.5}{7.5}\selectfont 
\setlength{\tabcolsep}{3pt}

\begin{longtable}{|l|r|r|r|l|l|r|l|r|r|}
\caption{PBC Circuit Statistics}\label{pbc_table}\\

\hline
\makecell{Algorithm \& Pipeline} &
\makecell{Raw\\Rotations} &
\makecell{Optimized\\Rotations} &
\makecell{Rotation\\Reduction} &
\makecell{Raw Avg $\pm$ Std\\Pauli\\Weight} &
\makecell{Optimized\\Avg $\pm$ Std\\Pauli\\Weight} &
\makecell{Avg Pauli\\Weight\\Reduction} &
\makecell{Avg $\pm$ Std\\Degree} &
\makecell{Graph\\Modularity} &
\makecell{Number of\\communities} \\ \hline
\endfirsthead

\multicolumn{10}{c}{{\bfseries \tablename\ \thetable{} — continued from previous page.}}\\
\hline
\makecell{Algorithm \& Pipeline} &
\makecell{Raw\\Rotations} &
\makecell{Optimized\\Rotations} &
\makecell{Rotation\\Reduction} &
\makecell{Raw Avg $\pm$ Std\\Pauli\\Weight} &
\makecell{Optimized\\Avg $\pm$ Std\\Pauli\\Weight} &
\makecell{Avg Pauli\\Weight\\Reduction} &
\makecell{Avg $\pm$ Std\\Degree} &
\makecell{Graph\\Modularity} &
\makecell{Number of\\Communities} \\ \hline
\endhead

\hline \multicolumn{10}{r}{{Continued on next page}} \\ \hline
\endfoot

\hline
\endlastfoot

        adder-10q-gs-5 & 56 & 32 & 42.86\% & 4.17 ± 2.79 & 5.40 ± 2.84 & -29.71\% & 9.00 ± 0.00 & 0.02 & 2 \\ \hline
        adder-10q-gs-8 & 56 & 32 & 42.86\% & 4.17 ± 2.79 & 5.40 ± 2.84 & -29.71\% & 9.00 ± 0.00 & 0.02 & 2 \\ \hline
        adder-10q-sk-1 & 56 & 32 & 42.86\% & 4.17 ± 2.79 & 5.40 ± 2.84 & -29.71\% & 9.00 ± 0.00 & 0.02 & 2 \\ \hline
        adder-10q-sk-2 & 56 & 32 & 42.86\% & 4.17 ± 2.79 & 5.40 ± 2.84 & -29.71\% & 9.00 ± 0.00 & 0.02 & 2 \\ \hline
        adder-28q-gs-5 & 168 & 96 & 42.86\% & 4.29 ± 2.76 & 5.56 ± 2.76 & -29.50\% & 10.07 ± 2.34 & 0.61 & 3 \\ \hline
        adder-28q-gs-8 & 168 & 96 & 42.86\% & 4.29 ± 2.76 & 5.56 ± 2.76 & -29.50\% & 10.07 ± 2.34 & 0.61 & 3 \\ \hline
        adder-28q-sk-1 & 168 & 96 & 42.86\% & 4.29 ± 2.76 & 5.56 ± 2.76 & -29.50\% & 10.07 ± 2.34 & 0.61 & 3 \\ \hline
        adder-28q-sk-2 & 168 & 96 & 42.86\% & 4.29 ± 2.76 & 5.56 ± 2.76 & -29.50\% & 10.07 ± 2.34 & 0.61 & 3 \\
\hline
\end{longtable}
\normalsize
\endgroup

%% file: tables/pbc_5trotter_stats_table_alphabetical.tex
\begingroup
\fontsize{6.5}{7.5}\selectfont 
\setlength{\tabcolsep}{3pt}

\begin{longtable}{|l|r|r|r|l|l|r|l|r|r|}
\caption{PBC 5-Trotter-Step Hamiltonian Circuit Statistics}\label{pbc_5trotter_table}\\

\hline
\makecell{Algorithm \& Pipeline} &
\makecell{Raw\\Rotations} &
\makecell{Optimized\\Rotations} &
\makecell{Rotation\\Reduction} &
\makecell{Raw Avg $\pm$ Std\\Pauli\\Weight} &
\makecell{Optimized\\Avg $\pm$ Std\\Pauli\\Weight} &
\makecell{Avg Pauli\\Weight\\Reduction} &
\makecell{Avg $\pm$ Std\\Degree} &
\makecell{Graph\\Modularity} &
\makecell{Number of\\communities} \\ \hline
\endfirsthead

\multicolumn{10}{c}{{\bfseries \tablename\ \thetable{} — continued from previous page}}\\
\hline
\makecell{Algorithm \& Pipeline} &
\makecell{Raw\\Rotations} &
\makecell{Optimized\\Rotations} &
\makecell{Rotation\\Reduction} &
\makecell{Raw Avg $\pm$ Std\\Pauli\\Weight} &
\makecell{Optimized\\Avg $\pm$ Std\\Pauli\\Weight} &
\makecell{Avg Pauli\\Weight\\Reduction} &
\makecell{Avg $\pm$ Std\\Degree} &
\makecell{Graph\\Modularity} &
\makecell{Number of\\communities} \\ \hline
\endhead

\hline \multicolumn{10}{r}{{Continued on next page}} \\ \hline
\endfoot

\hline
\endlastfoot

        fermi-hubbard-1d-128q-sk-1 & 40740 & 3040 & 92.54\% & 2.22 ± 1.36 & 1.45 ± 0.50 & 34.43\% & 14.44 ± 5.75 & 0.83 & 16 \\ \hline
        fermi-hubbard-1d-128q-sk-2 & 105840 & 38600 & 63.53\% & 2.28 ± 1.37 & 35.62 ± 28.92 & -1465.01\% & 14.44 ± 5.75 & 0.8 & 16 \\ \hline
        fermi-hubbard-1d-18q-sk-1 & 6083 & 587 & 90.35\% & 2.06 ± 1.28 & 1.93 ± 0.99 & 6.30\% & 11.78 ± 3.07 & 0.38 & 4 \\ \hline
        fermi-hubbard-1d-18q-sk-2 & 16073 & 6057 & 62.32\% & 2.18 ± 1.33 & 9.91 ± 4.41 & -355.36\% & 11.78 ± 3.07 & 0.37 & 4 \\ \hline
        fermi-hubbard-1d-200q-sk-1 & 64288 & 4560 & 92.91\% & 2.47 ± 1.68 & 1.45 ± 0.50 & 41.26\% & 16.46 ± 8.76 & 0.85 & 18 \\ \hline
        fermi-hubbard-1d-200q-sk-2 & 166178 & 60062 & 63.86\% & 2.50 ± 1.67 & 60.89 ± 51.24 & -2337.80\% & 16.46 ± 8.76 & 0.82 & 17 \\ \hline
        fermi-hubbard-1d-72q-sk-1 & 22968 & 1520 & 93.38\% & 2.23 ± 1.39 & 1.45 ± 0.50 & 34.88\% & 17.08 ± 8.23 & 0.71 & 9 \\ \hline
        fermi-hubbard-1d-72q-sk-2 & 59298 & 21222 & 64.21\% & 2.30 ± 1.46 & 30.29 ± 19.65 & -1217.70\% & 17.08 ± 8.23 & 0.67 & 9 \\
\hline
\end{longtable}
\normalsize
\endgroup